\documentstyle[aps,rmp,psfig]{revtex}

\begin{document}
\title{The World of the Complex Ginzburg-Landau Equation}
\author{Igor S. Aranson} 
\address{Materials Science Division, Argonne National Laboratory
9700 S. Cass Av., Argonne, IL60439}
\author{Lorenz Kramer} 
\address{Physikalisches Institut, University of Bayreuth,
Universit\"atstrasse 30, D-95440 Bayreuth, Germany
}
\maketitle
\begin{abstract}
The cubic complex Ginzburg-Landau equation
is one of the most-studied nonlinear
equations in the physics community.
 It describes  a vast variety
of phenomena
from nonlinear waves to second-order phase transitions,
 from superconductivity, superfluidity and Bose-Einstein condensation to liquid crystals and strings
in field theory. 
Our goal is to give an overview of various phenomena described the 
complex Ginzburg-Landau equation in one, two and three dimensions 
from the point of view of condensed matter 
physicists. Our approach is to study the relevant
solutions to get an insight into nonequilibrium phenomena 
in spatially extended
systems.
\end{abstract} 

\tableofcontents

\section{Preliminary Remarks}
\label{sec:intro}
\subsection{The equation}
The cubic complex Ginzburg-Landau equation (CGLe) 
is one of the most-studied nonlinear 
equations in the physics community.
It describes on a qualitative, and often even on a quantitative
 level a vast variety of phenomena
from nonlinear waves to second-order phase transitions,
 from superconductivity, superfluidity and Bose-Einstein condensation to liquid crystals and strings
in field theory 
(Kuramoto, 1984, Cross and Hohenberg, 1993, Newell {\it et al}, 1993,
Pismen, 1999, 
Bohr {\it et al}, 1998, Dangelmayr and Kramer, 1998).

Our goal is to give an overview of various phenomena described by the CGLe
from the point of view of condensed matter 
physicists. Our approach is to study the relevant 
solutions to get insight into nonequilibrium phenomena in spatially extended
systems.
More elementary and detailed introductions into the concepts underlying the 
equation can be found in Manneville (1990), 
van Saarloos (1993), van Hecke {\it et al} (1994), Nicolis (1995)
and Walgraaf (1997).

The equation is given by
\begin{equation}  \label{CGLe}
\partial_t A = A + (1+ib) \Delta A - (1+ic) |A|^2A,
\label{cgl}
\end{equation}
where $A$ is a complex function of (scaled) time $t$ and space $\vec x$ 
(often in reduced dimension $D=1$ or $2$) and the real
parameters $b$ and $c$ characterize linear and nonlinear dispersion. 
The equation
arises in physics in particular as a ``modulational'' 
(or ``envelope'' or ``amplitude'')
equation. It provides a reduced, universal description of
``weakly nonlinear'' spatio-temporal phenomena in extended (in $\vec x$) 
continuous media 
whose linear dispersion is of a very general type (see below)
and which are invariant under a global change of gauge (multiplication
of $A$ by $\exp(i\Phi)$). This symmetry typically arises when $A$
is the (slowly varying) amplitude of a phenomenon that is periodic
in at least one variable (space and/or time) as a consequence of 
translational invariance of the system. 

The assumptions of slow variation and  weak nonlinearity are valid in particular
near the instability of a homogeneous (in $\vec x$) basic state and Eq.
(\ref{CGLe}) can be viewed as a (generalized) normal form of the 
resulting "primary" bifurcation. 
Then, in analogy with phase transitions, $A$ is often called an order parameter.

To see more clearly 
the analogy with the order parameter concept 
we write the equation in the unscaled form,
\footnote{except maybe for a simple rescaling and rotation of the 
coordinate system, see below}
as derived for example from the underlying set of 
basic (e.g. hydrodynamic) equations
for a definite physical situation
%
\begin{equation}  \label{AGLE}
\tau (\partial_{\tilde t} \tilde A - {\vec v}_g \cdot \tilde \nabla \tilde A) 
= \epsilon (1+ia) \tilde A +\xi ^2 (1+ib) \tilde \Delta \tilde A 
- g(1+ic)|\tilde A|^2 \tilde A,\ \  .
\end{equation}
Equation (\ref{CGLe}) is obtained from (\ref{AGLE}) by the transformations
$\tilde A=(\epsilon/g)^{1/2} A \ exp(-i(\epsilon a/\tau) \tilde t)$, 
$\tilde t=(\tau/\epsilon) t$, and 
$\vec {\tilde x} -{\vec v}_g \tilde t=(\xi/\epsilon^{1/2}) \vec x$.
The case $\epsilon >0$ and $g>0$ was assumed. 
Otherwise the signs in front of the first
and/or last term on the right-hand side of (\ref{CGLe}) have to be reversed. 
The physical quantities ${\bf u} (\tilde t,\tilde r)$ (temperature,
velocities, densities, electric field etc.) are given in the form
\begin{equation} \label{physical}
{\bf u}=\tilde A e^{i({\vec q}_c \cdot 
\vec x-\omega_c t)}{\bf U}_l(\vec z)+c.c.+h.o.t.
\end{equation}
(c.c.$=$ complex conjugate, h.o.t.$=$ higher-order terms). 
If the phenomena occur in (thin) layers,  on surfaces, or in (narrow)
channels, then  ${\bf U}_l$, derived from the linear problem, describes the
spatial  dependence of the physical quantities in the transverse $\vec z$
direction(s). ${\bf U}_l$ is a 
linear eigenvector and $\omega_c,{\vec q}_c$ the corresponding eigenvalues.
In the case of periodically driven systems, ${\bf U}_l$ would include a periodic
time dependence.

In order to identify the character of the various terms in the linear 
part of (\ref{AGLE}) one may also consider the dispersion relation obtained from 
(\ref{AGLE}) and (\ref{physical}) for small harmonic perturbations  
of the basic state $A=0$
\begin{equation}  \label{dispersion1}
\tau \lambda =-i\tau \omega _c+i\vec v _g \cdot (\vec q-\vec q_c)
+\epsilon (1+ia) - \xi ^2 (1+ib)({\vec q}-\vec q_c)^2 .
\end{equation}
Here $\vec q$ is the wavevector in the physical system and 
$\lambda=\sigma-i\omega$ is the complex growth rate of the perturbation. 
$\tau$ is a characteristic time, $\xi$ is the coherence length, 
$\vec v_g$ a linear group velocity, and $\epsilon a/\tau$ a correction 
to the Hopf frequency $\omega_c$.
$\epsilon$ measures in a dimensionless scale the distance from threshold
of the instability, i.e. $\epsilon=(R-R_c)/R_c$, with $R$
the control parameter that carries the system through the threshold at $R_c$. 
Note that there is an arbitrary
overall factor in Eqs. (\ref{AGLE}) and (\ref{dispersion1}) which is
fixed by the ultimately arbitrary choice of the definition of $\epsilon$.
The value of the nonlinear coefficient $g$ in Eq. (\ref{AGLE}) depends on the 
choice of the normalization of the linear eigenvector ${\bf U}_l$.

Now we can proceed to summarize the conditions for validity of the CGLe.
The following four points are to some extent interrelated.

a) Correct choice of order parameter space, i.e. a single complex scalar: 
First of all this necessitates that 
$\omega_c$ and/or ${\vec q}_c$ are  nonzero, because otherwise one would 
expect a real order parameter as in simple phase transitions. 
An exception is the transition to superconductivity and superfluidity 
where the order parameter is complex for quantum mechanical reasons (see below).
Moreover, if $q_c=0$, one may run into problems with conservation laws which 
frequently exclude a homogeneous change of the system. 
In this case of long-wavelength instabilities often somewhat different 
order-parameter equations arise (see e.g. Nepomnyashchii, 1995a).
Secondly, a discrete degeneracy (or near degeneracy) of neutral modes is excluded, 
which may arise by symmetry (see below for an example) or by accident. 
If the eigenvectors of the different modes are different one would need several 
order parameters and a set of coupled equations.
If the eigenvectors coincide (or nearly coincide), which may happen at (or near) 
a co-dimension-2 point, one can again use one equation, which would now contain 
higher space or time derivatives.

b) Validity of the dispersion relation (\ref{dispersion1}): 
since there is no real contribution linear in $\vec q - \vec q_c$ and since
$\xi^2$ is a positive quantity the real growth rate $\sigma$ has a minimum at 
$\vec q_c$. In more than 1D this excludes an important class of systems, 
namely isotropic
ones with $q_c \ne 0$ like Rayleigh-B\'enard convection  in simple 2D fluid
layers. There one has a continuous degeneracy of neutral linear modes. The
neglect of terms of higher order in $\epsilon$ and in $\vec q - \vec q_c$ is
usually justified near the bifurcation.

c) Symmetries: translation invariance in $\vec x$ and $t$. 
Actually the CGLe incorporates translational invariance with respect to space and/or 
time on two levels. One is expressed by the global gauge invariance, which in 
the CGLe can be absorbed in a shift of $\vec x$ and/or time $t$. 
Note that this invariance excludes terms that are quadratic in $A$.
The other is expressed by the autonomy of the CGLe (no explicit 
dependence on space and time). The two invariances reflect the fact that 
the fast and the slow space and time scales are not coupled in
this description. 
This is an approximation which cannot be overcome by going to higher 
order in the expansion in terms of amplitude and gradients. 
The coupling effects are in fact nonanalytic in $\epsilon$ 
(``non-adiabatic effects'', see, e.g. Pomeau (1984), 
Kramer and Zimmermann (1985), Bensimon {\it et al} (1988)). 

d) Validity of the (lowest-order) weakly nonlinear approximation: 
we will deal mostly with the case of a supercritical (``forward'', 
or  ``normal'') 
bifurcation where  $g>0$  and  then higher-order nonlinearities in Eq. (\ref{cgl}) 
can be neglected sufficiently near threshold. 
If the nonlinear term in Eq. (\ref{cgl}) has the opposite sign,  which corresponds 
to a subcritical (``backward'' or ``inverse'') bifurcation, 
higher-order nonlinear terms are usually essential. 
However, even in this case, there exist for sufficiently large values of $|c|$
relevant solutions that bifurcate supercritically,
which will be discussed in Sec. \ref{smampsol}.

From the linear theory we can now distinguish three classes of primary 
bifurcations 
where Eq. (\ref{CGLe}) arises:

i) $\omega_c=0,{\vec q}_c \ne 0$: for such 
{\it stationary periodic} instabilities $\lambda$ is real, and in fact
all the imaginary coefficients (including the group velocity ${\vec v}_g$)
vanish. Generically, reflection symmetry is needed (see below).
Equation (\ref{CGLe}) then reduces to the ``real'' Ginzburg-Landau 
Equation (GLe) 
\begin{equation}  \label{GLe}
\partial_t A= A + \Delta A -|A|^2 A
\end{equation}
which one might also call the ``Complex Nonlinear Diffusion Equation''
in some analogy with the Nonlinear Schroedinger Equation (see below).
Examples that display such an instability are
Rayleigh-B\'enard convection in simple and complex fluids, 
Taylor-Couette flow, electroconvection in liquid crystals and many others.
%
In more than 1D there is the restriction mentioned under b). 
Thus in isotropic 2D systems 
the dispersion relation is changed and the Laplacian
in Eq.(\ref{GLe}) has to be substituted by a different differential
operator. The corresponding equation derived by Newell and Whitehead (1969) 
and by Segel (1969) was in fact the first amplitude equation that included
spatial degrees of freedom. It is applicable only for situations with nearly
parallel rolls, which is in isotropic systems an important restriction.

So in more than one dimension the system must be anisotropic, which is
the case in particular for convective 
instabilities in liquid crystals (Kramer and Pesch, 1995),
but holds also for Rayleigh-B\'enard convection 
 in an inclined layer (Daniels {\it et al}, 2000) 
or in
a conducting fluid in the presence of a 
magnetic field with an axial 
component (Eltayeb, 1971). 
Also the Taylor-Couette instability in the small
gap limit can be viewed as an anisotropic quasi-2D system.
In 2D Eq. (\ref{GLe}) was first considered in the context of
electrohydrodynamic convection in a planarly aligned nematic liquid
crystal layer
(Pesch and Kramer, 1986, Bodenschatz {\it et al}, 1988a, 
Kramer and Pesch, 1995,  for review see also Buka and Kramer,  1996) 
The Laplacian in Eq. (\ref{AGLE}) is obtained after a 
linear coordinate transformation.

ii) $\omega_c \ne 0,{\vec q}_c=0$: The prime example for such
{\it oscillatory uniform} instabilities are 
oscillatory chemical reactions
(see e.g. de Wit, 1999). 
In lasers (or passive nonlinear
optical systems) this type may also arise (Newell and Moloney, 1992). 
In hydrodynamic systems such instabilities  are often suppressed by mass
conservation (see, however, B\"orzs\"onyi {\it et al}, 2000).
Isotropy does not cause any problems here and the Laplacian applies directly. 
In the presence of reflection symmetry the group velocity term in 
Eq. (\ref{AGLE}) 
is absent.
The spatial patterns obtained from
$A$ reflect directly those of the physical system. The imaginary
parts proportional to $b$ and $c$ pertain to linear and nonlinear 
frequency change (renormalization) of the oscillations, respectively. 
In most systems the nonlinear frequency change is negative (frequency
decreases with amplitude), so that $c<0$ with our choice of signs. 
Coefficients for the CGLe have been determined e.g. from experiments
on the Belousov-Zhabotinsky (BZ) reaction (Hynne {\it et al}, 1993, 
Kramer {\it et al}, 1994).

iii) $\omega_c \ne 0,{\vec q}_c \ne 0$: This {\it oscillatory periodic}
instability occurs in hydrodynamic and optical systems. 
The best-studied example
is Rayleigh-B\'enard convection in binary mixtures, although here the 
bifurcation is in the accessible parameter range mostly 
subcritical (Sch\"opf and  Zimmermann,  1990, L\"ucke {\it et al}, 1992).
Also, in 2D, the system is isotropic, so that the simple CGLe 
(\ref{AGLE}) is not applicable. Other 1D examples are the oscillatory 
instability in Rayleigh-B\'enard convection 
in low Prandtl number fluids, which in 2D occurs as a 
secondary instability of stationary rolls. In a 1D geometry with just 
one longitudinal roll it can be treated as a primary bifurcation 
(Janiaud {\it et al}, 1992). 
Other  examples include the 
wall instability in rotating Rayleigh-B\'enard convection (Tu and Cross, 1992, 
van Hecke and van Saarloos, 1997, Yuanming and  Ecke, 1999) 
and hydrothermal waves, where the coefficients of the CGLe were 
determined from experiment (Burguette {\it et al}, 1999). 
In 2D the prime example is the electrohydrodynamic instability in 
nematic liquid crystals in thin and clean cells (otherwise one has the more 
common stationary rolls) (Treiber and Kramer, 1998).  Actually in such an
anisotropic 2D system one is lead to a generalization of Eq. (\ref{CGLe})
where the term $ib\Delta A$ is replaced by a more general bilinear form
$i(b_1 \partial_x^2+b_2\partial_y^2)$, see  Sec. \ref{subsec:aniso}. 

Most of the oscillatory periodic systems just mentioned have reflection 
symmetry and then one has to allow for
the possibility of counter-propagating waves which makes a description in 
terms of two coupled CGLes necessary (see Cross and Hohenberg, 1993).
The degeneracy between left and right traveling waves can be lifted
by breaking the reflection 
symmetry by applying additional fields or an additional flow. 
In this situation one roll system is favored over the other and, if the effect 
is strong enough, a single CGLe can be used.
Breaking reflection symmetry in stationary periodic instabilities (case i)) 
the rolls will generically start to travel and 
one indeed arrives at an oscillatory periodic instability (case iii)).
This has been studied experimentally by applying a through flow
in thermal convection (Pocheau and Croquette, 1984) 
or in the Taylor-Couette system 
(Tsameret and  Steinberg, 1994; 
Babcock,  Ahlers, and  Cannell, 1991) or by non-symmetric
surface alignment in electroconvection of nematics (pretilt or
hybrid alignment, see e.g. Krekhov  and  Kramer, 1996). 

Since the drift introduces a frequency, for sufficiently strongly broken 
reflection symmetry the 
distinction between cases i) and iii) is lost, as is obvious in open-flow systems 
(Leweke  and  Provansal, 1994, 1995; 
Roussopoulos and  Monkewitz, 1996).

The CGLe may also be viewed as a dissipative extension of the conservative
 nonlinear Schr\"odinger equation (NLSe)
\begin{equation} 
i\partial_t A = \Delta A \pm |A|^2 A \ ,
 \label{NLSe}
\end{equation}
which  describes weakly nonlinear wave phenomena (Newell, 1974). 
The prime examples are waves on deep water,
(Dias and Kharif, 1999) 
and nonlinear optics (Newell and Moloney, 1992). 
The conservative limit of Eq. (\ref{CGLe}) is
obtained by letting in Eq. (\ref{AGLE}) ${\vec v}_g ,\ \epsilon,\xi,\ g \to 0$ 
with $\xi^2 b,\ gc$ remaining nonzero, so 
that $b,\ c \to \infty$, and rescaling
space and the amplitude.

\subsection{Historical remarks}

Four key concepts come together in the CGLe philosophy:
\begin{itemize} 
\item  {\it Weak nonlinearity}, which amounts to an expansion in terms of the 
order parameter $|\tilde A|$. This concept goes back to Landau's theory
of second-order phase transitions (Landau, 1937a). 
Landau
also employed this type of expansion in his attempt to explain the 
transition to turbulence (Landau, 1944). In the context of stationary,
pattern-forming, hydrodynamic instabilities the weakly nonlinear expansion 
leading to a solvability condition at third order was introduced by 
Gorkov (1957) and  Malkus and   Veronis (1958). 
One should also mention the  work of Abrikosov 
in 1957 (for review see Abrikosov (1988))  
where he presents the theory of the mixed state 
of type II superconductors in a magnetic field based on 
the Ginzburg-Landau theory of superconductivity. The mixed 
state is a periodic array of flux lines (or vortices) corresponding to
topological defects (see below). Abrikosov introduced a weakly nonlinear 
expansion to describe this state valid near the upper critical field.

\item
{\it Slow relaxative time dependence}  was first used by Landau in the 
above-mentioned paper on turbulence in 1944. 
In the context of pattern-forming instabilities it 
goes back to Stuart (1960).

\item
{\it Slow nonrelaxative time dependence} 
 with nonlinear frequency renormalization 
in the complex-amplitude formulation was introduced by Stuart in 1960
using multi-scale analysis. Of course perturbation theory 
for periodic orbits (in particular conservative Hamiltonian systems) is 
a classical subject that was treated by
 Bogoliubov, Krylov and Mitropolskii in 1937  (for review, see
Bogoliubov and Mitropolskii, 1961).

\item 

{\it Slow spatial dependence}  was included already by Landau
(1937b)  in the context of $X$-ray scattering by crystals 
in the neighborhood of the Curie point.
However, the concept became known with the success of the (stationary)
phenomenological Ginzburg-Landau (GL) theory of superconductivity
[Ginzburg and Landau (1950)]. 
\end{itemize}

The  stationary GL theory for superconductivity has a particular 
resemblance to the modulational theories
of pattern-forming systems because the order parameter is complex,
although for a very different reason. Superconductivity being a 
macroscopic quantum state requires an order parameter that has the 
symmetries of a wave function. In spite of the different origin one has 
many analogies. 
However, in superconductors the time 
dependence is rendered nonvariational primarily through 
the coupling to the electric field due to local gauge invariance (see, e.g. 
Abrikosov, 1988),  
a mechanism that has no analog in
pattern-forming systems. 

The time-dependent GL theory for superconductors was 
presented (phenomenologically) only in 1968 by Schmid,  (derived from
microscopic theory shortly afterwards by Gorkov and Eliashberg (1968)),  when 
the first modulational theory was derived in the context of 
Rayleigh-B\'enard convection by Newell and  Whitehead (1969) 
and Segel (1969).
Eq. (\ref{GLe}) with additional noise term has been studied intensively 
as a model of phase transitions in equilibrium systems, see e.g. 
Hohenberg and Halperin (1977). 

The full CGLe was introduced phenomenologically by 
Newell and Whitehead (1971).  
It was derived by Stewartson and Stuart (1971) and 
DiPrima,  Eckhaus and Segel (1971)
in the context of the destabilization of plane shear flow, where its 
applicability is limited by the fact that one deals with a strongly
subcritical bifurcation. In the context of chemical systems the CGLe was 
introduced by Kuramoto and Tsuzuki  (1974).

There exists an extended mathematical literature on the
CGLe, which we will touch rather little, see e.g.
Doering {\it et al} (1987, 1988), Levermore and  Stark (1997),
Doelman (1995),  van Harten (1991), Milke and Schneider (1996), 
Schneider (1994), Melbourne (1998), Milke (1998).

\subsection{Simple model -- vast variety of effects}

Clearly the CGLe (\ref{CGLe}) may be 
viewed as a very general normal-form type
equation for a large class of bifurcations and nonlinear wave phenomena
in spatially extended systems,
so a detailed investigation of its properties is well justified.
The equation "interpolates" between the two opposing limits of the
conservative NLSe and the purely relaxative GLe.
The CGLe world lies between these limits
 where new phenomena and scenarios arise, 
like sink and source solutions (spirals in 2D and filaments in 3D), 
various core and wave instabilities, 
nonlinear convective versus absolute instability, screening of
interaction and competition between sources, various types of spatio-temporal 
chaos and glassy states.


\section{General Considerations}
\label{sec:gencons}
In this Section we will study general properties of Eq. (\ref{CGLe}) relevant in
all dimensions.
\subsection{Variational case}

For the case of $b=c$ it is useful to transform into a ''rotating`` frame 
$A \to A \exp( i b t) $. Then Eq. (\ref{CGLe}) goes over into 
 \begin{equation}  \label{RGLe1}
  \partial_t A = (1+i b) \left(A + \Delta A - |A|^2 A \right).
 \end{equation}
Eq. (\ref{RGLe1}) can be obtained by variation of the functional
 \begin{equation}  \label{potential2}
  {\cal V}= \int{U d^D r},\qquad U=- |A|^2 + \frac{1}{2}|A|^4 + 
  |\nabla A|^2 .
 \end{equation}
 leading to $\partial_t A= - (1+ib) \delta {\cal V}/\delta A^*$ and
\begin{equation}
 \partial_t {\cal V} = -\frac{2}{ 1+b^2}  \int | \partial_t A|^2 d^D  r. 
\end{equation} 
One sees that for all non-infinite $b$ the value of ${\cal V}$ decreases,
so the functional (\ref{potential2}) 
plays the role of a global Lyapunov functional
or generalized free energy (${\cal V}$ is bounded from below). 
The system then relaxes towards local minima 
of the functional. In particular the stationary solutions 
$A({\vec r})$ of the GLe (b=0) correspond in the more general case to $A e^{-i b t}$ 
with corresponding stability properties.

In the NLSe limit  $b \to \infty$ the functional
 becomes a Hamiltonian, which is conserved. More generally, 
the NLSe is obtained from the CGLe by taking the limit
$b,c \to \infty$ without further restrictions. After rescaling one obtains
Eq. (\ref{NLSe}). 
One sees that the equation comes in two variants, 
the focusing ($+$ sign) and defocusing ($-$sign)
case (the notation comes from nonlinear optics.
In 1D it is completely integrable (Zakharov and Shabat, 1971). 
In the focusing case it has a two-parameter family of ``bright'' solitons 
(irrespective of space translations and gauge transformation).
In D$>1$ solutions typically exhibit finite-time singularities (collapse)
(Zakharov, 1984, for recent review see Robinson, 1997).
In the defocusing case one has in 1D a three-parameter family of 
"dark solitons" that connect asymptotically to plane waves and vortices in 2 and 3D.
We will here not discuss the equation since there exists a vast literature
on it (see e.g. Proceedings of  Conference on 
The Nonlinear Schr\"odinger Equation,1994). 
It is useful to treat the CGLe in the limit of large $b$ and $c$
from the point of view of a perturbed NLSe. 

\subsection{The amplitude-phase representation}
Often it is useful to represent the complex function $A$  by its real amplitude and 
phase in the form $A=R \exp(i\theta)$. Then Eq. (\ref{CGLe}) becomes
\begin{eqnarray}  \label{modpha}
\partial_t R&=&[\Delta-(\nabla \theta)^2]R
-b(2\nabla \theta \cdot \nabla R +R\Delta \theta)+(1-R^2)R \ ,\\ \nonumber
R\partial_t \theta &=& b[\Delta-(\nabla \theta)^2]R
+ 2\nabla \theta \cdot \nabla R +R\Delta \theta - c R^3 \ .
\end{eqnarray}
For $b=0$ this corresponds to a class of reaction-diffusion equations called 
$\lambda-\omega$ systems, which are generally of the form 
\begin{eqnarray}  \label{lambda_omega}
\partial_t R=[\Delta-(\nabla \theta)^2]R + R \lambda(R) \ ,\\ \nonumber
R\partial_t \theta =
 2\nabla \theta \cdot \nabla R +R\Delta \theta + R \omega(R) \ .
\end{eqnarray}
Such equations have been studied in the past extensively by applied mathematicians
(see Hagan, 1982, and references therein). 
Clearly, one can combine Eqs. (\ref{modpha}) in such a way that the right-hand sides
are those of a $\lambda-\omega$ system.

\subsection{Transformations, coherent structures, similarity}
\label{transform}

The  obvious symmetries of the CGLe are time and space translations,
spatial reflections and rotation, and global gauge (or phase) symmetry 
$A \to A e^{i \phi}$.
The transformation $A,b,c \to A^*,-b,-c$ leaves the equation invariant so that
only a half plane within the $b,c$ parameter space has to be considered.

Other transformations hold only for particular classes of solutions.
To see this it is useful to consider the following transformation 
 \begin{eqnarray}  \label{trafo}
  A(\vec{x},t)&=& e^{i(\vec{Q}\cdot \vec{x}-\omega t)} 
  B(\vec{x}-\vec{v}t,t), 
 \end{eqnarray}
leading to
 \begin{eqnarray}  \label{coherent}
  \partial_t B &=& [\sigma  +\vec{w} \cdot \nabla  + (1+i b)\Delta  
  - (1+i c)|B|^2] B
 \end{eqnarray}
with 
$\sigma=1+i\omega-(1+i b)Q^2,\ \vec{w}=\vec{v}-i(1+i b)\vec{Q}$.
Most known solutions are either of the {\em coherent-structure} type,
where $B$ depends only on its first argument (i.e. with properly chosen $\omega$
it is time independent in a moving frame), or are disordered in the sense of 
spatio-temporal chaos. Coherent structures can be localized or extended. 
The ``outer wavevector'' $\vec{Q}$ could be absorbed in $B$ (then $\omega$
would be replaced by $\omega-Qv$). 
It may be useful to introduce $\vec Q$  when the gradient of
the phase of $B$, integrated over the system, is zero (or at least small).

With a little bit of algebra one can now derive a useful {\em similarity 
transformation} that connects coherent structures along the lines 
$(b-c)/(1+bc)=const $ in parameter space. 
Defining
 \begin{equation}   \label{rescale}
  \vec{r}=\beta \vec{r'},\ \ |B|=\gamma |B'|,
 \end{equation}
the transformation relations between unprimed and primed quantities 
can be written as
 \begin{eqnarray}
   \beta^{-2}&=&\frac{1}{1+b^2} \left(1+b b' +(b-b')(\omega-Qv)
    +\frac{1}{4} (b-b')^2 v^2 \right) \ , \nonumber \\
     \omega'&=&b'+ \beta^2 \frac{1+b'^2}{1+b^2} (\omega - b)\ , \
      v'=\beta \frac{1+b'^2}{1+b^2} v \ ,\
     Q'=\beta \left( Q-\frac{1}{2}\frac{b-b'}{1+b^2} v \right) \ , \label{linear-sim} \\
    \frac{b-c}{1+bc}&=&\frac{b'-c'}{1+b'c'} \ , \ \
   \gamma \beta=\left[\frac{1+bc}{1+b'c'} \frac{1+c'^2}{1+c^2}\right]^{1/2}\ ,
  \label{nonlinear-sim}
\end{eqnarray}

The relations (\ref{linear-sim}) are independent of the nonlinear part 
of (\ref{CGLe}) and therefore survive generalizations.
Solutions with $\vec{v}=0$ remain stationary (for $b < \infty$) and that
transformation was given by Hagan, 1982.
Clearly one can very generally transform to $b=0$, where the CGLe represents
a $\lambda - \omega$ system.
Note that for $v \ne 0$ one cannot have $Q=0$ and $Q'=0$. 
The similarity line $b=c$ (vanishing group velocity, see below) includes the 
real case. 

Note, that the stability limits of coherent states are in general not expected to
conform with the similarity transform.
By taking in Eqs. (\ref{nonlinear-sim}) the limit $b \to \infty$, one finds
$c'=c$. Then the $1$ in the factor $1+ib$ can be dropped
and changes in $b$ can be absorbed in a rescaling of length.
The similarity transformation then connects solutions with arbitrary velocity,
which is a manifestation of a type of Galilean invariance (van Saarloos
and Hohenberg, 1992).
Thus in this limit solutions appear as continuous families moving at
arbitrary velocity.

Similarly, by letting $c \to \infty$, $b$ tends to a constant.
Then, in addition to the $1$ in the factor $1+ic$, the linear growth term
can be dropped, and then changes in $c$ can be absorbed in
a rescaling of B as well as length and time.
The similarity transformation then turns into a scaling
transformation.
Thus in this limit solutions appear as continuous families of rescaled
functions.

For $b$ and $c \to \infty$ one has both transformations together, so solutions
appear generically as two-parameter families.
Indeed, one is then left with the NLSe.

\subsection{Plane-wave solutions and their stability}
\label{plw}
The simplest coherent structures are the plane-wave solutions
\begin{equation}  
A=\sqrt{1-Q^2} \exp \left[i ({\bf Q\cdot r} -\omega_p (Q)t+ {\phi})\right], 
\qquad F^2=1-Q^2,
\qquad \omega_p (Q)=c(1-Q^2)+bQ^2. 
\label{plane_wave}
\end{equation}
%
(${\phi}$ is an arbitrary constant phase)
which exist for $Q^2<1$. To test their stability one considers 
the complex growth rate $\lambda$ of the modulational modes.
One seeks the perturbed solution in the form 
\begin{equation} 
A=( F + \delta a_+ \exp[\lambda t + i {\bf k}\cdot {\bf r}] + 
\delta a_-\exp[\lambda^* t-i {\bf k}\cdot {\bf r}]) 
 \exp [i({\bf Q}\cdot {\bf r}-\omega t)]
\label{pers}
\end{equation}
where ${\bf k}$ is a modulation wave vector and $\delta a_{\pm}$ are the 
amplitudes of the small perturbations. 
One easily finds the expression for the growth rate $\lambda$
(Stuart and DiPrima, 1980): 
\begin{eqnarray}  \label{dispersion}
\lambda^2 +2(F^2 +2ib{\bf Q}\cdot {\bf k}+k^2)\lambda & &\nonumber \\ 
+(1+b^2)(k^4-4{(\bf Q}\cdot {\bf k})^4)&+&2F^2[(1+bc)k^2 + 2i(b-c){\bf Q}\cdot {\bf k}]=0.
\end{eqnarray}
By expanding this equation for small $k$ one finds
 \begin{equation}  \label{long-wave}
  \lambda= - i V_g k-D_2k^2 + i \Omega_g k^3 -D_4 k^4 + O(k^5),
 \end{equation}
with
 \begin{eqnarray} \label{long-wavepar}
  V_g &=& 2(b-c) Q_k \nonumber \\
   D_2 &=& 1+bc -\frac{2 (1+c^2)Q_k^2}{1-Q^2} \nonumber \\
    \Omega_g & = & {2[b (1-Q^2)-2c Q_k^2](1+c^2)Q_k\over(1-Q^2)^2} \nonumber\\
   D_4 &=&\frac{1+c^2}{2(1-Q^2)^3}\left[b^2(1-Q^2)^2-12bc(1-Q^2)Q_k^2+
4(1+5c^2)Q_k^4 \right]\nonumber \\
 \end{eqnarray}
where $Q_k={\bf Q}\cdot{\bf \hat k}$ is the component of ${\bf Q}$ parallel 
to ${\bf k}$. The quantities $V_g, D_2, \Omega_g$ and $D_4$ for ${\bf k} 
 \parallel  {\bf Q}$ will be denoted by $V_{g \parallel}, D_{2 \parallel} ,
 \Omega_{g \parallel}$ and $D_{4 \parallel}$. Similarly, for
${\bf k} \perp  {\bf Q}$  we may use the notation 
$ V_{g \perp}, D_{2 \perp} ,
 \Omega_{g \perp}$ and $D_{4 \perp }$.
Clearly the longitudinal perturbations with ${\bf k} \parallel {\bf Q }$ are
the most dangerous ones.
The solutions (\ref{plane_wave}) are long-wave stable as long as the
phase diffusion constant $D_{2 \parallel}$
is positive.
Thus one has a stable range of wave vectors with $Q^2<Q_E^2=(1+bc)/(3+2c^2+bc)$ 
enclosing the homogeneous ($Q=0$) state as long as the 
``Benjamin-Feir-Newell criterion'' $1+bc>0$ holds. This criterion
conforms with the similarity transform (\ref{nonlinear-sim}).
The condition $D_{2\parallel}>0$ is the (generalized) Eckhaus criterion. 
For $b=c$ it reduces to the classical Eckhaus criterion $Q^2<Q_E^2=1/3$
for stationary bifurcations. 
We will call the 
quadrants in the $b,\ c$ plane with $bc>0$ the ``defocusing quadrants''.
Otherwise we will speak of the ``focusing quadrants''.

From Eqs. (\ref{dispersion},\ref{long-wave}) one sees that 
for $b-c\ne 0$ and $Q\ne 0$ the destabilizing modes have a group velocity
$V_g=\nabla_Q \Omega=2(b-c){\bf Q}$, 
so the Eckhaus instability then is of convective nature and does not 
necessarily lead to destabilization of the pattern (see next subsection). 

The Eckhaus  instability signalizes bifurcations to quasiperiodic 
solutions (including the solitary limit), which are of the form 
(\ref{coherent})
with periodic function $B(\vec{x}-\vec{V}t)$.
It becomes supercritical before the Benjamin-Feir-Newell criterion
is reached  (Janiaud {\it et al}, 1992) 
and remains so in the unstable range. 
The bifurcation is captured most easily in the long-wave limit
by phase equations, see below. A general analysis of the bifurcating solutions
has recently been done (Brusch et al 2000), see subsection \ref{other_cs}.

It is well known that the Eckhaus instability in the CGLe
is not in all cases of long-wave type. Clearly, a necessary condition for it  
to be the case is that $D_{4 \parallel}$ is positive where $D_{2 \parallel}$ changes sign. 

From the above expressions one deduces that this fails to be the case for 
$|b|>b_4(c)$, where 
\begin{equation}   \label{short wave}
    b_4=\left[2|c|(1-c^2)-\sqrt{4c^2(1-c^2)^2-(1-3c^2)(1+5c^2)} \right]
    /(1-3c^2),\ \ bc>0.
\end{equation}
Thus the range is in the defocusing quadrant, far away from the 
Benjamin-Feir-Newell stability limit.
We are presently not aware of effects where this phenomenon is of relevance.


\subsection{Absolute versus  convective instability of plane waves}
\label{abs}

For  a nonzero group velocity 
$V_{g}= \nabla_{Q} \Omega = 2(b-c) {\bf Q}$ the Eckhaus criterion can
be taken only as a test for convective instability. 
In this case a localized  1D initial
perturbation $S_{0}(x)$ of the asymptotic plane wave, although amplified in
time, drifts away and does not necessarily
 amplify at a fixed position (Landau and Lifshitz, 1959). 
For absolute instability localized perturbations
 have to amplify at fixed position. 
The time evolution of a localized perturbation is in the linear range given by
\begin{eqnarray}
S(x,t)= \int_{-\infty}^{\infty} dk/(2\pi ) \hat{S}_{0}(k) \exp(ikx+\lambda
(k) t)
\label{evol}
\end{eqnarray}
where $\hat{S}_{0}(k)$ is the Fourier transform of $S_{0}(x)$.
\footnote{It can be shown strictly that the destabilization occurs at first
for purely longitudinal perturbations ${\bf Q}\parallel{\bf k}$}
The integral can be deformed into the complex k-plane.  In the limit
$t \rightarrow \infty $ the  integral is dominated by the largest saddle
point $k_0$ of $\lambda(k)$ (steepest descent method, see e.g., 
Morse and Feshbach, 1953) 
and the test for absolute instability is
\begin{eqnarray}
Re[\lambda(k_{0})] >0 \;\;\; {\rm with}\;\;\;
\partial_{k} \lambda (k_{0}) =0
\label{instab}
\end{eqnarray}
The long-wavelength expansion  (\ref{long-wave}) indicates that at the
Eckhaus instability, where $D_{2}$ becomes negative, the system 
remains stable in the above sense. When $D_{2}$ vanishes and
$Q\ne 0$
the main contribution comes from  the term linear in $k$ that then can 
suppress instability. 

\begin{figure}
\centerline{\hspace{.0cm}  \psfig{figure=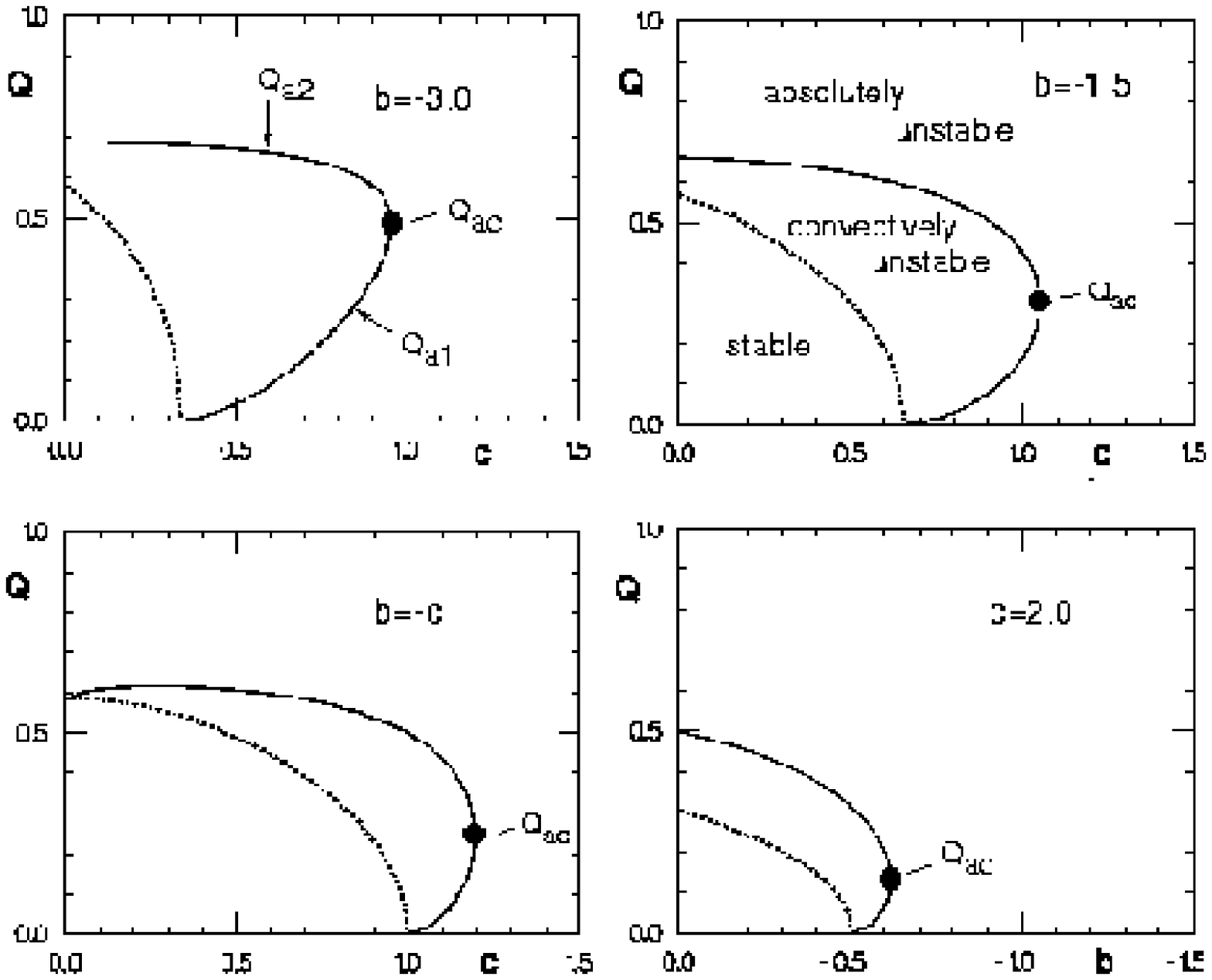,height=3.in}}
 \caption{Absolute stability limits $Q_{a1,2}$ for four cuts in the $b,\ c$
 plane. Also included are the convective (=Eckhaus) stability limits that
 separate the stable (light shadowing) from the convectively unstable
 (dark shadowing) regions.}
 \label{figqbc}
\end{figure}

In the following, results of the analysis are shown in the $b-c$ plane
(Aranson {\it et al},  1992; Weber {\it et al},  1992).
In Fig.~\ref{figqbc} the scenario is demonstrated for four cuts in the 
$b,\ c,\ Q$ space.
The stable region (light gray) is limited by the Eckhaus curve, which
terminates at $Q=0$ on the  curve.
To the right of it there exists a convectively unstable wavenumber band
 $0<Q_{a1}<|Q|<Q_{a2}$ (dark grey).
 $Q_{a1}$ goes to zero on the Benjamin-Feir-Newell curve as $(1+bc)^{3/2}$, which can be seen from
the long-wavelength expansion~(\ref{long-wave}) with the fourth-order term
included.
Moving away from the Benjamin-Feir-Newell line (into the unstable regime) $Q_{a1}$ increases
and $Q_{a2}$ decreases until they come together in a saddle-node-type process 
at a value
$Q_{ac}$. Beyond $Q_{ac}$ there are no convectively unstable plane waves.
The saddle nodes $Q_{ac}$ are shown in the $b,\ c$ plane in
Fig.~\ref{fig-1CGLE} (curve $AI$).
Thus, convectively unstable waves exist up to this curve.
The Benjamin-Feir-Newell criterion  curve up to which convectively unstable waves exist is also included.
The other curves include the Eckhaus instability  and the absolute stability
limit  for waves with wavenumber selected by the stationary hole solutions 
(see Sec. \ref{NB_hole}).

Eq. (\ref{instab}) by itself gives only a necessary  
condition for absolute stability. However, it is also sufficient, as  
long as one of the two roots $k_{1,2}$ of the dispersion relation,  
which collide at the saddle point when $Re [\lambda]$ is decreased  
from positive values to zero, say $k_1$, is the root that produced  
the convective instability, and if $k_2$ does not cross the real  
$k-$axis before $k_1$ does, when $Re[\lambda]$ is increased from  
zero. (Note that $k_1$ has to cross the axis when $Re[\lambda]$ is  
increased. Note also that, because of the symmetry $\lambda,k  
\rightleftharpoons  \lambda^*,-k^*$ one has parallel processes with  
opposite sign of $Im[\lambda]$). For a discussion of the underlying  
``pinching condition'', see e.g.   Brevdo and Bridges  (1996). In  
the parameter range considered here the sufficient condition is  
fulfilled.


Clearly the absolute stability boundary can be reached only 
if reflection at the boundaries of the system is sufficiently weak,
so in general one should expect the system to lose stability
before the exact limit is reached. 
In many realistic situations in the CGLe  the interaction of the emitted 
waves with the boundaries leads to 
sinks (shocks) in analogy to the situation where different waves collide. 
These shocks are strong perturbations of the plane
wave solutions but they absorb the incoming perturbations. Thus here, even for
periodic boundary conditions, the absolute stability limit 
is relevant. Moreover, 
in infinite systems, states with a cellular structure made up of sources 
surrounded by sinks can exist in the convectively unstable regime.
However, the convectively unstable states are very susceptible to noise, which
is exponentially amplified in space. The amplification rate goes to zero
at the convective stability limit and diverges at the absolute stability
limit.

The concept of absolute stability is relevant in particular for the waves 
emitted by sources and possibly also for some characteristics of 
spatio-temporally chaotic states. 
When boundaries are considered one also has to allow for a linear group 
velocity term in the CGLe (see Eq.~(\ref{AGLE})). 
Such a term does not change the convective stability threshold, but clearly 
the absolute stability limit is altered,
and this is important in particular in the context of open flow systems.
The saddle point condition in (\ref{instab}) ensures the existence of bounded 
solutions of the linear problem that satisfy nonperiodic boundary conditions 
(irrespective  of their precise form) at well separated side walls 
(sometimes called ``global modes''), 
because for that purpose one needs to superpose neighboring (extended)
eigenmodes, which are available precisely at the saddle point.
This is an alternative view of the absolute instability
(Huerre and Monkowitz, 1990, 
Tobias and Knobloch, 1998, Tobias, Proctor,  and  Knobloch, 1998). 

Often the condition (\ref{instab}) coincides with the condition that a front
invades the unstable state in the upstream direction according to the linear
front selection criterion (marginal stability condition, see, e.g.
van Saarloos (1988)).

\subsection{Collisions of plane waves and effect of localized disturbances}
\label{collisions} 

The nonlinear waves discussed above have very different properties 
from linear waves. 
In particular, when two waves collide they almost do not interpenetrate.
Instead a ``shock'' (sink) is formed along a point (1D), line (2D), or surface 
(3D).
When the frequency of the two waves differs the shock moves with the average
phase velocity, provided there are 
no phase slips in 1D, or its equivalent in
higher dimensions (creation of vortex pairs
in 2D and inflation of vortex loops in 3D), for review see
Bohr {\it et al} (1998).  
A stationary shock is formed most easily when a wave impinges on an
absorbing boundary. Here we will consider the general situation 
where a plane wave is perturbed by a stationary, localized disturbance. 
This concept will be particularly useful in the 
context of interaction of defects.

Sufficiently upstream (i.e. against the group velocity) from the disturbance the perturbation will be small and we can linearize around the 
plane-wave solution. In general one obtains exponential behavior with exponents
$p=ik$ calculated from the dispersion relation (\ref{dispersion}) with 
$\lambda=0$. This gives
 \begin{equation} \label{asdisp}
  p\{(1+b^2)(4Q^2+p^2)p-2F^2[(1+bc)p-2(b-c)Q]\}=0 \ .
 \end{equation}
After separating out the translational mode $p=0$ one is left with a 
cubic polynomial. To discuss the roots $p_1,\ p_2,\ p_3$ choose the group 
velocity $V_{g\parallel} 
=2(b-c)Q \ge 0$ (otherwise all signs must be reversed). 
For $V_{g \parallel} =0$ 
one has $p_1=-p_3<0, p_2=0$ in the Eckhaus stable range. 
For $V_{g\parallel}  >0 $ one finds that $p_2$ increases with increasing $|Q|$.
Before $|Q|=1$ is reached $p_2$ and $p_3$ collide and
become complex conjugate. 

The existence of roots with positive real part is an indication of screening of
disturbances in the upstream direction because one needs the solutions that grow exponentially to match to the disturbance.
The screening length is given by $(Re p_2)^{-1}$, the root with the 
smaller real part.
So the screening is in general exponential. The length diverges
for $V_g \to 0$ and then one has a crossover to power law.

In the focusing quadrants the root collision always occurs before the Eckhaus 
instability 
is reached. (It fails to hold only in part of the defocusing quadrant
away from the origin and restricted to $c^2>b^2$.) 
We will refer to the situation where $p_{2,3}$ are real to the monotonic case
and otherwise to the oscillatory case, because this characterizes the
nature of the asymptotic interaction of sources that emit waves,
see Sec. \ref{sec:1D}  and \ref{sec:2D}.
In Fig.~\ref{fig-1CGLE} the transition from monotonic to oscillatory behavior 
for the waves emitted by standing hole solutions is also shown (curve MOH).
A generalization to disturbances that move with velocity $v$ is straightforward 
by replacing in Eq. (\ref{dispersion}) the growth rate $\lambda$ by $-vp$.

\subsection{Phase equations}
\label{phaseequations}

The global phase invariance of the CGLe leads to the fact that, 
starting from a coherent state (in particular a plane wave), 
one expects solutions where 
the free phase ${\phi}$ becomes a slowly varying function,
and one can construct appropriate equations for these solutions.
(In the case of localized structures one then also has to allow for a variation
of the velocity.) 
In the mathematical literature these phase equations are sometimes called
``modulated modulation equations''.
There are several ways to proceed technically in their derivation, but the 
simplest is to first establish the linear part of the equation 
by a standard linear analysis and then construct nonlinearities by separate reasoning. 
It is helpful to include symmetry considerations to exclude terms from the
beginning on.

By starting from a plane wave state (\ref{plane_wave}) with wave number $Q$
in the $x$ direction
one may perform a gradient expansion of ${\phi}$ leading in 1D to 
 \begin{equation} \label{general1Dphase}
  \partial_t {\phi}= V_{g\parallel} \partial_x { \phi} +
   D_{2\parallel} \partial_x^2 { \phi} - \Omega_{g\parallel}
 \partial_x^3 { \phi} 
  - D_{4\parallel} \partial_x^4 { \phi} + h.o.t.
 \end{equation}
Note the absence of a term proportional to ${ \phi}$ which would violate
translation invariance.
The prefactors of the gradient terms have to be chosen according to 
(\ref{long-wavepar}) in order to reproduce the 
linear stability properties of the plane wave (with the wavenumber 
$Q_k$ replaced by $Q$).

Nonlinearity can be included in Eq. (\ref{general1Dphase}) by substituting  
in the coefficients of Eq. (\ref{long-wavepar}) 
$Q \to Q+\partial_x \phi$.
Expanding $V_{g\parallel}(Q)$ and $D_{2\parallel}(Q)$ one generates the leading
nonlinear terms, so that eq. (\ref{general1Dphase}) goes over into
(Janiaud {\it  et al}, 1992)
 \begin{equation} \label{special1Dphase}
  (\partial_t - V_{g\parallel} \partial_x)\phi = 
D_{2\parallel} \partial_x^2 \phi- D_{4\parallel} \partial_x^4 \phi 
   -g_1(\partial_x \phi)^2 - \Omega_{g\parallel} \partial_x^3 \phi 
  -g_2\partial_x \phi \partial_x^2 \phi+ h.o.t.
 \end{equation}
with the linear parameters from (\ref{long-wavepar}) and
\begin{eqnarray}
g_1 &=& -\frac{1}{2} \partial_Q  V_{g\parallel} = (b-c) \nonumber \\
g_2 &=& -\partial_Q D_{2\parallel} =  \frac{4 Q_0  ( 1+c^2)}{1-Q_0^2}
\label{param}
\end{eqnarray}
Because of translation invariance the lowest relevant nonlinearities 
are $(\partial_x \phi)^2$ and $\partial_x \phi \partial_x^2 \phi$ [Kuaramoto, 1984].

In the stationary case ($c=b$) one has $V_{g \parallel}=\Omega_{g \parallel}=g_1=0$. The remaining
nonlinearity $\partial_x \phi \partial_x^2 \phi$ does not saturate 
the linear instability and one recovers the 
results for the nonlinear Eckhaus instability (Kramer and Zimmermann, 1985). 
For $b\ne c$ and $Q = 0$ the dominant nonlinearity is 
$(\partial_x \phi)^2$ which does saturate the linear instability. 
The resulting equation with $\Omega_{g \parallel}=g_2=0$
\begin{equation}
\partial_t \phi - D_{2\parallel} \partial_x^2 \phi 
+ D_{4\parallel } \partial_x^4 \phi+ g_1 (\partial_x \phi)^2 =0
\label{pde1}
\end{equation}
describes the (supercritical) bifurcation at the 
Benjamin-Feir-Newell instability and is known as 
the Kuramoto-Sivashinsky equation. It has stationary periodic solutions, 
which are stable in a small wavenumber range $0.77 p_0<p<0.837 p_0$,
where $p_0= \sqrt{D_2/D_4}$ 
(Frisch et al. 1986, Nepomnyashchii, 1995b). 
It also has spatio-temporally
chaotic solutions, which are actually the relevant 
attractors, and it maybe represents the simplest partial differential equation 
displaying this phenomenon. 
Since it is a rather
general equation for long-wavelength instabilities, independent of the present
context of the CGLe, it has attracted much attention in recent years
(see e.g. Bohr {\it et al} (1998)).

In the general case $Q \ne 0$ both nonlinearities are important (and also the term 
proportional to $\Omega_g$).
Then the bifurcation to modulated waves, which are represented by periodic
solutions of (\ref{general1Dphase}), can be either forward or backward 
depending on the values of $b,c$. 
In fact the Eckhaus bifurcation always becomes 
supercritical before the Benjamin-Feir-Newell instability is reached. Modulated
solutions can be found analytically in the limit $D_{2 \parallel} \to 0$, 
where Eq. (\ref{general1Dphase})
reduces to
\begin{eqnarray}
\partial_t \phi -V_{g \parallel} \partial_x \phi  +
 \partial_x^3 \phi + (\partial_x \phi)^2 + O(\epsilon) =0.
\label{pde2}
\end{eqnarray}
with $\epsilon=\sqrt{D_4 D_{2}}/|\Omega_g| \ll 1$.
This limit is complementary to that leading to the 
Kuramoto-Sivashinsky equation and one is in fact left 
with a perturbed  Korteweg de Vries equation.
This equation is (in 1D) completely integrable and the perturbed case was 
studied by 
Janiaud {\it  et al} (1992) and by 
Bar and Nepomnyashchii (1995). They showed that a 
fairly broad band of the periodic solutions (much larger than in the 
Kuramoto-Sivashinsky equation  
equation) persists the perturbation of 
the Korteweg de Vries equation and is stable. Outside this 
limit, and in particular in 
the Kuramoto-Sivashinsky equation regime, one has in extended systems 
spatio-temporally chaotic solutions. 
It is believed that this chaotic state is a 
representation of the phase chaos observed in the 
CGLe, although this view has been challenged (see Sec. \ref{stc_1d}). 
Sakaguchi (1992) 
introduced higher-order nonlinearities into the 
Kuramoto-Sivashinsky equation, which allowed to capture
the analog of the 
transition (or cross over) to amplitude chaos manifesting itself
by finite-time singularities in $\partial_x \phi$.

Under some conditions, namely when $D_{2 \parallel}$ is positive and $Q$ is small, 
the first nonlinearity in Eq. (\ref{special1Dphase})
 is dominant and then it reduces  
to Burgers equation
\begin{eqnarray}
\partial_t \phi  - D_{2 \parallel} \partial_x^2 \phi +
 g_1 (\partial_x \phi)^2 =0
\label{pde3}
\end{eqnarray}
which is completely integrable within the space of functions that do not
cross zero, because it can be linearized by a Hopf-Cole transformation
$\phi=(D_{2 \parallel}/g_1) \log W$. 
The equation allows to describe analytically sink solutions (shocks),
(see e.g., Kuramoto and Tsuzuki (1976), 
Malomed (1983) and is (in 2D) useful for the qualitative
understanding of the interaction of spiral waves in the limit $b,c \to 0$
(see Biktashev (1989), Aranson {\it et al} (1991)).

Clearly generalization of the phase equations to higher dimensions is possible. 
Using the full form of Eq. (\ref{long-wavepar}) one can easily generalize
the phase equation  to 2D
(for details see also Kuramoto (1984); Lega (1991)). The equation is 
most useful in the range where $D_{2\parallel}$ goes through zero and becomes negative.
Also, they can be derived outside the range of applicability of the CGLe 
(far away from threshold) and for more general nonlinear long-wavelength
phenomena, where $\phi$  does not represent the phase of a periodic function
(see e.g. Bar and Nepomnyashchii (1995)). 

\subsection{Topological defects}

Zeros of the complex field $A$ result in singularity of the phase 
$\theta=\arg A$. In 2D point singularities
 correspond to quantized vortices 
with so-called topological charge $n=1/(2 \pi) 
\oint _L \nabla \theta {\bf d l}$, 
where  $L$ is a contour encircling the zero of $A$. 
Although for $b \ne c$ they represent wave emitting spirals,  
they are analogous to vortices in
superconductors and superfluids and represent topological defects because
a small variation of the field will not eliminate the phase circulation condition.
Clearly, vortices with topological charge $n=\pm 1$ are topologically stable. 
The vortices with multiple topological 
charge can be split into single-charged 
vortices \footnote{In small  samples vortices with multiple charge can be dynamically stable, 
see for detail Geim {\it et al} (1998); Deo {\it et al} (1997).}.  
2D point defects become line defects in 3D. 
Then, one can close such a line to a loop, which can shrink to zero.
Some definitions of topological 
defects from the point of view of
 energy versus topology are given by Pismen (1999).

Topological arguments do not guarantee the 
existence of a stable, coherent solution
of the field equations. In particular, the stability of the topological defect 
depends on the
background state it is embedded in. For example,  spiral waves
are stable 
in a certain parameter range, where they select the
background state (see Sec. \ref{sec:2D}).
Simultaneously charged sinks coexist with spirals 
and play a passive role. 
Also, defects can become unstable 
against spontaneous acceleration of their cores.

\subsection{Effects of boundaries}

Boundaries may play an important role in nonequilibrium 
systems. Even in large systems the boundary may provide 
restriction or even selection of the wavenumber. 
We will not discuss this topic (see, e.g. Cross and Hohenberg, 1993)
and consider situations where such effects are not important.

\section{Dynamics in 1D}
\label{sec:1D}

In this Section we will consider the properties of 
various 1D solutions of the CGLe. We will discuss 
the stability and interaction of coherent structures 
and the transition to and characterization of spatio-temporal chaos.

\subsection{Classification of coherent structures, counting arguments} 
\label{count_arg}

Coherent structures introduced in the previous chapter can be characterized
using simple  counting arguments put forward by  
van Saarloos and Hohenberg (1992). It should be mentioned, however,  that 
the counting arguments cannot account for all the circumstances, for example, 
hidden symmetries, and may fail for certain class of solutions, see below. 
1D coherent structures can be written in the form
 \begin{eqnarray}  \label{coherent1}
   A(x,t)&=& e^{-i \omega_k t + i \phi (x-vt)} a(x-vt),
 \end{eqnarray}
with the real functions $\phi,a$ satisfying a set of 3 ordinary differential 
equations 
\begin{eqnarray} 
 a_x & = &  s ; \nonumber \\
 s_x & = &  \psi^2 a
- \gamma^{-1} [(1+b \omega_k)a+v(s+b\psi a)-(1+bc )a^3];  \nonumber  \\ 
\psi_x &=& 
-2s \psi/a+\gamma^{-1} [b-\omega_k+v(s b/a-\psi)-(b-c) a^2].
\label{count} 
\end{eqnarray} 
Abbreviations used are $\gamma = 1 + b^2$, $a_x =da/dx, s_x = ds/dx$ and
$\psi = d\phi/dx$. 
These ordinary differential equations 
constitute a dynamical system with three degrees of freedom.

The similarity transform of Sec. \ref{transform} can be adapted to these 
equations by absorbing the external wavenumber in the phase $\phi$, 
which leads to
$\omega_k=\omega-Q v$, $\omega'_k=\omega'-Q' v'$, and
 \begin{eqnarray} 
  \omega'_k=b'+\beta^2 \frac{\gamma'}{\gamma}\left[\omega_k-b +\frac{1}{2}
   \frac{b-b'}{\gamma} v\right] \ , \
  \psi'=\beta \left( \psi -\frac{1}{2}\frac{b-b'}{\gamma} v \right) \ .
 \end{eqnarray} 
The other relations (\ref{rescale}-\ref{nonlinear-sim}) remain.

The counting arguments allow 
to establish necessary conditions for the existence of localized 
coherent structures, which correspond to homo- or heteroclinic orbits of
Eqs. (\ref{count}).
Consider, e.g. the trajectory of Eqs. (\ref{count}) flowing from fixed point
 $N$ to fixed point $L$. If  $N$ has $n_N$ unstable directions, 
there are $n_N -1$ free parameters characterizing the flow on the 
$n_N$-dimensional subspace spanned by the unstable eigenvectors. 
Together with the parameters $\omega_k$ and $v$ this will yield $n_N+1$ 
free parameters. If $L$ has $n_L$ unstable directions, the 
requirement that the trajectory should come in orthogonal to these 
yields $n_L$ conditions. The multiplicity of this type of trajectory will, 
therefore, 
be $n=n_N-n_L+1$, and, depending on $n$, it will give either a
$n$-parameter family ($n \ge1$), a discrete set of structures
($n=0$) or no structure ($n<0$). In addition one may have symmetry arguments
that reduce the number of conditions.

The asymptotic states can either correspond  to non-zero 
steady states (plane waves), or to the trivial state 
$a=0$. Accordingly, the localized coherent structures can be 
classified as {\it pulses, fronts}, {\it domain boundaries},
and {\it homoclons}. 
A pulse corresponds to 
the homoclinic orbit connecting to the trivial state $a=0$.
They come in discrete sets (Hohenberg and van Saarloos, 1992). 
For a full description of the pulses see Akhmediev {\it et al} 
(1995,1996,2001), 
Afanasjev {\it et al} (1996).
Fronts are heteroclinic orbits connecting on one side to a plane-wave state 
and to the unstable trivial state on the other side. 
They come in a continuous family, but sufficiently rapidly decaying initial 
conditions evolve into a  ``selected'' front
that moves at velocity $v^*=2 \sqrt{1+b^2}$ and generates a plane wave with wave 
number $Q^*=b/\sqrt{1+b^2}$ (``linear selection'').
For a discussion, 
see Hohenberg and van Saarloos (1992), Cross and Hohenberg (1993).

Domain boundaries are heteroclinic orbits connecting 
two different plane-wave states. The 
domain boundaries can be active (sources, often called holes) or 
passive (sinks or shocks) 
depending on whether in the co-moving frame the group velocity is directed 
outward or inward, respectively. 
They will be discussed in the next subsection.
Homoclons (or homoclinic holes, phasons) connect to the same 
plane-wave state on 
both sides. They are embedded in solutions representing periodic arrangements 
of homoclons, which represent quasiperiodic  solutions satisfying 
ansatz (\ref{coherent1}) and
correspond to closed orbits  of Eqs. (\ref{count}). 
They will be discussed in Subsec. \ref{other_CGLe}.
There also exist chaotic solutions of Eqs. (\ref{count}), which
correspond to nonperiodic arrangement of holes and shocks, or homoclons,
see Subsecs.\ref{per_sol}, \ref{other_CGLe}.

\subsection{Sinks and sources, Nozaki-Bekki  hole solutions} 
\label{NB_hole}

Sinks (shocks) conform with the counting arguments of 
Hohenberg and van Saarloos (1992): there is a two-parameter family of sinks.
However, there are no exact analytic expressions for the sink solution 
connecting two traveling waves with arbitrary wavenumbers. 
The exact sink solution
found by  Nozaki and Bekki (1984)
corresponds to some special choice of the wavenumbers
and is not therefore typical.
For small difference in wavenumber a phase description of sinks is possible
(Kuramoto, 1984).

However, the counting arguments (which are, strictly speaking, neither 
necessary  nor sufficient, and 
cannot account for specific circumstances, such as a hidden symmetry)  
fail for the source solution. 
According to the counting arguments there 
should be only
a discrete number, including in particular the symmetric standing 
hole solution  
that has a zero at the center and emits plane waves of 
a definite wavenumber (for given $b,c$). However, the standing hole is 
embedded in a continuous  family of {\it analytic} moving sources, the 
Nozaki-Bekki  hole solutions. They are characterized by a 
localized dip in $|A|$ 
that moves with constant speed $v$ and emits plane waves with wave numbers 
$q_1 \ne q_2$. 

This is a special feature of the cubic CGLe as demonstrated by the 
discovery that the moving holes do 
not survive a generic perturbation of the CGLe, 
e.g. a small quintic term (see below). Thus they 
are not structurally stable 
(Popp {\it et al}, 1993, 1995, Stiller {\it et al}, 1995a,b).
With the perturbation the holes either accelerate or decelerate 
depending on the sign of the perturbation and other solutions appear, see also
Doleman (1995). 
Clearly, the 1D 
CGLe possesses a ``hidden symmetry'' and has retained some 
remnant of integrability from the NLSe. Apparently this non-genericity 
has no consequences for other coherent structures.

The Nozaki-Bekki hole solutions are of the form (Nozaki and Bekki, 1984)
\begin{eqnarray}
\label{NB}
A^{NB}_v 
    &  = &
           [\hat B \partial_{\zeta}\varphi_v(\kappa\zeta) +\hat A v] \times
           \exp[i \varphi_v(\kappa\zeta) + i\hat \alpha v - i\Omega t]
\\ \mbox{where}&&\quad \varphi_v(\kappa\zeta)
= \hat\kappa^{-1}\ln \cosh(\kappa\zeta)
\; 
\nonumber
\end{eqnarray}
and  $\zeta=x-vt$, $v$ is the velocity of the hole, and $q_{1/2}$ are the
asymptotic wavenumbers. 
Symbols with a ``hat'' denote real constants depending 
only on  $b$ and $c$, e.g. $\hat \alpha = 1/(2(\!b\!-\!c\!))$ .
The frequency $\Omega$ and $\kappa^2$ are linear functions of
$v^2$. The emitted plane waves have wavenumbers
\begin{eqnarray}
\label{q}
q_{1/2}=
\pm\beta+ \alpha
\end{eqnarray}
where $\beta = \kappa/\hat\kappa $ and $\alpha = v\hat\alpha$.
One easily derives the relation
 \begin{equation}  \label{phasecons}
  v=(\omega(q_2)-\omega(q_1))/(q_2-q_1)
 \end{equation}
where $\omega(q)$ is the dispersion relation for the plane waves.
This relation can be interpreted as phase conservation and is valid also for 
more general equations possessing phase invariance.
For the cubic CGLe it reduces to $v=(b-c)(q_1+q_2)$, i.e. the hole moves simply
with the mean of the group velocities of the asymptotic plane waves.
The exact relations between the parameters can be derived  by inserting
ansatz (\ref{NB}) into the CGLe.
The resulting algebraic equations (8 equations for 8 parameters)
turn out to be not independent, yielding the one-parameter family.
$\kappa^2$ becomes zero at a maximal velocity $\pm v_{max}$ and
here the hole solution merges
with a plane wave with wavenumber
$q_1=q_2 = v_{max}/(2(\!b\!-\!c\!))$.
In a large part of the $b,c$ plane these bifurcations occur in the range 
of stable plane waves.
This is the case outside a strip around the line $b=c$ given by
$|b-c| > \delta $ with $\delta$ varying from $1/2$ for large values
of $|b|,|c|$ to $0.55$ for small values.
The region extends almost to the Benjamin-Feir-Newell   curves.
The bifurcation cannot be captured by linear perturbation around a plane 
wave, since the asymptotic wavenumbers of holes differ.
Presumably the bifurcation can be captured by the phase equation 
(\ref{special1Dphase}).
For $b \to c$ only the standing hole survives (the velocities of the moving 
holes diverge in this limit).

\begin{figure}
\centerline{  \psfig{figure=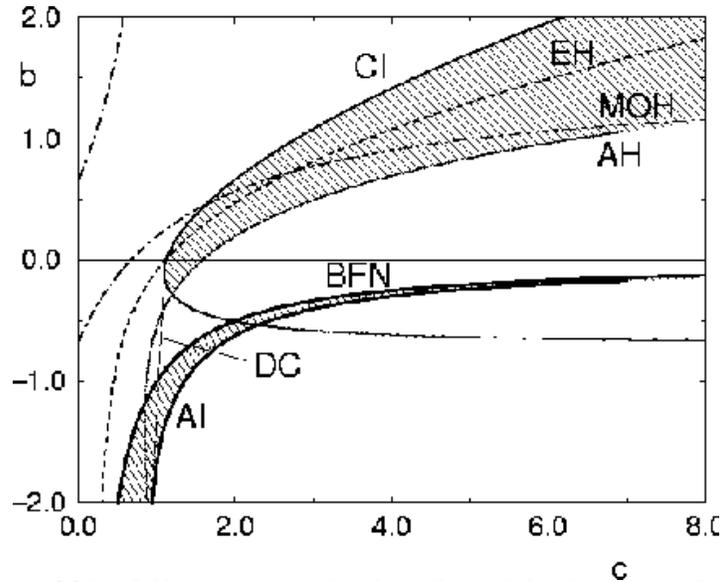,height=3.in}}
 \caption{Phase diagram of the 1D CGLe.
 BFN = Benjamin-Feir-Newell instability line up to which one has
 convectively stable plane-wave solutions;
 AI= absolute instability line up to which one has convectively
 unstable plane-wave solutions;
 DC = boundary of existence of DC towards small $|c|$.
 The other lines pertain to standing Nozaki-Bekki hole solutions:
 CI=core instability line,
 EH = convective (Eckhaus) instability of the emitted plane waves,
 AH = absolute instability of the emitted plane waves, MOH = boundary
 between monotonic and oscillatory interaction.}
 \label{fig-1CGLE}
\end{figure}

The interest in the hole solutions comes particularly from the fact
that they are  dynamically stable in some range. 
The stability 
has first been investigated by Sakaguchi (1991), 
in direct simulations of the CGLe.
The stability problem was then studied
by Chat\'e and Manneville (1992),  numerically for $v=0,v=0.2$
and by Sasa and Iwamoto (1992), 
semi-analytically for $v=0$ and by Popp {\it et al}. (1995),
essentially analytically. 
As a result, hole solutions were found to be stable in
a narrow region of the $b$-$c$ plane which is
shown in Fig. \ref{fig-1CGLE} for the standing hole ($v=0$) 
(upper shaded region).
From below,  the region is bounded
by the border of (absolute) stability of the
emitted plane waves with wavenumber $q(b,c)$ (see Eq.(\ref{NB}))
corresponding to the continuous spectrum of the linearized problem,
see curve AH in Fig.\ref{fig-1CGLE}.
From the other side,  the stable range is bounded by
the instability of the core with respect to localized
eigenmodes corresponding to a discrete spectrum of $ {\cal L}_{v} $
see curve CI in Fig.\ref{fig-1CGLE}.
For  $c \to \infty$, one has 
 \begin{eqnarray}  \label{CGLENLS}
  b_{CI}^2 &\to&  \frac{3}{4} c +O( \sqrt c) \;\;\;  
\mbox{for} \;b>0 \\
 b_{CI} & \to &  -\frac{1}{\sqrt 2} + O(1/c) \;\;\; \mbox{for} \;b<0   \nonumber 
 \end{eqnarray}
The result could be reproduced fully analytically by perturbing around the 
NLSe limit, where the
Nozaki-Bekki
holes emerge as a subclass of the 3-parameter family of dark solitons
and is supported by detailed numerical simulations and shows 
excellent agreement (Stiller {\it et al}, 1995a). 
It differs from that obtained 
by Lega and Fauve (1997) and by Kapitula and Rubin (2000).
Lega and Fauve (1997),   Lega (2000) claim a 
larger stability domain which is in disagreement with 
simulations (Stiller {\it et al}, 1995a).

The core instability turns out to be
connected with a stationary bifurcation where the destabilizing mode
passes through the neutral mode related to translations of the hole. 
This degeneracy is specific to the cubic CGLe and is thus structurally unstable
(see below).
When going through the stability limit the standing hole transforms into 
a moving one.
Indeed, the cores of moving holes were found to be more stable
than those of the standing ones (Chat\`e and Manneville, 1992).


\subsubsection{Destruction of Nozaki-Bekki holes by   small perturbations} 
Consider the following ``perturbed cubic CGLe''

\begin{eqnarray}
\partial_t A = \left[1 + (1+ib) \partial_x^2 - (1+ic)|A|^2
+ d |A|^4 \right] A
\label{eq1}
\end{eqnarray}
where a quintic term with a small complex prefactor
$ d = d' + i d ''$ ($ |d| \ll 1 $) is included, which will be treated as a
perturbation.
There are, of course, other corrections to the cubic CGLe, but
their ``perturbative effect'' is expected to be similar.

\begin{figure}
\centerline{\psfig{figure=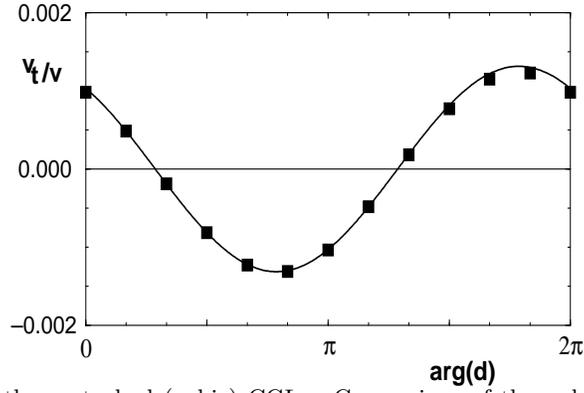,height=2.in}}
\caption{
Acceleration instability in the perturbed (cubic) CGLe.
Comparison of the reduced acceleration $\partial_t v /v$
from theory (full line, Eq.(\protect \ref{phen_a})) and simulations (squares)
for $b=0.5, \; c=2.0, \; |d|=0.002$
and varying phase $arg(d)$, Stiller {\it et al} (1995)a,b.
}
\label{fig_h2}
\end{figure}

Simulations with small but finite $  d $ 
show that stable,
moving holes are in general either accelerated and eventually
destroyed or slowed down and stopped to the standing
hole solution depending on the phase of $ d$ 
(Popp {\it et al} 1993, 1995; Stiller {\it et al} 1995a,b).
In particular, for real $  d =  d' $ one has
\begin{eqnarray}
\label{slowed}
\left[\frac{\partial_t v}{v}\right]
  {>\atop <}  0 \Leftrightarrow   d' {>\atop <} 0 \ .
\end{eqnarray}
One finds that the relations (\ref{phasecons}) connecting
the core velocity and the emitted wavenumbers
are (almost) satisfied at each instant
during the acceleration process.
The acceleration thus occurs approximately
along the Nozaki-Bekki hole family
and it can be described by taking  $ v = v(t) $ as a slowly
varying variable while
other degrees of freedom follow adiabatically.
The semianalytic matching-perturbation approach gives the reduced acceleration,
which agrees with the simulations, see Fig. \ref{fig_h2}.

Of special interest is the case where the core-stability line
is crossed while  $d \neq 0 $
(which corresponds to the typical 
situation with higher-order corrections to the CGLe).
Then the two modes which cause the acceleration instability
and the core instability (which is stationary for $  d =0 $)
are coupled which in the decelerating case leads to a Hopf bifurcation.
As a result, slightly above the (supercritical) bifurcation, one has solutions 
with oscillating hole cores (Popp {\it et al} 1993), see Fig. \ref{fig_h5}d.  
The normal form for this bifurcation -- valid for small $ |d|,v,u $ -- is
     \begin{eqnarray}   \label{phen}
\dot{u} &=& (\lambda - s v^2)u + d_1 v     \\
\dot{v} &=&  \mu u             + d_2 v \quad ,
     \end{eqnarray}
which is easy to analyze. Here $u$ and $v$ are the amplitudes of the
core-instability and acceleration-instability modes respectively.
$s$ and $\mu$ are of order $d^0$
while $d_{1,2}$ must be of order $d^1$, since in the absence
of a perturbation holes with nonzero velocity exist.
$\lambda$ can be identified
with the  growth rate  of the core-instability mode at $d=0$.
The nonlinear term in Eq.(\ref{phen}) takes care of the
fact that moving holes are more stable
than standing ones and at the same time saturates the instability
($s>0$).

Far away from the core-instability threshold where $\lambda$ is
strongly negative
$u$ can be eliminated adiabatically from Eqs.(\ref{phen})
which for $v \to 0$ yields
\begin{eqnarray}
  \label{phen_a}
  \dot{v} = (d_2 - d_1\frac{\mu}{\lambda}){v}
\end{eqnarray}
The term in brackets can be identified with the growth rate of the
acceleration instability.
The parameters of Eqs. (\ref{phen_a}) were calculated fully
analytically by Stiller {\it et al} (1995a) 
for $b,c \to \infty$, $b c >0$.

\subsubsection{Arrangements of holes and shocks}
\label{per_sol}

When there are more than one hole they have to be separated by shocks. 
The problem of several holes is analogous to that of interacting
conservative particles, which is difficult to handle numerically.
The solutions resulting from a periodic arrangement of holes and shocks are
actually special cases of the quasiperiodic solutions -- ``homoclons'' in
the limit of large periodicity -- to be discussed in the next subsection.
In the situation discussed here the solutions depend sensitively on 
perturbations of the CGLe

\begin{figure}
\centerline{\psfig{figure=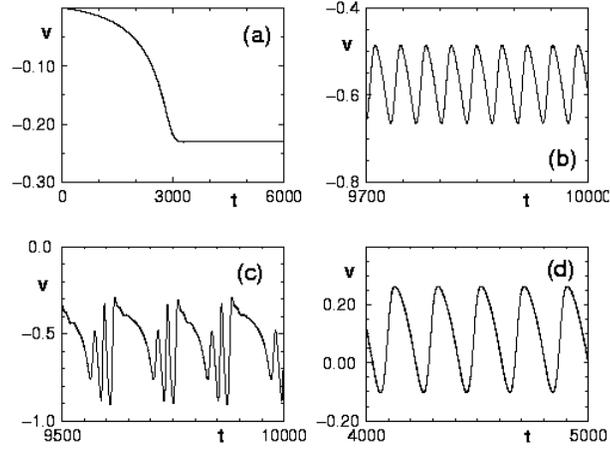,height=2.5in}}
\caption{
Simulations showing the velocity of the hole
center $v=v(t)$ of interacting
hole--shock pairs (periodic boundary conditions).
In Figures (a),(b),(c) the CGLe parameters were
$b=0.5, c=2.3, d'=+0.0025$
(i.e. far away from the core instability line).
(a) relaxation into a constantly moving solution
    for period $P=48.4$.
(b) selected final state with (almost) harmonic oscillating
    velocity for period $P=37.0$.
(c) selected final state with anharmonic oscillating
    velocity for period $P=40.0$.
(d) CGLe parameters $b=0.21, c=1.3,d'=-0.005$
    (near the core instability line), $d^{\prime \prime} =0$;
    state with oscillating velocity
    (including change of the direction)
    for period $P=50$.
}
\label{fig_h5}
\end{figure}

\begin{figure}
\centerline{\psfig{figure=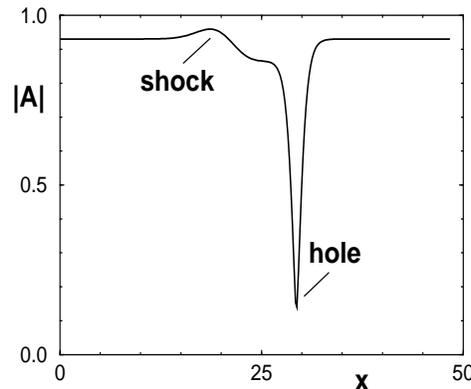,height=2.in}}
\caption{
Snapshot of the modulus $|A|=|A(x)|$ of a stable
uniformly moving hole--shock pair in a simulation for
$b=0.5,\; c=2.3,\; d'=+0.0025$.
The solution is space periodic with period $P=48.4$.
}
\label{fig_h4}
\end{figure}

Such states are frequently observed in
simulations with periodic boundary conditions
(see e.g. Manneville and Chat\'e (1992), Popp {\it et al} (1993), 
Stiller {\it et al}
(1995)a,b).
Figure \ref{fig_h4} shows the modulus $|A|=|A(x)|$ of a typical solution
found in a simulation.  
\footnote{For $d=0$ one expects to  observe conservative dynamics 
for holes and shocks, similar to interacting particles for zero friction.}
As shown in Figure \ref{fig_h5} one finds uniform as well as (almost)
harmonic and strongly anharmonic oscillating hole velocities.
(Slightly) beyond the core instability line the direction of the
velocity is changed in the oscillations (see Figure \ref{fig_h5} d).
The solutions are seen to be very sensitive to
$d$--perturbations of the cubic CGLe. 

The uniformly moving solutions can be
well understood from the results of the last subsections.
First, they are not expected to
exist (and we indeed could not observe them in simulations)
in the range of monotonic interaction, below curve MOH in Fig. \ref{fig-1CGLE},
\footnote{The boundary between monotonic and oscillatory interaction
          depends on the hole velocity (see Subsec. \ref{collisions}).}
since here the asymptotic hole--shock interaction is always attractive. 
In the oscillatory range, and away from the core instability line 
uniformly moving periodically modulated solutions can then
be identified as fixed points of a first-order ordinary differential equation
for the hole velocity $v$ which can be derived by the matching-perturbation
method (Popp {\it et al}, 1995).
In addition, solutions with oscillating hole velocities
were  found coexisting (stably) with uniformly moving solutions
 (see Figures \ref{fig_h5}b, \ref{fig_h5}c).
They can (in a first approximation)
be identified as stable limit cycles of a two-dimensional dynamical system with
the hole velocity $v$ and the hole--shock distance $L$ as active variables.
These oscillations are of a different nature than those found slightly
beyond the core instability line (Fig. \ref{fig_h5}d), see above.

By an analysis of the perturbed equations (\ref{count}) Doelman (1995) has
shown that the quintic perturbations create large families of traveling 
localized structures which do not exist in the cubic case.

\subsubsection{Connection with experiments}
Transient hole-type solutions were observed experimentally by
Lega {\it et al} (1992), and 
Flesselles {\it et al} (1994)    in the (secondary) oscillatory instability in
Rayleigh-B\'enard convection in an annular geometry.
Here one is in a parameter range
where holes are unstable in the cubic CGLe, so that small
perturbations are irrelevant.
Long time stable stationary holes (``1d spirals'') were
observed in a quasi-1d chemical
reaction system (CIMA reaction) undergoing a Hopf bifurcation by
Perraud {\it et al} (1993).
The experiments were performed in the vicinity of the
cross-over (codimension-2 point) from the
(spatially homogeneous) Hopf bifurcation
to the spatially periodic, stationary Turing instability.
Simulations
of a reaction-diffusion system (Brusselator) with appropriately
chosen parameters exhibited the hole solutions
(and in addition more complicated
localized solutions with the Turing pattern appearing in the core
region).
Strong experimental evidence for Nozaki-Bekki holes in hydrothermal nonlinear 
waves is given by Burguette {\it et al} (1999). 

Finally we mention experiments by Leweke and Provensal (1994), 
where the CGLe is used to describe results of open-flow experiments
on the transitions in the wake of a bluff body in an annular
geometry.
Here the sensitive parameter range is reached and in the observed
amplitude
turbulent states holes should play an important role.

\subsection{Other coherent structures} \label{other_cs}

Coherent states (\ref{coherent1}) with periodic functions
$a$ and $\psi= d\phi/dx$ (same period) have been of particular interest.
In general, when the spatial average of $\psi$ is nonzero, the complex amplitude
$A$ is quasiperiodic, i.e. $A$ can be written in the form (\ref{trafo}) with
$B$ a periodic function of $\zeta=x-vt$. The associated wavenumber $p$ will be 
called the ``inner wavenumber'', in contrast to the outer wavenumber $Q$,
which is equal to the spatial average of $\psi$.
%
%
These quasiperiodic solutions 
bifurcate from the traveling waves (\ref{plane_wave}) in the
Eckhaus unstable range $Q_E^2<Q^2<1$ at the neutrally stable positions obtained
from Eq. (\ref{dispersion}) with $Re \lambda=0$ and $k$ replaced by $p (|p|<2)$.
In fact, the long-wave Eckhaus instability is signalizes in particular
by bifurcations of the solitary (or homoclinic) limit solution $p \to 0$.
In this limit the velocity $v$ is at the bifurcation equal to the group
velocity $V_g$, see Eq. (\ref{long-wavepar}).

Well away from the Benjamin-Feir-Newell instability (on the stable side) the
bifurcation is subcritical (as for the GLe).
However, it becomes supercritical before the Benjamin-Feir-Newell
line is reached (Janiaud {\it et al} (1992))
and remains so in the unstable range.
The bifurcation is captured analytically most easily in the long-wave limit
by phase equations, see Subsec. \ref{phaseequations}.
The supercritical nature of the bifurcation allows to understand the
existence of ``phase chaos''  that is found when the Benjamin-Feir-Newell
line is crossed (see below).

\subsubsection{The GLe and NLSe}
\label{other_GLe}
For the GLe a full local and global bifurcation analysis is possible and the 
quasi-periodic solutions
can be expressed in terms of elliptic functions 
(Kramer and  Zimmermann, 1985; Tuckerman and Barkley, 1990).
Indeed, for $b=c=0$ Eqs. (\ref{count}) 
lead to the second-order system
\begin{equation}
a_{xx}= \partial_a U \mbox{   where   }
U= \frac{1}{2}[-a^2 +\frac{1}{2} a^4 - h^2/a^2].
\end{equation}
with $h=a^2 \psi$ an integration constant.
This allows to invoke the mechanical analog of a point particle
(position $a$, time $x$ ) moving in the potential $-U(a)$.
On sees that for $|h|<h_E=\sqrt{4/27} \approx 0.385$ the potential $-U$
has extrema at $a_{1,2}$ with $a_{1,2}^2\sqrt{1-a_{1,2}^2} = h$
corresponding  to the plane-wave solutions (\ref{plane_wave}) with
$F_{1,2}=a_{1,2}$, $|Q_{1,2}|=\sqrt{1-a^2_{1,2}}$.
The maximum corresponds to $|Q_1|<Q_E:= 1/\sqrt{3}$.
It is stable since it correspond to a minimum of ${\cal V}$.
The solution  $|Q_2|>Q_E$ is unstable.
In this way the Eckhaus instability is recovered.
The bifurcations of the quasi-periodic solutions 
are subcritical and the solutions are all unstable.
They represent the saddle points separating (stable) periodic solutions of
different wavenumber, and thus characterize the barriers against
wavelength-changing processes involving phase slips.

As was already mentioned in the Sec. \ref{sec:gencons}, all stationary solutions
and their stability properties of the
1D  (defocusing) NLSe coincides with those of the GLe.
In addition, the NLSe has more classes of coherent structures due to
the additional Galilean and scaling invariance absent in GLe.
We will not touch this question since the 1D NLSe is a fully integrable
system (Zakharov and  Shabat, 1971) 
that has been studied in great detail
(see, e.g. Proceedings of  Conference on
The Nonlinear Schr\"odinger Equation (1994)).
 
\subsubsection{The CGLe} \label{other_CGLe}

For the CGLe the local bifurcation analysis has been first carried out in
the limit  of small $p$ using the phase equation (\ref{special1Dphase}) 
(Janiaud {\it et al} 1992) and subsequently for arbitrary $p$ 
(Hager and Kramer, 1996).
It shows that the Eckhaus instability becomes supercritical slightly before the
Benjamin-Feir-Newell curve is reached.
The bifurcation has in the long-wave limit a rather 
intricate structure, since the limits $p \to 0$ and $Q \to Q_E$ do not
interchange (this can already be seen in the GLe).
When taking first the solitonic limit $p \to 0$ (while $Q>Q_E$) the 
Eckhaus instability becomes supercritical slightly later than in
the other (``standard'') 
case, which corresponds to harmonic bifurcating solutions.
These features are captured nicely by the phase equation description.

\begin{figure}
\centerline{\psfig{figure=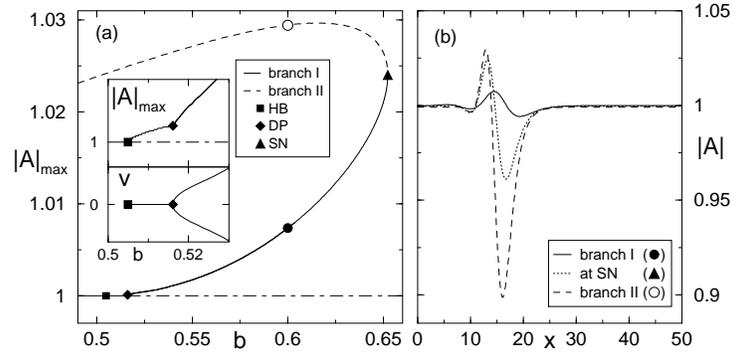,height=2.in}}
\caption{
(a) Example of the bifurcation diagram of the quasiperiodic solutions
 for $c=-2.0,
p=2 \pi/50$ (see text). The inset illustrates the drift pitchfork bifurcation
($v\!=\!0$ branch not shown beyond bifurcation). (b) QPS profiles at lower
(full circle) and upper (open circle) branch, and at the saddle-node (SN)
(triangle).
HB and DN mean  Hopf bifurcation and drift pitchfork bifurcation,
Brusch {\it et al}, 2000
}
\label{fig2_br}
\end{figure}

Recently Brusch {\it et al} (2000) carried out a systematic 
numerical bifurcation 
analysis based on Eqs. (\ref{count}) for the case $Q=0$, 
where $A$ is periodic (nevertheless we will usually refer to these 
solutions as quasi-periodic solutions).
It shows that the supercritically bifurcating quasi-periodic 
branch ({\em shallow quasi-periodic solutions})
terminates in a saddle-node  bifurcation and merges there with 
an ``upper'' branch ({\em deep quasi-periodic solutions}), 
see Fig. \ref{fig2_br}. 
For $p \ne 0$ the bifurcating solution has velocity $v=0$ 
(in Eqs. (\ref{count})  the bifurcation is of Hopf type). 
It is followed by a drift-pitchfork bifurcation generating a nonzero
velocity.
The separation between the two 
bifurcations tends to zero for vanishing $p$, so 
that the solitary solutions develop a drift from the beginning on (at second
order in the amplitude).
\footnote{for $Q \ne 0$ there is drift already after the first bifurcation, so
the drift-pitchfork gets unfolded (Brusch,  private communication).}
These features can be reproduced by the Kuramoto-Sivashinsky equation, which only exhibits the shallow
homoclons.
By adding higher-order (nonlinear) terms 
(Sakaguchi,  1990) the other branch
can be generated.
Brusch {\it et al} (2000)
 give evidence that the existence of the two branches
provides a mechanism for the stabilization of phase chaos
(see below).

Deep quasi-periodic solutions
(in the solitary limit) were first studied by van Hecke (1998) and 
then by van Hecke and Howard (2001) 
in the context of spatio-temporal chaos in the intermittent regime (see next
subsection). Following this author we have adopted the name ``homoclons'' for the
localized objects.

The stability properties of quasi-periodic solutions
were also analyzed by Brusch {\it et al}., 2000. 
Both branches of quasi-periodic solutions
have neutral modes corresponding to translation and
phase symmetries. The eigenvalue associated with the 
saddle-node bifurcation
is positive for deep homoclons and negative for shallow ones. 
Apart from these three purely real eigenvalues, 
the spectrum consists of complex conjugated pairs.

For not too small wavenumber $p$ all the eigenvalues of the 
shallow branch are stable within a system of length $L=2 \pi/p$, 
but when $L$ increases, the quasi-periodic solutions  may become 
unstable with respect finite-wavelength instabilities (near the 
bifurcation they certainly do). 

Shallow quasi-periodic solutions 
can  be observed in simulations of the CGLe (away from 
the bifurcation from the plane waves),  
in particular for $Q \ne 0$ (Janiaud {\it et al}, 1992;
Montagne, {\it et al} 1996, 1997;
Torcini,  1996;
Torcini, Fraunkron,  and Grassberger, 1997).

Also more complex coherent structures, 
corresponding to nonperiodic arrangements 
of shallow homoclons were found numerically.
They can be understood from the fact that the interaction between shallow
homoclons can be  of oscillatory nature depending on parameters.
This  is suggested by the fact
that in the supercritical regime the spatial exponents introduced in
Sec.  \ref{collisions} are complex.
Clearly chaotic solutions may be expected to exist in Eqs. (\ref{count}).

Recently, quasi-periodic solutions have been observed experimentally in the
form of modulations of spiral waves excited by core meandering 
(see Sec. \ref{corins},\ref{break-up})
in an oscillatory reaction-diffusion system by Zhou and Ouyang (2000).
\subsection{Spatio-temporal chaos}
\label{stc_1d}

\subsubsection{Phase chaos  and the transition to defect chaos}
\label{stc_1da}

When crossing the Benjamin-Feir-Newell
line with initial conditions $A \approx 1$ 
(actually any nonzero, spatially constant $A$ is equivalent) one 
first encounters phase chaos, which persists approximately in the lower
dashed region between  Benjamin-Feir-Newell and 
absolute instability  curves in Fig. \ref{fig-1CGLE}.
In this spatio-temporally chaotic state $|A|$ remains saturated 
(typically above about $0.7-0.9$) so there
is long-range phase coherence and the global phase difference (here equal 
to $0$) is conserved.
In this parameter range one also has stable, periodic solutions, 
but they obviously have a small domain of attraction which is not reached 
by typical initial conditions.
Beyond this range phase slips occur and a state with a nonzero (average) rate of
phase slips is established (defect or amplitude chaos).
Since in phase chaos only the phase is dynamically active it can be described by phase
equations (in the case of zero global phase the the Kuramoto-Sivashinsky equation, otherwise
Eq. (\ref{special1Dphase})).

In the CGLe, phase chaos
and the transition to defect chaos (see below), was first studied by 
Sakaguchi (1990) and subsequently systematically in a large parameter range by
Shraiman {\it et al} (1992) and for selected parameters and in particular 
very large systems by Egolf and Greenside (1995).
One of the interests driving these studies was the question if in phase chaos
the rate of phase slips is really zero, so it could represent a separate
phase (in the thermodynamic sense), or if the rate is only very small.
The recent studies by Brusch {\it et al} (2000) on the quasi-periodic solutions  with small modulation
wavenumber $p$ (see subsec. \ref{other_CGLe}) 
point to a mechanism that prevents
phase slips in a well-defined parameter range.

\begin{figure}
\centerline{\psfig{figure=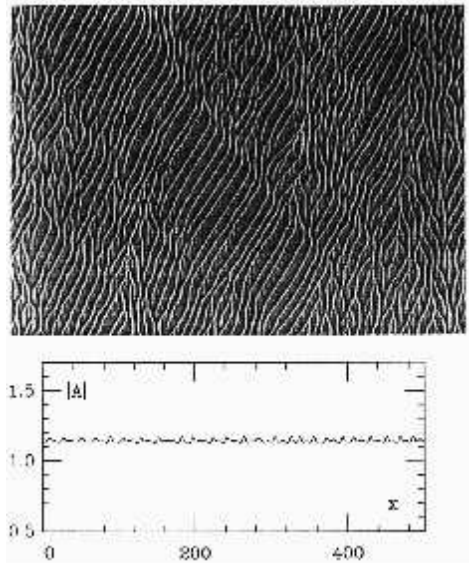,height=3.in}}
\caption{
Phase turbulence observed for $b=-1$ and $c=1.333$. Part (a) is
a spatio-temporal plot of $A$, part $b$ shows snapshot $|A|$,
Chat\'e, 1994.}
\label{fig_ph}
\end{figure}

In fact, phase chaos ``lives'' on a function space spanned by the forwardly bifurcating 
branches of the quasi-periodic solutions.
Thus, snapshots of phase chaos can be characterized (roughly) as a disordered array of 
shallow homoclons (or ``disordered quasi-periodic solutions''), see Fig. \ref{fig_ph}. 
The dynamics can be described in terms of birth and death processes of 
homoclonal units.
When these branches are terminated by the saddle-node bifurcation,
phase-slip processes are bound to occur. 
Brusch {\it et al} (2000) have substantiated this
concept by extensive numerical tests with
system sizes ranging from $L\!=\!100$ to $L\!=\!5000$ and integration
times up to $5 \times 10^6$. 
For a given $L$ (not too small) phase slips occurred only past
the saddle-node for the quasi-periodic solutions 
with inner wavenumber $p=2 \pi/L$.
Tracking of (rare) phase-slip events corroborated the picture.
Thus, the authors conjectured that  the saddle-node line for 
$p \to 0$ provides a strict lower boundary for the transition from
phase to defect chaos. 

One expects existence
of a continuous family of two types of 
phase chaos with different background wavenumbers $Q$.
For $Q \ne 0$ the state should arise when crossing the Eckhaus boundary for
the plane wave with wavenumber $Q$, before the Benjamin-Feir-Newell limit. 
Such states were studied
by Montagne {\it et al} 
(1996,1997) and Torcini (1996) and Torcini {\it et al} (1997).
It was found that the parameter range exhibiting phase chaos decreases with 
increasing $Q$. 
Thus, at fixed parameters $b,\ c$, one has a band of phase chaotic  states which is bounded 
from above by some $Q_{max}(b,c)$. Approaching the limit of phase chaos, 
$Q_{max}$ decreases 
smoothly to zero, i.e. the phase chaotic state with $Q=0$ is the last to loose stability,
similar to the situation for the plane waves.
As mentioned before, for not too small $Q$, one finds coexistence with 
quasi-periodic solutions and
spatially disordered coherent states.
\subsubsection{Defect chaos}

The transition (or crossover) between phase  and defect chaos
is reversible only for 
$|b|$ larger than about $1.9$ (near the lower edge of Fig. \ref{fig-1CGLE})
[Shraiman {\it et al}, (1992)].
There, when approaching the transition from the side of defect chaos,
 the phase slip rate goes smoothly to zero. 
On the other hand, for $|b|< 1.9$, there is a region where 
phase and defect chaos coexist
(``bichaos'', lower dashed region to the right of line DC), and
in fact defect chaos even persists 
into the Benjamin-Feir-Newell stable range (the limit towards small
$c$ is approximated by the dashed line DC in Fig. \ref{fig-1CGLE}) 
(Shraiman {\it et al}, 1992; Chat\'e, 1993; Chat\'e and Mannneville, 1994).
There the defect chaos takes on the form of spatio-temporal intermittency.

A qualitative understanding of the parameter region where one has
bistable defect chaos can be obtained as follows: 
In this region, starting from the ("saturated") state with $|A| \approx 1$ 
and forcing a large excursion of $|A|$ leading to a phase slip, makes 
it easier for other phase slips to follow. 
This memory effect is suppressed with increasing $|b|$ as can be seen from
the increase of the velocity $v^*=2 \sqrt{1+b^2}$ with $|b|$ of fronts that 
tend to restore a saturated state (see Sec. \ref{count_arg}).

From above (towards small values of $|b|$) the chaotic state joins up
with the region of stable Nozaki-Bekki holes. 
The characteristics of these holes is
influenced by small perturbations of the CGLe, and this in turn
affects the precise boundary of of spatio-temporal chaos, see Subsec. \ref{NB_boundary}.

\subsubsection{The intermittency regime}
\label{homoc-intermit}

This regime where defect chaos coexists with stable plane waves has been studied 
numerically in detail by Chat\'e (1993, 1994)
and he has pointed out the relation with spatio-temporal intermittency.
There, typical states consist of patches of plane waves, separated
by various localized structures characterized by a depression of $|A|$.
The localized structures can apparently be divided into two groups depending on
the wavenumbers $q_l$ and $q_r$ of the asymptotic waves they connect.
On one hand one has slowly moving structures which can be related to 
Nozaki-Bekki holes, 
which in this regime are either core stable or have long lifetime.
They become typical as the range of stable Nozaki-Bekki holes is
approached (Chat\'e, 1993, 1994).

\begin{figure}
\centerline{\psfig{figure=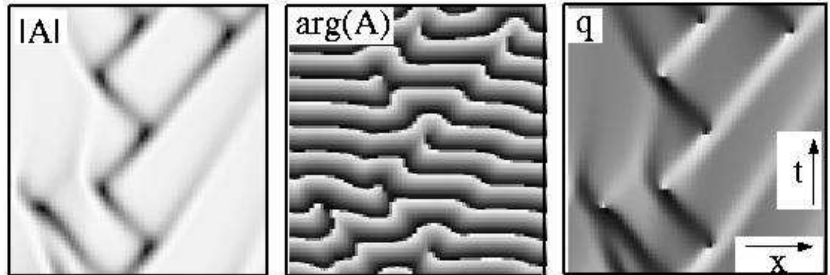,height=1.5in}}
\caption{
 Space-time plots (over a range of $60\times50$)
of $|A|$ (black corresponds to $|A|\!=\!0$)
showing chaotic states in the spatiotemporal intermittent regime,
for coefficients $(b,c)=(0.6,-1.4)$, van Hecke, 1998.
}
\label{fig2_vh}
\end{figure}

\begin{figure}
\centerline{\psfig{figure=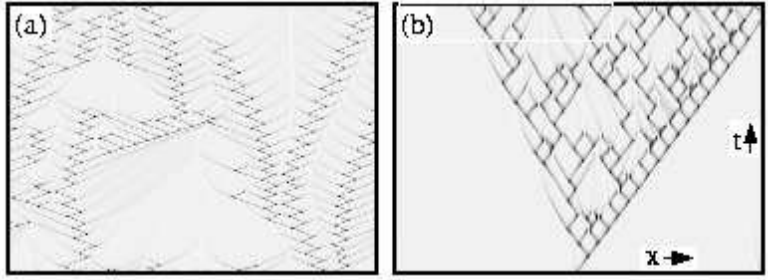,height=1.5in}}
\caption{Space-time plots (over a range of $512 \times 1000$) of $|A|$
showing  zigzagging holes near the transition
to plane waves for $b=-0.6,\ c=1.2$ (a);
evolution of a homoclons in a background state
with wavenumber 0.05 for $b=-0.6, c=1.4$,
space $\times $ time $=512 \times 250$ (b),
van Hecke, 1998. }
\label{fig5_vh}
\end{figure}

However, in a larger range the dominant local structures have velocities and
asymptotic wavenumbers that are incompatible with the Nozaki-Bekki
holes 
and which, according to van Hecke (1998), must be associated with the 
deep homoclons. 
Space-time plots of the amplitude $|A|$, the phase ($\arg(A)$), and
the local wavenumber $q\!:=\!\partial_x \arg(A)$ in such a regime
are shown in Fig. \ref{fig2_vh} (the plot is a zoom in of a larger simulation 
as shown in Fig. \ref{fig5_vh}b).
The wavenumbers of the laminar patches are quite close to zero,
while the cores of the local structures are
characterized by a sharp phase-gradient (peak
in $q$) and dip of $|A|$).  The holes propagate with a
speed of $0.95\pm 0.1$ and either their phase gradient spreads out
and the hole decays, or the phase-gradient steepens and the
hole evolves to a phase slip. The phase slip then sends out one (or maybe
two) new localized objects which repeat the process.
Van Hecke (1998) provides evidence that the dynamics evolves for 
much of the time
around the one-dimensional unstable manifold of homoclons with 
background wavenumber near zero.
They provide the saddle-points that separate dynamical processes which 
lead to phase slips from those that lead back to the (laminar) plane-wave state.
When the parameters $b$ and $c$ are  quenched in the direction of
the transition to plain waves, these zigzag motions
of the holes becomes very rapid (Fig. \ref{fig5_vh}a).

Additional consideration of homoclons and their role 
in spatio-temporal chaos  is presented by 
van Hecke and Howard, 2001. 
The simulations of the CGLe show that when 
an unstable hole invades a plane wave state,
defects are nucleated in a regular, periodic fashion,  and 
new holes can then be born from these defects. 
Relations between the holes and defects obtained from a detailed numerical
study of these periodic states are incorporated into a 
simple analytic description of isolated ``edge'' holes, which are seen in 
Fig. \ref{fig5_vh}. 

Recently Ipsen and van Hecke (2001) found in long-time simulations
that in restricted parameter range composite zigzag patterns formed by 
periodically oscillating homoclinic holes represent the attractor.

\subsubsection{The boundary of defect chaos  towards Nozaki-Bekki holes}   
\label{NB_boundary}

As the boundary of stability of Nozaki-Bekki holes is approached 
(curve AH in Fig. \ref{fig-1CGLE}) one can increasingly observe Nozaki-Bekki
hole-like
structures that emit waves and 
thereby organize (``laminarize'') their neighborhood. 
In this regime the state becomes sensitive to small perturbations of the 
cubic CGLe, as discussed in Sec. \ref{NB_hole}.

\begin{figure}
\centerline{\psfig{figure=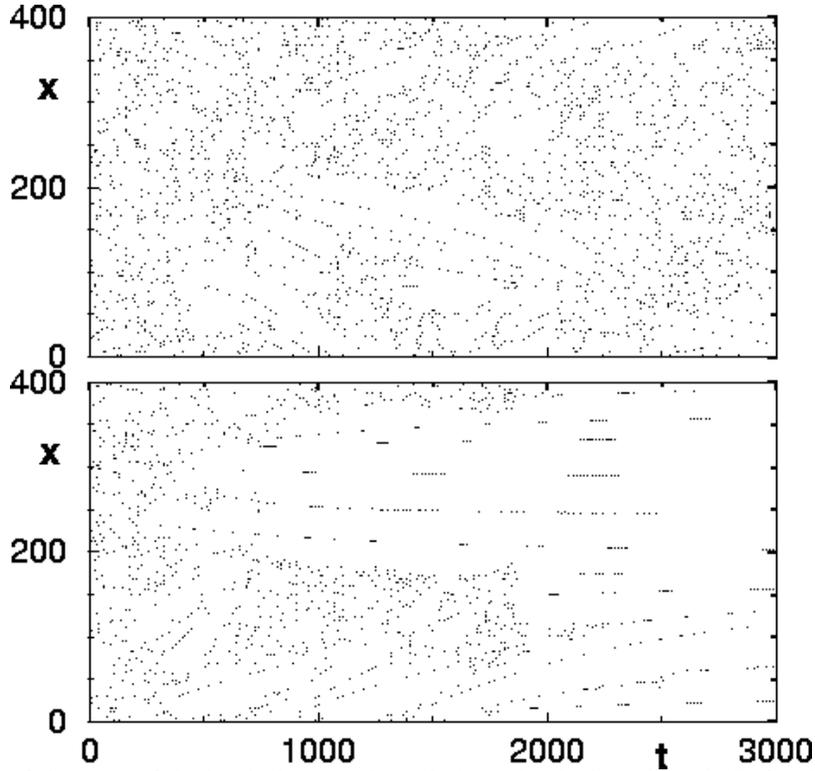,height=4in}}
\caption{
Space-time plot of the zeros of the local
phase gradient $\partial_x \phi=0$ (with $A=|A| \exp [ i \phi|]$)
for the perturbed cubic CGLe. The
parameters are $b=0.9, c =5.55$ (i.e. slightly below the absolute instability
line for the plane waves AH of the Figure \protect \ref{fig-1CGLE}) and
$d'=0.005$ (upper figure), $d'=-0.005$ (lower figure), 
$d ^{\prime \prime } =0$. Periodic
boundary conditions and small amplitude noise as initial conditions.
Isolated lines can be identified as holes and shocks (alternating)
which are spontaneously formed out of the chaotic state.
In their neighborhood one has approximately plane waves.
For decelerating $d'$ the holes suppress the chaotic state
after some transient (lower figure) and one is left with a 
``1D vortex glass'',  while in the accelerating case
($d'>0$, upper figure) , the chaotic state persist.
}
\label{fig_s1d}
\end{figure}

It is found that stable hole solutions suppress
spatio-temporal chaos and as a consequence, for a decelerating ($d^\prime <0$
if $d^{\prime \prime}=0$)
perturbation, the (upper) boundary of spatio-temporal chaos is simply
given by the stability boundary AS in Fig. \ref{fig-1CGLE} of the Nozaki-Bekki
hole solutions,
whereas for an accelerating  perturbation spatio-temporal
chaos is also observed further up (Popp {\it et al} 1993; 
Stiller {\it et al} 1995b).
For $d'<0$ random initial conditions lead to an irregular grid 
of the standing holes, separated by shocks from each other, 
see Fig. \ref{fig_s1d}b. 
Such grids are the 1D analog of the ``vortex glass'' state found in 2D
(see Subsec. \ref{vgs}), see also Chat\'e, 1993. 

Clearly, for accelerating $d'>0$ such a grid is unstable. In this case 
Nozaki-Bekki holes are created from random initial conditions, too. In their 
neighborhood they suppress small-scale variations of $|A|$ 
typical for amplitude chaos in the major part of the chaotic regime. 
However, since they are accelerated they only  have  finite life time,
see Figure \ref{fig_s1d}a. For parameters $b,c$
below about curve EH in Fig. \ref{fig-1CGLE} (the coincidence with the
Eckhaus limit presumably  fortuitous) the destruction of 
the holes leads to creation of new 
holes and shocks, yielding a chaotic scenario 
of subsequent acceleration, destruction and creation processes.


\section{Dynamics in 2D}
\label{sec:2D}
\subsection{Introduction}  \label{3_intr}

\begin{figure}
\centerline{\psfig{figure=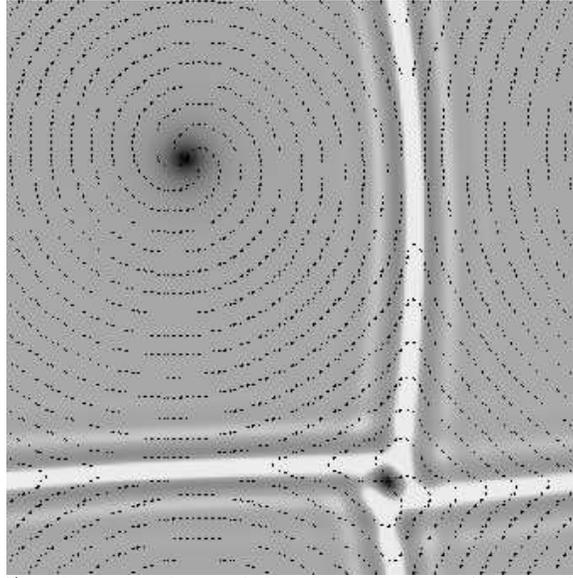,height=3in}}
\caption{
Image of the amplitude $|A|$ and contour lines
of the  phase $\phi=\arg A=0, \pi$ for
single-charged spiral solution in a domain
with periodic boundary conditions. Note  edge vortex in the corner.
Image is   coded in the grey scale,
maximum of the field corresponds to black, minimum to white.}
\label{fig_s}
\end{figure}

The 2D CGLe has a variety of coherent structures.  
In addition to the quasi-1D solutions derived from the coherent structures 
in 1D discussed above the CGLe possesses localized sources in 2D known 
as spiral waves. An isolated spiral solution is of the form
\begin{equation}
A(r, \theta, t) = F(r) \exp [ i (-\omega t + m \theta + \psi(r))]
\label{spiral}
\end{equation}
where $r, \theta$ are polar coordinates.
The 
(nonzero) integer $m$ 
is the topological charge,
and $\omega$ is the (rigid) rotation frequency of the spiral, 
$F(r)>0$ is  the amplitude and $\psi(r)$ is the phase of the spiral, 
$Q=\partial_r \psi$ for $r \to \infty $ is the  asymptotic wavenumber
selected by the spiral (Hagan 1982), 
and  $\omega=c+(b-c)Q^2$  is the spiral frequency 
\footnote{Spirals with zero topological charge $m=0$ (targets) are unstable
in the CGLe. However, inhomogeneities can stabilize the target.
Hagan (1981) have found stable targets in the inhomogeneous CGLe 
in the limit of small $c$. Stationary  and breathing targets 
in the wide parameter range of the CGLe were studied by 
Hendrey {\it et al} (2000)}. 
Since  spirals emit asymptotically plane waves 
(group velocity $V_g=2 (b-c) Q >0 $), they are source 
solution. 
In addition to spirals there exist sinks which absorb waves. 
The form of sinks is determined by the configuration of surrounding sources. 
Sinks  with topological charge appear as edge vortices (see below).  
Spiral solutions with $m \ne \pm 1$ are unstable (Hagan, 1982).
Single-charged spiral solutions are dynamically stable in certain regions of
parameter space. The grey-coded image  of a spiral 
and an  edge vortex is shown in Fig. \ref{fig_s}.
For $b-c \to 0$, and also for $b,c \to \infty$, 
the asymptotic wavenumber of the spiral
vanishes and the solution goes over into 
the well-known vortex solution of the 
GLe and NLSe, known in the context of
superfluidity (Pitaevskii,1961, Gross 1963, Donnelly, 1991)
and somewhat similar to that found in superconductivity in the London limit
(Abrikosov, 1988, Blatter {\it et al} 1994).
In periodic patterns spirals and vortices become dislocations.

The asymptotic interaction is very different for the case $b=c$, where it
it is long range decaying like $r^{-1}$ with some corrections,  
whereas for $b-c\neq 0$ it is short range 
decaying exponentially (see below). 
Interaction manifests itself in a motion of each spiral. 
The resulting velocity can have a 
radial (along the line connecting the spiral cores) 
and a tangential component. 
 
Spirals may form regular lattices and/or disordered 
quasi-stationary structures called vortex glass or frozen state. 
When individual spirals become unstable (spiral break-up), 
the typical spatio-temporal behavior is chaotic. 

\subsection{Spiral stability}

\subsubsection{Outer stability}
\label{outstab}

For spirals to be stable the wavenumber of the asymptotic plane wave 
has to be in the absolutely stable range (see Sec. \ref{abs}).
In order to find this stability limit we have to evaluate the condition
(\ref{instab}) for the wavenumber $Q$ emitted by the spiral, see Fig.
\ref{fig_stab}.
Existence of an absolutely stable spiral solution guarantees that small 
perturbations within the spiral will decay, but does not assure that 
it will evolve from generic initial conditions. For further discussion see 
Sec.\ref{break-up}.

\subsubsection{Core Instability} 
\label{corins}

The spiral core may become unstable in a 
parameter range where diffusive effects are 
weak compared to dispersion (large $b$ limit).  
Then it  is convenient to rewrite CGLe in the form 
\begin{eqnarray}
\partial_t A = A  + (\varepsilon +i) \Delta A -(1 + ic)|A|^2 A .
\label{CGLEL}
\end{eqnarray}
where $\varepsilon = 1/ | b|$ and length has been
rescaled by $\sqrt \varepsilon$. 
This situation is typical in nonlinear optics (transversely 
extended lasers or passive nonlinear media). 
In this case a systematic derivation of  
the CGLe from the Maxwell-Bloch equations in the "good cavity limit"  for 
positive detuning between the cavity resonance and the atomic 
line leads to very
small values of $\varepsilon>0$ 
(Coullet {\it et al} 1989a; Oppo {\it et al} 1991;
Newell and Moloney, 1992; Newell, 1994). 
Representative values are $\varepsilon \sim 10^{-3} - 10^{-2}$ 
(Coullet et. al, 1989a).

For $\varepsilon=0$ one has the Galilean invariance mentioned in 
Sec.\ref{transform}, (see Saarloos and Hohenberg, 1992,  for the 1D case)
and then in addition to the stationary spiral there exists a family of spirals 
moving with arbitrary constant velocity ${\bf v}$
\begin{equation}
A(r ,t)=  F(r^\prime) \exp i[ -\omega^\prime t + \theta + \psi(r^\prime)
-\frac{ {\bf r}^\prime \cdot {\bf v}}2]
\label{mov}
\end{equation}
where ${\bf r}^\prime = {\bf r} + {\bf v} t$, $ 
\omega^\prime= \omega -v^2/4$,
 and the functions $F, \psi $  are those of
Eq.\ (\ref{spiral}) (this invariance holds for any stationary solution). 
For $\varepsilon \ne 0$ the diffusion  term $\sim \varepsilon \Delta A$ 
destroys the family and in fact leads to ac- or deceleration of the spiral 
proportional to $ \varepsilon v$.  
The natural assumption is that one has deceleration 
so that the stationary spiral is
stable (Coullet et. al, 1989a). In fact this is not
the case. Stable spirals exist only above some critical value  
$\varepsilon_c$.
Below $\varepsilon_c$ the stationary spirals are unstable with respect 
to spontaneous acceleration (Aranson {\it et al}, 1994).

For small values of $\varepsilon$ one may expect  
the solution (\ref{mov}) to be slightly perturbed and have 
a slowly  varying velocity ${\bf v}$ which obeys
an equation of motion of the form
$\partial_t {\bf v}+ \varepsilon \hat{K} {\bf v}  = 0$.
Because of isotropy the elements of the tensor $\hat{K}$ must satisfy 
$K_{xx}=K_{yy}$ and $K_{xy}=-K_{yx}$, so the relation can also be
written as 
\begin{eqnarray}
\partial_t \hat{v} +  \varepsilon \kappa \hat{v} = 0 
\label{new1}
\end{eqnarray}
where $ \hat{v} = v_y + i v_x$ and $ \kappa = K_{xx}-i K_{xy}$. 
Since in general the 
friction constant $\kappa$ is complex the spiral core moves on a 
(logarithmic) spiral trajectory. 

The acceleration instability of the spiral core has a well-known counterpart
in excitable media, where
the spiral "tip" can perform a quasiperiodic motion 
leading to meandering (see, e.g. Barkley, 1994).
The main difference between the two cases 
can be understood by considering the nonlinear extension of 
Eq.\ (\ref{new1})
$\partial_t \hat{v} + \varepsilon \kappa \hat{v} = f(|\hat{v}|^2) \hat{v}$
with $f(0)=0$. In excitable media one has $Re f(|\hat{v}|^2)<0$,
which provides the saturation of the 
instability. On the other hand, in CGLe
the sign of $Re f(|\hat{v}|^2)$ is opposite and destabilizes the core
according to simulations.
Thus one has an alternative scenario
of the meandering instability. 
The scenario appears to
be generic  and is not destroyed by small perturbations of the  CGLe.

Now going to larger $\varepsilon$ one finds that 
$Re \kappa$ increases with $\varepsilon$
and finally changes sign at a value $\varepsilon_c$. The result obtained from  
extensive numerical simulations is shown in 
Fig.\ \ref{fig10}.

The acceleration instability 
can also be interpreted as the 
destabilization of a localized core mode similar to the situation of the 
standing hole solution in the 1d CGLe (Popp {\it et al}, 1993, 
Sasa and Iwamoto, 1993; Chat\'e and  Manneville,  1992,
see Sec. \ref{sec:1D}). 

The case of
negative diffusion ( $ \varepsilon  < 0 , \; | \varepsilon| \ll 1$ )
is important especially for lasers with negative detuning (Oppo, 1991; 
Staliunas, 1993; Lega {\it et al}, 1994).
Then higher-order corrections to the diffusion (fourth derivative)
have to be included
to stabilize the short-wave instability. The resulting complex Swift-Hohenberg
equation is 
\begin{eqnarray}
\partial_t A = A  + i \Delta A -\varepsilon ( q_c^2+\Delta)^2 A  -(1 + ic)|A|^2 
A .
\label{CGLEL1}
\end{eqnarray}
This equation exhibits in some range of parameters 
$\varepsilon , q_c$ a similar core instability.  
In a certain parameter 
range one has stable meandering of the spiral 
(Aranson and Tsimring, 1995; Aranson, Hochheiser, and  Moloney, 1997)

\subsection{Dynamics of vortices in the GLe, NLSe, and for $b=c$}
\label{3_disl}

\subsubsection{Dynamics of vortices in GLe}
In the GLe the spirals become vortices, i.e in Eq. (\ref{spiral}) one has 
$\omega=0$ and $\psi=0$. 
$F(r)$ is a
monotonic function and $F \sim \alpha r $ for $r \to 0$, 
$ F^2(r) \to 1-1/r^2$ for $
r \to \infty$.

A more general isolated moving vortex solution with wavevector ${\bf Q} $ away from 
band center is described by :
\begin{equation}
A=B({\bf r -v} t) \exp[i{\bf Q r} ]
\label{stat1}
\end{equation}
where $B$ fulfills the boundary conditions $|B|^2 \to 1-Q^2$ for 
$r \to \infty$
with the appropriate phase jump of $\pm 2\pi$. 
\footnote{It is important to mention that such a problem is
not well defined for spiral waves ($b-c \ne 0$). Since the spiral waves are 
active sources they can in 
general not coexist with the plane waves.
Depending on the relation between
the two frequencies either the spiral would invade the entire plane or
the plane wave would push the spiral away. For details see Sec. \ref{symb}}

For large systems such that $R v \gg 1 $, where $R$ is the system size 
or the distance to another defect,  
the drift velocity $v$ is perpendicular  to the background 
wavenumber ${\bf  Q}$, such that the free energy decreases. 
The speed $v$ is given by the expression 
\begin{equation} 
\frac{1}{2} {\bf v} \log (v_0/v) = {\bf U Q} [1-0.35 Q^2] 
\label{mob1} 
\end{equation} 
with the constant $v_0=3.29$
(see Bodenschatz {\it et al}, 1988b, 1991, Kramer {\it et al},  1990). 
Here ${\bf U}$ is the $\pi/2$ rotation  matrix 
$${\bf U} = {0 \;\; 1 \choose  -1\;\;0}$$
The term in the brackets on the
rhs has been fitted to numerical results (see Bodenschatz {\it et al}. 1991). 
The formula can be used 
throughout the stable range and is consistent with experiments of
Rasenat {\it et al} (1990)
in electroconvection in nematics. 
In small systems $vR \ll 1 $, $R \gg 1$  the velocity is given by 
$v = 2Q/  \log (R/\xi_0)$ with $\xi_0=1.13$. 

The above relations were rederived by Pismen and Rodriguez 
(1991),  Rodriguez and Pismen (1991), Ryskin and Kremenetsky (1991), 
and in part by Neu (1990a) using matching asymptotics techniques.

The result shows that for an 
isolated defect the limits of the system size going
 to infinity 
and wavenumber mismatch 
(or defect velocity) going to zero are not interchangeable. This is
analogous to the 2D-analog of Stokes' law 
(drag of an infinite cylinder through an 
incompressible fluid), where one also has 
a logarithmic divergence of the mobility at
vanishing velocities. The similarity was 
pointed out by Ryskin and  Kremenetsky (1991). 
It is also useful to note that the force exerted on dislocations in a
phase gradient expressed by a wave vector ${\bf Q}$ is the analog of
the well-known Peach-Koehler force on dislocations in crystals under stress.


In the following we consider 
linearly stable periodic states, where $Q^2 <Q^2_E
$.
All periodic states, except $Q=0$ (band center), are metastable. 
Evolution to the band center can occur by the motion of vortices, once
they have been nucleated by finite fluctuations or 
disturbances (see Bodenschatz
{\it et al}., 1988b,
and  Hernandez-Garcia {\it et al}, 1993,  
for a discussion of the problem of homogeneous nucleation of
vortices).

The asymptotic interaction between two vortices  can be
described approximately  by Eq. (\ref{mob1}) 
with the  r.h.s. 
replaced by the gradient of the resulting phase of other  vortex 
at the center of the vortex. 
Since the gradient is perpendicular to the line connecting the vortex cores,
the resulting force is directed along this line.
As a result, oppositely-charged vortex attract each other and eventually 
annihilate, whereas like-charged repel each other. 

The analysis of an isolated moving vortex shows that the 
deformation of the roll pattern decays in front exponentially over a distance
$R \sim 1/v$, while behind the dislocation 
the decay is proportional to $R^{-1/2
}$.
Thus, if the background wavenumber is nonzero and two vortices with 
opposite topological charge are approaching each other, the velocity will be
constant for $R\le 2/v \sim 2/Q$. Subsequently 
the motion will accelerate and eventually 
the attraction will dominate over the 
Peach-Koehler force  
and lead to annihilation. 
\footnote{In this discussion we have tacitly considered the situation where 
the Peach-Koehler force and the  force due to
interaction are parallel to each other. 
If this is not the case the vortices will
 move on curved trajectories.}
%
In experiments by Brawn and Steinberg (1991),
the annihilation process was studied in detail near the band center ($Q \ll 1$) 
where accurate determination of $Q$ is not possible. The quantitative comparison
with theory  is shown 
in Fig. \ref{fig1d} where the distance $L$ 
between two vortices of opposite polarity 
approaching each other along a straight line versus time $T$ is 
shown for different $Q$ in scaled units, (Bodenschatz {\it et al}, 1991). 
At time $T=0$ the vortices annihilate. The solid line 
gives the numerical results, and the squares, 
circles and diamonds are the experimental data. 

The case $Q=0$, where the Peach-Koehler force vanishes so that $v \to 0$ for 
$L \to \infty$, deserves special attention. 
The analysis of Bodenschatz {\it et al} (1988)b,
leads to $v L = 2/C$, where $C\sim \ln(L/2.26)$ 
for $v L \ll 1$. Clearly for this to hold 
$L$ needs to be exponentially large. 
Otherwise one is caught in the intermediate region 
$v L \sim 1$ where no accurate relation for $C$
was found. For 
that reason the so-called self-consistent 
approximation was  proposed by Pismen 
and Rodriguez (1990), Rodriguez and Pismen (1991), Pismen (1999): 
\begin{eqnarray}
\ln (v_0/ v) =\exp(\pm v r/2)[K_0(vr/2) \pm K_1(vr/2)], v_0 = 3.29
\label{mot}
\end{eqnarray}
where the $+$ sign corresponds to a 
like-charged pair, the $-$ sign to an oppositely-charged
one, and $K_0$ and $K_1$ are modified Bessel 
functions. The formula is not very accurate, as
shown by
comparison with full simulations 
(Weber, 1991, 1992a). 

Allowing for a very small  wavevector displacement $Q$, 
one finds $v(L+1/Q)=2/C$.
One can then get to the limit $v L \gg 1$ and 
then, for sufficiently large systems, 
$C=\ln(3.29/v)$. 

For small distances, shortly before annihilation of an oppositely charged pair, 
the gradient terms in the GLe 
become dominant and one may argue that 
the dynamics is governed by self-similar solutions of the diffusion equation 
so that $L \sim \sqrt T$, which appears consistent with the numeric results. 

\subsubsection{Dynamics  of vortices in the NLSe}
\label{nse0}

Although the stationary vortex solution of the NLSe coincides with that of 
GLe, the dynamic behavior of NLSe  is very different from that of the GLe
and is similar
in the limit of small velocities, large separation, and properly quantized 
circulation to the motion of 
point vortices in an ideal incompressible fluid
(Lamb, 1932).
The problem of the vortex motion in the context of the 
NLSe was first considered 
by Fetter (1966), and then many researchers rederived 
this result more accuracy 
(see, e.g. Creswick and Morrison, 1980; Neu, 1990b; Lund, 1991, 
Pismen and Rubinstein, 1991a, Rubinstein and Pismen, 1994). 

Introducing the amplitude-phase representation $A= R \exp [i  \theta]$
and the ``superfluid density'' $\rho=R^2$ as well as the superfluid velocity 
${\bf V} = -2 \nabla \theta$ one obtains 
from Eq. (\ref{NLSe}) an Euler equation
\begin{equation}
\partial_t {\bf V } + ({\bf  V \cdot  } \nabla ) {\bf V}  = \nabla P 
\end{equation} 
and the continuity equation for the ``density'' of superfluid 
\begin{equation} 
\partial_t \rho + \mbox{div} \rho {\bf V} = 0. 
\end{equation} 

Here $P= 2( - \nabla^2 \sqrt \rho /\sqrt \rho + \rho -1)$ is the 
 effective ``pressure'',    whereas the first 
term  is called ``quantum pressure''.  
The quantum pressure is irrelevant 
for large-scale perturbations of the density, 
i.e. $ \nabla^2 \rho  \ll \rho$. 
In the Euler  equation 
incompressibility 
approximation $\rho=\rho_0 =const \ne 0$ 
is reliable for the velocities much smaller than the sound velocity  
$c_s=\sqrt {2 \rho_0}$. 
One expects the analogy with 
classical incompressible vortex tubes in a liquid 
to hold for the well-separated vortices  moving with the 
velocity $v \ll c_s$. 
Thus, according to classical hydrodynamics  ( Lamb, 1932), 
one obtains that the   
oppositely charged vortex pair drifts with the velocity $v = 2/L $ normal to 
the line connecting the cores of the vortices, and like-charged pair rotates around the center of 
the symmetry, where $L$ is the 
distance between vortex center. This result up to higher order corrections 
holds for well separated vortices ($L \gg 1$)  in the NLSe, as 
it was rederived later on by Neu (1990)b,  
Pismen and Rubinstein (1991),  and by Rubinstein and Pismen  
(1994),  using matching asymptotic technique. Numerical investigation of 
vortex dynamics in the  NLSe was carried out also by Nore {\it et al} (1993), 
Frisch {\it et al} (1992), Abraham {\it et al} (1995), 
Josserand and Pomeau (1995). 

For $N$ well-separated 
vortices located at the points ${\bf R}_j=(X_j,Y_j)$ one has the 
more general formula
\begin{equation}
{\bf v_j} = 2 \sum_{j=1}^N n_j \frac{{\bf U(R_j-R_i)}}{|{\bf R_i - R_j}|^2}
\label{solh}
\end{equation}
According to Eq. (\ref{solh}), obtained from the corresponding incompressible 
Euler equation,  the vortex velocities 
diverge as the intervortex distances vanish.
However,
the above expression becomes incorrect when the inter-vortex distance $d$
becomes of the order of the coherence length $d \sim O(1)$ (one in 
our scaling). For small separations the  interaction of two 
like-charged vortices can be considered as a small perturbation of a 
double-charged vortex. Aranson and Steinberg (1995) have shown that 
for $d \to 0$ the
velocity of vortices 
vanishes, and the frequency of rotation $\Omega$ for like-charged vortex pair 
approaches a constant  as $d \to 0$ 
(the ``classical formula'' (\ref{solh}) results in
infinite frequency). 
Similarly,  for an oppositely charged pair 
the velocity remains finite for $d \to 0$.

An important problem in the context of NLSe  is the 
nucleation (or generation) of vortices  
from  zero-vorticity flow. 
Frisch, Pomeau, and  Rica (1992) considered  
the stability of flow passing an obstacle 
and found nucleation of vortex pairs above some critical velocity.
Yet another mechanism of vortex generation  is related to the 
instability of quasi-1D
dark solitons, predicted by Kuznetsov and Turitsyn (1988). 
Josserand and Pomeau (1995) studied the instability of dark 
solitons numerically and
found that the nonlinear stage of this instability results in 
creation of vortex pairs.

\subsubsection{Dynamics of vortices for   $b=c$}
The general case $b=c$, including the NLSe limit, can be related 
to the GLe limit by generalizing the ansatz (\ref{stat1}) to
\begin{equation}
A=B({\bf r -v} t) \exp[i{\bf Q r}-ibt ].
\label{stat2}
\end{equation}
Eq. (\ref{CGLe})  can be written as
\begin{equation} 
-\frac{1} {1+b^2} {\bf v} \nabla B = \Delta B + 2 i \left(
{\bf Q} +\frac{ b} {2(1+b^2)} {\bf v} \right) \nabla B 
+ (1-Q^2) B - |B|^2 B .
\label{bec} 
\end{equation}
After neglecting  the higher-order term $Q^2 B$,  
one observes that Eq. (\ref{bec}) coincides with the corresponding 
equation for $b=0$ if one replaces ${\bf Q}$ by 
${\bf Q} +\frac{ b} {2(1+b^2)} {\bf v}$.
\footnote{The calculations can be generalized to arbitrary $Q$.}

As a result we may extend the expression (\ref{mob1}) for arbitrary $b$
leading to
\begin{equation} 
\frac{1}{2} {\bf v} \log ((1+b^2) v_0/v) =  {\bf U} \left[
 (1+b^2){\bf Q} +\frac{ b} {2} {\bf v} \right]
\label{mob2} 
\end{equation} 
One obtains oblique motion with respect to 
${\bf Q}$ with the angle depending on $b$ and $Q$. In the NLSe limit the motion
becomes parallel to ${\bf Q}$.
Concerning the interaction between vortices one obtains oblique repulsion or 
attraction: for $b \ne 0$ 
two vortices never move along the line connecting their 
centers but instead spiral with respect each other (Pismen, 1999). 

\subsection{Dynamics of spiral waves for $b \ne c$} 
\label{spdyn}
\subsubsection{General} 
As we mentioned above, in the generic  case  of 
$b \ne c $ the asymptotic interaction  of topological 
defects (spirals) is very different from the corresponding dynamics of 
vortices in GLe  and in NLSe. 
The problem of spiral wave interaction was considered by  
Biktashev (1989), Rica and Tirapegui (1990),(1991)a,b,(1992), 
Elphick and Meron (1991),
Aranson {\it et al} (1991),(1993)a, Pismen and  Nepomnyashchii (1991)a,b.
The approach of Rica and Tirapegui (1990),(1991)a and  Elphick 
and Meron (1991), 
explored a type of global phase approximation 
which in essence gave long-range interaction  $\sim 1/r$
between the spirals, in analogy to the cases treated above. 
The results were in obvious disagreement with numerical simulations 
done by Aranson {\it et al} (1991), 
revealing exponentially weak interaction between the spirals. 

A more adequate solution of the problem was obtained later by 
Aranson {\it et al} (1991), and Pismen and  Nepomnyashchii (1991)a,b,
on the basis of the phase diffusion equation (see also Biktashev, 1989).
The results demonstrated  an exponentially weak asymptotic 
interaction, although
they 
failed to describe the numerically observed bound states of spiral waves. 
The abovementioned  approach is adequate for  the case
$|b-c| \ll 1$. 
Biktashev (1989) considered a the very restrictive problem of 
spiral motion in an almost-circular domain.  
Due to some unphysical boundary conditions the results of Pismen and  
Nepomnyashchii (1991),(1992), 
were at conflict with numerical simulations (Aranson {\it et al} 1993a). 

In later papers Rica and Tirapegui (1991)b,(1992) attempted to improve their 
interaction approach by combining it with that of 
Aranson {\it et al} (1991). However, in our
view, the results remain unsatisfactory because of an 
uncontrolled  perturbation technique. 

\begin{figure}
\centerline{\psfig{figure=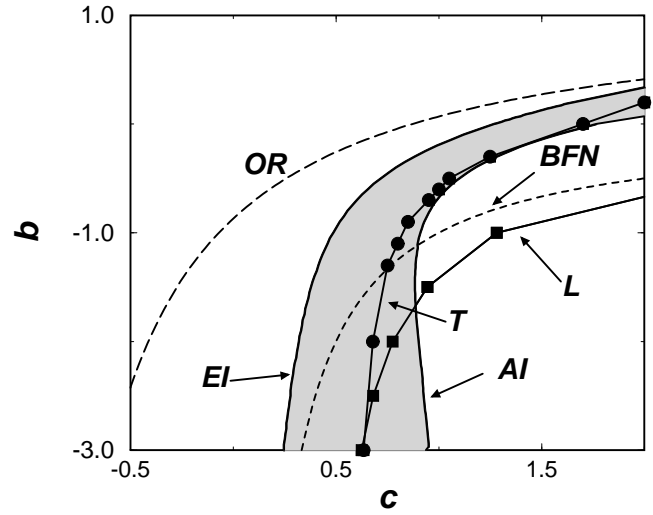,height=3in}}
\caption{
Stability limits of spiral wave solution.
The
boundary of convective ($EI$) and
absolute instability ($AI$) for the waves emitted by spiral
are also shown (for explanation see (Aranson {\it et al}, 1992)),
Bound states exist right of $OR$ line $(c-b)/(1+bc)=0.845$.
BFN indicates Benjamin-Feir-Newell limit $1+bc$, $L$ shows limit of
2D phase turbulence, and line $T$ corresponds to transition
to defect turbulence for random initial conditions
(Chat\'e and Manneville, 1996).
}
\label{fig_stab}
\end{figure}

A consistent theory with quantitative predictions 
of asymptotic spiral wave interaction 
including bound-state formation was presented by Aranson {\it et al} (1993)a,b
using the matching perturbation method.
The idea of the method  is straightforward: 
in the full solutions for a spiral pair (or a spiral and a wall, or
more complicated aggregates of spirals) the spirals move with certain
velocities, and thus solutions exist only
with the ``correct'' velocities. Such solutions may be constructed
approximately by starting with isolated spirals, 
each one restricted to the region in
space filled by its emitted waves 
and moving with (small) velocities to be determined.
For a symmetric spiral pair one simply has two half planes.
In a first step the corrections are assumed to 
be determined to sufficient accuracy by
the linearized (perturbational) problem with boundary conditions that take into
account the neighboring spiral (or wall etc.). 
The velocity comes in as an inhomogeneity.
The occurrence of bound states can already be seen from the behavior 
of stationary perturbations of the asymptotic plane waves emitted by
the spirals, as given by Eq. (\ref{asdisp}).
Bounds states occur in the oscillatory regime, outside the curves
$(c-b)/(1+bc)= \pm c_{cr},\ c_{cr} = 0.845$ (one must choose the curves inside
the Benjamin-Feir-Newell-stable region).
The oscillatory range is 
plotted  in  Fig.~\ref{fig_stab} (curve $OR$). 
Also included in Fig.~\ref{fig_stab}
is the convective 
Eckhaus-stability boundary (curve $EI$, see Sec. \ref{plw}) and
the boundary of absolute stability (curve $AI$, see Sec. \ref{abs}) for 
the waves emitted by the spirals.  

The inhomogeneities involve the velocity linearly and as a consequence
the solution for arbitrary
(but not too small) distance can be expressed in terms
of one inhomogeneous and a homogeneous solution. Once these are
determined numerically and used to match the
boundary conditions the velocity versus distance relation comes out.
For $b=0$ (for simplicity) the resulting velocities of the spiral at distance 
$X$ from a plane boundary (or, equivalently, a pair of  oppositely-charged 
spirals at the points $(\pm X,0)$) are of  the form 
\begin{eqnarray}
v_x &=&Im (\frac{-Q \sqrt{1-Q^2} \exp(-p X)}
{\delta C_y \sqrt{2\pi p X }} X^{- \mu}) / Im (C_x/C_y)  ,\nonumber \\
v_y &=&Re (\frac{-Q \sqrt{1-Q^2} \exp(-p X)}
{\delta C_y \sqrt{2\pi p X }} X^{- \mu}) - v_x Re (C_x/C_y)  .
\label{velxy}
\end{eqnarray}
where $Q$ is the spiral wavenumber, $v_x$ is the component of the 
velocity along the line connecting spiral cores (radial velocity) 
and $v_y$ is the velocity perpendicular to this line (tangential velocity). 
The constants $C_x$ and $C_y$ are obtained numerically
and the parameter $\mu$ is derived from the linearized problem 
(Aranson {\it et al}, 1993a).  
The bound states correspond to the case
of $v_x =0$.
The bound state will drift with the velocity $v_y$. 
The equilibrium distance $2 X_e$ between the spirals 
is found from the
equation
\begin{eqnarray}
Im (\frac{-Q \sqrt{1-Q^2} \exp(-p X_e)}
{\delta C_y \sqrt{2\pi p X_e }} X_e^{- \mu})  = 0.
\label{eq.dist}
\end{eqnarray}
Since only $p$ and $\mu$ are complex this gives
\begin{eqnarray}
Im [ p X_e + \mu \ln X_e] = -\phi + \pi l
\label{eq.dist1}
\end{eqnarray}
where $l = 1, 2 , 3 ...$ and $\phi = -\arg [1/(\delta C_y p^{1/2})]$. 
Even values of $l$ correspond to stable bound states, odd $l$ to unstable ones.
Usually  the lowest bound state with $l=2$ is relevant.
Approaching the boundary of the oscillatory range the equilibrium distance diverges.

Generalization to arbitrary $b$ can be done along the lines of the similarity 
transformation (\ref{nonlinear-sim}). The result is at leading order unchanged. 
\footnote{Note that the additional term $-ib {\bf v} \nabla A$ arising 
in the corresponding equations of motion for each spiral  can be absorbed
in the nonsingular part of the linear problem (Aranson and Pismen, 2000).}

Note that  
the symmetries $A(x,y) = A(-x,-y)$ (like-charged) and $A(x,y) = A(-x,y)$
(oppositely-charged) have nearly the same effect on the boundary conditions 
at $x=0$ for large spiral separation.
Thus for large separation $X$ the interaction of like-charged
spirals is similar to the interaction
of oppositely-charged ones. The only difference is
that for the like-charged case both components
of the velocities of the spirals have opposite sign whereas
for the oppositely-charged case the tangential components have the same
sign. This causes the rotation of the
spirals around the common center of the symmetry in the like-charged case.

Like-charged spirals may form more complicated bound states or aggregates.
In contrast to the two-spiral bound states, which are simply
rotating with constant velocity, each spiral in the aggregate performs a more
complicated motion (possibly nonperiodic) 
on the background of a steady-state rotation, perhaps similar to the "dance"
of spiral aggregates observed in a variant of the Belousov-Zhabotinsky reaction
(Steinbock and  M\"uller, 1993). 
Certain problems of interaction between spiral waves and hole solutions were 
considered by Bazhenov and Rabinovich (1993,1994). 
Interaction of spiral with external periodic perturbation and 
response functions of the spiral core were studied by
Biktasheva {\it et al} (1998,1999,2000).

\subsubsection{Comparison with results of numerical simulations}
Bound states  of spirals are 
shown in Fig.~\ref{fig2}. 
In Fig.~\ref{fig4}a the dependence of the
velocities on the spiral separation $2 X$
is plotted
for $b=0, c=1$ and compared with results from full numerical simulations. 
There 
is reasonable
agreement, particularly for the radial velocity $v_x$.
 The first two zeros of $v_x$
at $2 X_e\approx 11.5$ and $22.8$ correspond to $l=1,\ 2$
 in Eq.~(\ref{eq.dist1}).
Here the velocities are already extremely small. 
From the simulations no other bound state could be resolved.
The equilibrium distance obtained from the theory for the stable bound state
($l=2$ in Eq.~(\ref{eq.dist1})) is in a very good
agreement with the results of the simulations of the full CGLe (see
Fig.~\ref{fig4}a). There is a discrepancy between the values of
the
tangential velocity $v_y$ by a factor of about $1.5$ (see Fig.~\ref{fig4}b)
although the functional dependencies on $c$ are very similar. 
This discrepancy results from the fact that  
the shock is not described well in linear treatment. 

More
accurate estimates can be obtained by treating the shock fully nonlinearly. 
Far away from the core the motion of the spiral can be neglected and one may 
consider the 
stationary shock produced by the waves emitted by the spirals.
Thus one has to solve the stationary CGLe 
with boundary conditions $A \sim r \exp [i \theta]$  for $x \to X$, and 
$\partial_x A= 0$ at $x \to 0$. This is a rather complicated 2D problem. 
Numerically
one could  extract
more accurate  values of the constants $C_{1,2n}$, 
characterizing reflection at the shock, see Fig.~\ref{fig4}b. 

\subsubsection{Interaction in the monotonic range. }
In the case $0 < c < c_{cr}$
there exist two real positive roots of Eq.~(\ref{asdisp}).
Using the analogous numerical procedure one can also determine the 
constants $C_{x,y}$, but for small $c$ it is technically very 
difficult in this form. 
The results can be simplified considerably for the case
$c \to 0$ and $ |cQ| X \gg 1$. Then one
can neglect the coefficients $C_{2n}$ because
for $c \to 0$ one has
$0< p_1 \approx  -2cQ \ll 1, p_2 \approx  \sqrt{2},\mu_1 \to 0$
and $\delta_1 \to -1/(2 Q^2)$.
One
obtains $ v_{x,y} \approx C_{x,y}^{-1} \exp(-2|cQ|X)/\sqrt{X}$, 
which exhibits explicitly the exponential decay of the interaction
and reproduces the earlier analysis using a phase diffusion equation
(Aranson {\it et al}, 1991; Pismen and Nepomnyashchii, 1991,1992; Biktashev, 1989).

Numerical simulations for oppositely charged spirals 
for $b=0$ and $c=0.5$ indicate  
asymptotic repulsion (Aranson {\it et al}, 1993a). 
This result can be inferred
already from the work of Biktashev (1989), who used a phase-diffusion
equation and asymptotic matching to treat the interaction of spirals with
a boundary.  
It appears to be at conflict with the works of
Pismen and Nepomnyashchii (1991)a,b where 
in the limit $c \to 0$ matching with the internal solution was done
analytically and asymptotic attraction was found.
The repulsive range is expected to move to larger $X$ roughly as
$|c Q |^{-1}$ for smaller $c$. In this way the crossover to the long-range 
attraction within the GLe is recovered.
Interaction becomes
attractive at smaller distance leading to final annihilation of the spiral
pair.  

Like-charged spirals for $c<c_{cr}$  have  repulsion at large distance
(as in the oppositely-charged case) and also at small distance. So it is 
quite clear that the interaction is repulsive everywhere.

\subsection{Interaction of spirals with a inhomogeneity}
\label{interl}
Spirals may interact with inhomogeneities of the medium.
This problem is particularly interesting  in the context of Bose-Einstein 
condensation as described  by the NLSe in a parabolic potential 
(for a review see Dalfovo {\it et al}, 1999). 
Weak inhomogeneities are sometimes included into the CGLe in the form
\begin{eqnarray}
\partial_t A  =
(1 + \nu({\bf r})) A + (1 + i b) \Delta A -(1 + ic)|A|^2 A .
\label{CGLEN}
\end{eqnarray}
with an appropriate function $\nu({\bf r})$ .
An interesting problem is the drift of a spiral in the gradient created by
a localized, radially symmetric inhomogeneity 
(see Staliunas, 1992, Gil {\it et al}, 1992).
The spiral may be trapped (pinned) by the
inhomogeneity or perform a stationary rotation 
at some distance from the inhomogeneity
as observed in optical systems (Arecchi, 1990, 1991; Brambilla {\it et al}, 1991).

Using the method described above one arrives 
at the following expression for the spiral velocity due to interaction 
with weak axisymmetric inhomogeneity
(Aranson {\it et al}, 1993c) 
\begin{equation} 
{\bf v } = {\bf G(r)}
\label{velxy_i}
\end{equation} 
where $r$ is the distance from the inhomogeneity to the spiral core,
$v$ is the spiral core velocity
and the function $G$ is obtained by a convolution of the 
inhomogeneity profile $\nu(r)$ with the corresponding  solution
of linearized CGLe. 
Equation (\ref{velxy_i}) is the analog 
to the motion of a "massless particle" in 
a radially-symmetric field. Numerical simulations with 
CGLe shows that Eq. (\ref{velxy_i}) describes the motion of 
the spiral fairly well for not too large values of $|b|$, 
see Fig. \ref{fig3_14}a. 

For large $|b|$ the complex trajectories of the spiral core are not captures by 
Eq. (\ref{velxy_i}), 
see Figure  \ref{fig3_14}b. 
In order to describe the numerical simulations one has to
include the effective ``mass'' term into the equation of motion:
\begin{equation}
\frac{d {\bf  v}}{d t }+ \varepsilon \hat{K} {\bf  v}  = {\bf G}
\label{New1}
\end{equation}
where $\hat{K}$ is the friction tensor (or mobility tensor),
and $\varepsilon =1/b$, see Sec. \ref{corins}.
Although these features were obtained in the stable
range $Re \kappa >0 $ ($\varepsilon>\varepsilon_c$),
the external forces created by the inhomogeneity  can
excite the weakly damped core mode.
This  can explain the complex motion of the spiral
core observed in the presence of
obstacles (Sepulchre and Babloyantz, 1993).

Spiral motion in a slowly varying 
(on the scale of the coherence length) inhomogeneity 
was considered by Hedrey {\it et al}, (1999). 
In this situation additional simplifications are possible. 
A related problem is the motion of the spiral in the presence of small thermal 
noise. As was shown by Aranson {\it et al} (1998), the spiral core
has finite mobility and performs Brownian motion.

\subsection{Symmetry breaking}
\label{symb}
Symmetric bound states of spirals are not necessary stable. 
There is numerical evidence that 
symmetric bound states, after a sufficiently long
evolution, spontaneously  break the symmetry such that one spiral
begins to dominate, pushing away other spirals~(Lega, 1990; 
Weber {\it et al}, 1991a,b; Aranson {\it et al}, 1993b). 
Symmetry-broken states are often  
produced directly from random initial conditions (Aranson {\it et al}, 1993b).
To understand the
symmetry-breaking instability one must consider the perturbation of the
relative phase, or the frequency $\omega$ 
of the waves emitted by each spiral, caused by the interaction with
the other spiral. Indeed, from the analysis of CGLe it is known, that the
shock (or sink) where two waves with different frequencies $\omega_i$
collide moves in the
direction of smaller frequency, due to conservation of phase. 
This means that
after sufficiently long time only the larger frequency 
(or equivalently the larger wavenumber $|Q|$,
because of the dispersion relation $\omega= c (1-Q^2) +b Q^2 $) 
dominates in a bounded system. The velocity of the motion 
is given by
$ v_f \sim  (c-b) (Q_1 + Q_2)  = \frac{1}{2}(v_{g1}+v_{g2}) $ where 
$ v_g =  d\omega / dQ $ is the group velocity of a plane-wave state.
If due to the interaction the frequencies of the spirals
become different,
one can expect a drastic breaking of the symmetry of the system. This effect 
appears to be very important for the generic long-time evolution of large 
systems containing spirals (vortex glass).

The symmetry breaking instability of spiral pairs can be understood from the
dynamics of the relative phase of the spirals $\phi=\phi_1-\phi_2$. 
Aranson {\it et al} (1993b) derived an equation governing the phase difference
between two spirals separated by the distance $2X$
\begin{eqnarray}
\partial_t \phi =2Im \left(\frac{-Q \sqrt{1-Q^2} \exp(-p X)}
{\delta \sqrt{2\pi p X }} X^{- \mu} C_0^* \sinh(\frac{p\phi}{2Q})\right)/
{ Im (C_{10} C_0^*)} \sim \zeta \phi
\label{freq1}
\end{eqnarray}
where $C_{10},  C_0$ are the constants obtained from the 
linearized CGLe. 
The last term represents the lowest-order term of an expansion in $\phi$.
The constant $\zeta$ determined numerically from Eq. (\ref{freq1})
turns out to be positive for the (first) symmetric bound state
(see Fig. \ref{fig7_b}), so that the state is unstable with respect to 
$\phi$. According to Aranson {\it et al} (1993b),  bound states
are unstable with respect to symmetry breaking in the whole oscillatory range.
Due to the  fact that 
$\zeta$ is very small,  the symmetric
bound state is rather long lived. 
As a result of the symmetry breaking 
only one ''free'' spiral
will remain, whereas the other spiral 
is pushed away to the boundary.
Depending on the boundary conditions the second spiral will finally
either annihilate at the boundary (non-flux boundary conditions, i.e.
zero normal derivative on the boundary), or, 
with periodic boundary conditions the defect will persist for topological 
reasons, but reduced to a sink and enslaved in the corner of the shock
structure of the free spiral (edge vortex)
(see Fig. \ref{fig8}). This leads to an
asymmetric lattice of topological defects, which
appears to be stable in the oscillatory case.

The absence of symmetry breaking  in the limit
$|c-b| \to 0$ and $ |(c-b)Q| X \gg 1$
can be extracted already 
from the paper of Hagan (1982). From this
work one finds that decreasing the radius
increases the spiral wavenumber, which ultimately means stability of the shocks.
The stability of symmetric states together with the repulsion of 
the spirals indicates  the possibility for the existence of symmetric
(``antiferromagnetic'') lattices of spirals (or ``Wigner crystal''),
i.e. lattices made up of
developed spirals with alternating topological charge. 
Such lattices
were obtained in numerical simulations in this range 
by Aranson {\it et al} (1993b).
They appear to exist in the whole non-oscillatory range
\footnote{Aranson {\it et al} (1993b) have found stable 
square lattices in rather small system and for relatively 
small number of spiral ($4 \times 4$). There is no guarantee that similar 
lattices are stable in arbitrary large systems.}

Since for large separation $X$ the properties of 
like-charged spirals are analogous 
to those of oppositely-charged spirals, one expects the same mechanism of 
symmetry breaking. Indeed 
for $|b-c|$ above the critical value such a breaking
was observed in numerical simulations.

Certain aspects of the  spiral pair dynamics  
were studied numerically 
by Komineas  {\it et al} (2001). It was shown that for $c>c_{cr}$ 
(symmetry breaking range) the spirals interchange partners and form new 
pairs. For $c<c_{cr}$ (monotonic range) symmetric spiral states were found. 
\subsection{Vortex glass}
\label{vgs}
A fascinating problem concerns the effect of the symmetry breaking on the
long-time behavior of large systems that evolve from
random initial conditions in the parameter range where one has the oscillatory 
interaction and spirals. 
Numerical simulations,
carried out by Huber {\it et al} (1992), Chat\'e and Manneville (1996),  
have shown a remarkable phenomenon: 
from an initially strongly turbulent state eventually spirals with a 
developed far field evolve. The spirals grow until the entire space is filled 
(see Figure \ref{fig_gr}), each spiral occupying a certain domain.
The domains have various  sizes and typically 
have a four or five-sided near polygon structure. Locally each domain 
boundary, represented by a shock, is nearly hyperbolic. The shocks often
contain ``enslaved'', or edge vortices. 
This state appears to persist indefinitely, although 
sometimes domain boundaries break and edge vortices  annihilate. 
The state was called ``vortex glass'' by Huber {\it et al} (1992). 
Some aspects of relaxation to the vortex glass  were 
considered by Braun and Feudel (1996).

The existence of the  ``vortex glass'' is connected 
with the nonmonotonic 
dependence of the spiral frequency on the domain size.
As one sees from Figure \ref{fig_om} the spiral frequency 
oscillates as the radius of the domain increases  and approaches the asymptotic 
value for $r \to \infty$. (The computations were done for a
radially symmetric domain, but we expect the result to be qualitatively
correct for domains with arbitrary convex  shape.) 
As a result, one has sets of domain radii $r_i$ corresponding to 
the same frequency. Neighboring domains can only coexist without (further) 
``symmetry breaking'' if they correspond to the same
spiral frequency. Ultimately one can conceive of a network of 
fully synchronized domains
of various shapes and sizes.
These are likely  the ingredients necessary to explains the variability 
of the domain sizes observed  in the quasi-stationary vortex glass.
The vortex glass is presumably the global attractor also in the full 
convectively stable oscillatory range. 

Bohr {\it et al} (1996), (1997) considered the shapes  of spiral domains
in the vortex glass. They found that they differ from 
Voronoi Polygons and obtained the form of the domain boundaries 
from the condition of phase continuity across the 
shocks. In particular, far away from the center of an unperturbed spiral, 
the phase is given by $\phi_i = \pm \theta_i + Q r_i +C_i$, where 
$r_i, \theta_i$ are the polar coordinates measured from the center of the
spiral and $C_i$ is the phase constant of the spiral.
If one assumes that the distance between two neighboring spirals is much 
larger than  the wavelength 
$2 \pi/Q$, one obtains simply  a hyperbola describing the shock shape: 
$r_i - r_j= (C_j - C_i) /Q$. This simple formula reproduces 
the structure of the spiral domains with high accuracy (see Figure 
\ref{fig3_bohr}).

\subsection{Phase and defect turbulence in two dimensions} 

On a rough scale, two types of turbulent  behaviors can be 
identified in 2D: phase and defect turbulence. Defect turbulence is 
characterized by persistent creation and annihilation of point defects. 
By contrast, 
in phase turbulence no defects occur. In the system with periodic boundary 
conditions the total phase gradient across the system (the 
``winding number'') is conserved. 
The simulations of Manneville and Chat\'e (1996) provide some evidence 
that phase turbulence breaks down in the infinite size, infinite time limit. 
However, it could be that the transition is characterized 
by a ``sharp'' bifurcation scenario similar to the situation in 1D 
(see \ref{stc_1da}). 

\subsubsection{Transition lines} 
The most detailed survey of various regimes occurring in 2D is 
given by Chat\'e and Manneville (1996) and Manneville  and Chat\'e  (1996). 
According to Chat\'e and Manneville (1996), the transition from vortex glass to 
defect turbulence starting from 
random initial conditions occurs at the numerically 
determined line ``T'', see Fig. \ref{fig_stab}. 
The transition occurs somewhat   
{\it prior} to the absolute instability limit  given from the linear 
stability analysis of plane waves emitted by spirals. 
However, starting from carefully prepared initial conditions 
in the form of large spirals one can approach the absolute instability limit.

Before the line ``T''one 
finds transient defect turbulence which  finally exhibits 
spontaneous nucleation of spirals from the
``turbulent sea''. At the line ``T''  the nucleation time presumably diverges.
Before the line ``T'' the entire space will finally be filled 
by ``large spirals'' separated by shocks, forming a vortex
glass state.

As in 1D, persistent 
phase turbulence exists also in 2D between the Benjamin-Feir-Newell line and 
the line ``L'' (Chat\'e and Manneville, 1996; Manneville and  Chat\'e, 1996). 
The range is somewhat smaller than in 1D.
Beyond the line ``L'' defects are created spontaneously leading to defect chaos.
Actually, in 2D, phase turbulence is always metastable with respect to 
defect turbulence or vortex glass. 

\subsubsection{Spiral break-up}
\label{break-up} 

Spiral break-up was observed in  experiments on chemical
oscillatory media by  Ouyang and  Flesselles (1996), 
Zhou and Ouyang (2000), in simulations of
CGLe  by Chat\'e and Manneville (1996) and 
model reaction-diffusion  systems by B\"ar and Or-Guil (1999). 
The instability likely  originates from the outer region of the spiral
and thus is expected to occur beyond the line ``T''.
Quenching a large  spiral to
this domain will usually trigger a breakup scenario.
On the other hand, a careful adiabatic procedure allows, in principle,
to avoid breakup and to preserve one large spiral up to the (linear)
 {\it absolute instability threshold}.
(In the presence of noise,
the convective instability actually imposes a maximal noise-dependent
radius $R_{\rm noise}$.)
Not too far from the absolute instability
limit one is left with  a smaller spiral surrounded
by strong chaos (Fig. \protect \ref{fig_gr}, left) whose well-defined radius 
$R$ {\em does  not}
depend on system size and/or distance to the boundaries
but vanishes as one approaches the
absolute instability threshold.

In the experiments and in some model systems, a spiral need not
be introduced initially, as the chaotic phase is only metastable to the
spontaneous nucleation of spirals whose radius grows
up to $R$. Thus,
the asymptotic configuration on a long time scale is
actually  a quasi-frozen
cellular state (Fig. \protect \ref{fig_gr}, right).

The above concept of   spiral break-up 
in large systems 
is in contrast to arguments by
Tobias and Knobloch (1998), who
stated that the spiral wave breakup 
occurs in the regime of {\em absolute} Eckhaus instability
via ``a globally unstable wall-mode confined to the outer boundary'' whose
front structure is at the origin of the stable ``laminar'' spirals immersed
in a turbulent sea. 
Whereas a stationary front can indeed be associated with the absolute stability 
limit (see Sec.\ref{abs}) it can only give an upper limit to the existence of
the convectively unstable state.

In the break-up scenario studied by Zhou and Ouyang (2000) spirals emitting
modulated waves in the presence of stable meandering of the core were observed.
This behavior can be understood in terms of spatial amplification of 
periodic perturbations due to core meandering in the regime of convective 
instability of the background waves (Brusch {\it et al.}, 2001). 
Although the saturated meandering
does not occur in the framework of the CGLe, a properly perturbed CGLe may 
exhibit both, saturated meandering and convective instability, see
Aranson, Hochheiser and Moloney (1997). 
\subsubsection{Defects statistics} 
Defect turbulence is the most chaotic state in 2D. It is characterized 
by exponential  decay of correlations, with short correlation 
lengths and times. The density of defects varies with 
$b$ and $c$. However, it can be argued that, at least in the most studied 
region with $b,\ c$ of order one, the defects do not play the 
role of ``particle-like'' excitations in this case. 
Indeed, the defects in the defect turbulence are very different from  spiral
waves since they don't emit waves. They behave as passive objects 
and  are merely advected by the surrounding chaotic fluctuations. 
According to 
Chat\'e and Manneville (1996) ``amplitude turbulence'' is 
a more appropriate name for such spatio-temporally chaotic state.

One may argue that the stationary
distribution for the number of defects  is described by 
$p(n) \sim \exp[-(n-\bar{n})^2/\bar{n}]$
where $\bar{n}$ is
the average number of defects. The formula follows in the limit $n \to \infty$
from treating defect pairs as statistical
independent entities (Gil {\it et al}, 1990).
Note that this assumption is strictly
valid only for large domains. Otherwise the statistics is influenced by
defects entering/leaving the subsystem.
When these  processes dominate the exponent
acquires a factor $1 \over 2$.

Egolf (1998) extracted the degrees of freedom associated with 
defects from those of phase fluctuations by using the concept of finite-time
Lyapunov dimension.
He found that each defect ``carries'' from one to two degrees of freedom. 
Although one may argue that the number of 
defects can be a convenient characterization of some types of 
spatio-temporal chaos, 
the method of separation does not appear fully convincing because
the relation between the finite-time dimension and number of
defects is not examined for different  system sizes and duration intervals.

It is interesting to mention that a similar analysis performed by 
Strain and Greenside (1998) for a reaction 
diffusion system results in a much higher dimension per/defect, namely
between 3 and 7. 

Mazenko (2001) studied defect statistics in the CGLe in the 
``defect-coarsening regime''
(presumably the region of small $b$ and $c$ and mean distance between defects
smaller than the screening length $1/(|b-c|Q)$) using 
an expression for the defect velocity similar to that of Rica and Tirapegui
(1990),(1991a), which is not applicable for defects separated by more than
the core size of O(1).

\subsubsection{Core instability and spiral turbulence for large $b$}
\label{core-inter}

One expects that in the full 
core-unstable range $\varepsilon =1/b<\varepsilon_c$, 
a state with persistent defects should be 
typically  spatio-temporally chaotic. 

\begin{figure}
\centerline{\psfig{figure=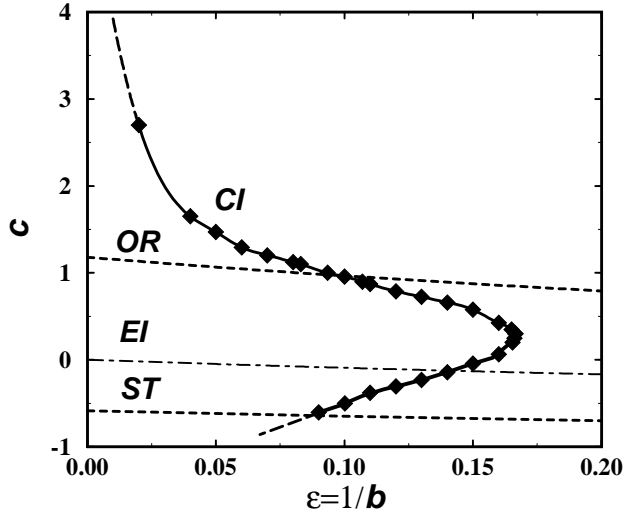,height=3in}}
\caption{
Stability limits of spiral wave
in the large $b$ limit, $\varepsilon=1/b$.
Here  $CI$ is the core stability limit $\varepsilon=\varepsilon_c$
(unstable to the left), below
$EI$ is the Eckhaus unstable region, $ST$ designates
the transition line to strong turbulence (see text), below  $OR$ is
the oscillatory range (Aranson {\it et al}, 1994).
}
\label{fig10}
\end{figure}

\begin{figure}
\centerline{\psfig{figure=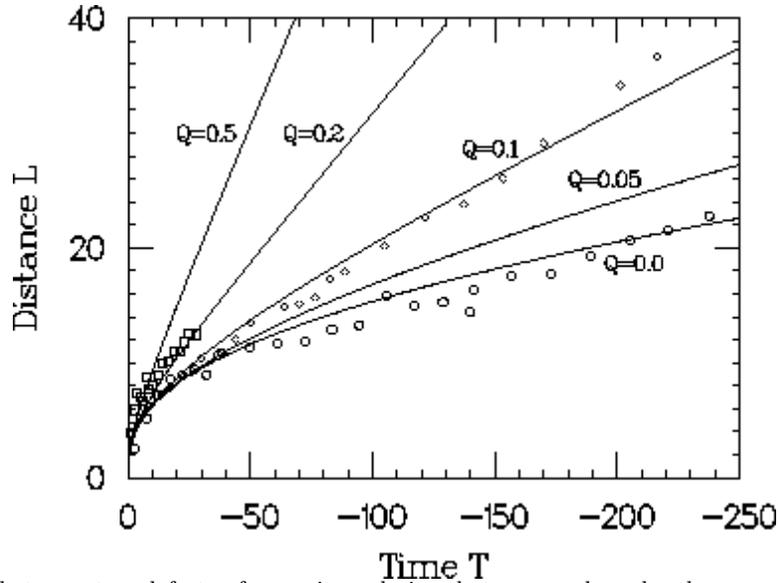,height=3in}}
\caption{
The distance $L$ between two defects of opposite polarity that approach each
other on a straight line is plotted for different $Q$ versus time $T$
(Bodenschatz {\it et al}, 1991,
For comparison the experimental data of Braun and Steinberg (1991) are included.The different symbols denote different distances $\varepsilon$
from the threshold (circles: $\varepsilon=0.02$; squares $\varepsilon=0.04$;
diamonds $\varepsilon=0.005$)
}
\label{fig1d}
\end{figure}

\begin{figure}
\centerline{a) \psfig{figure=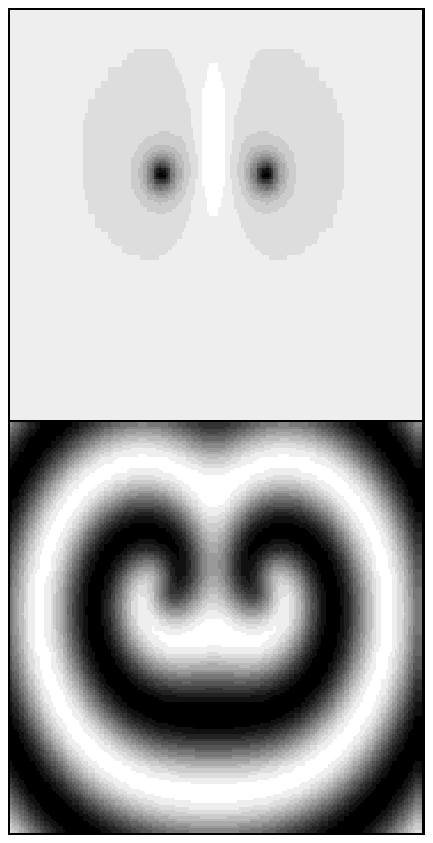,height=3in}
b) \psfig{figure=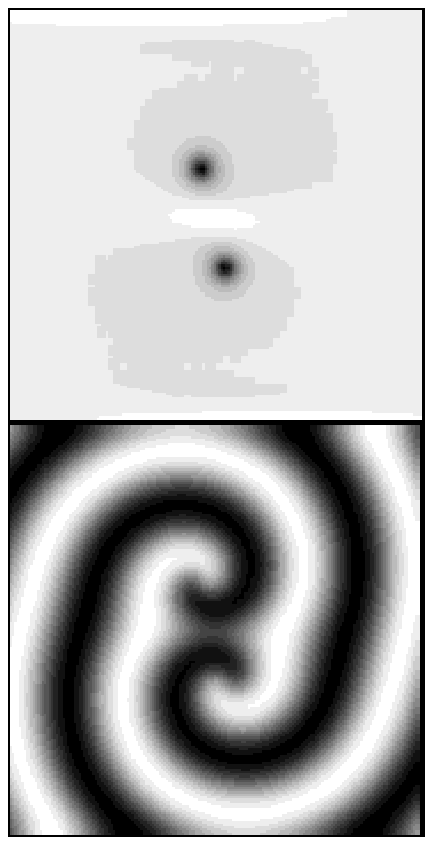,height=3in} }
\caption{
The bound state of oppositely-charged spirals (a) and like-charged
spirals (b) for $b=0, c=1.5$ in a
$100 \times 100$ domain.
Images show  $|A(x,y)|$ (top) and  $Re A(x,y)$ (bottom) correspondingly.
}
\label{fig2}
\end{figure}

\begin{figure}
\centerline{ (a) \psfig{figure=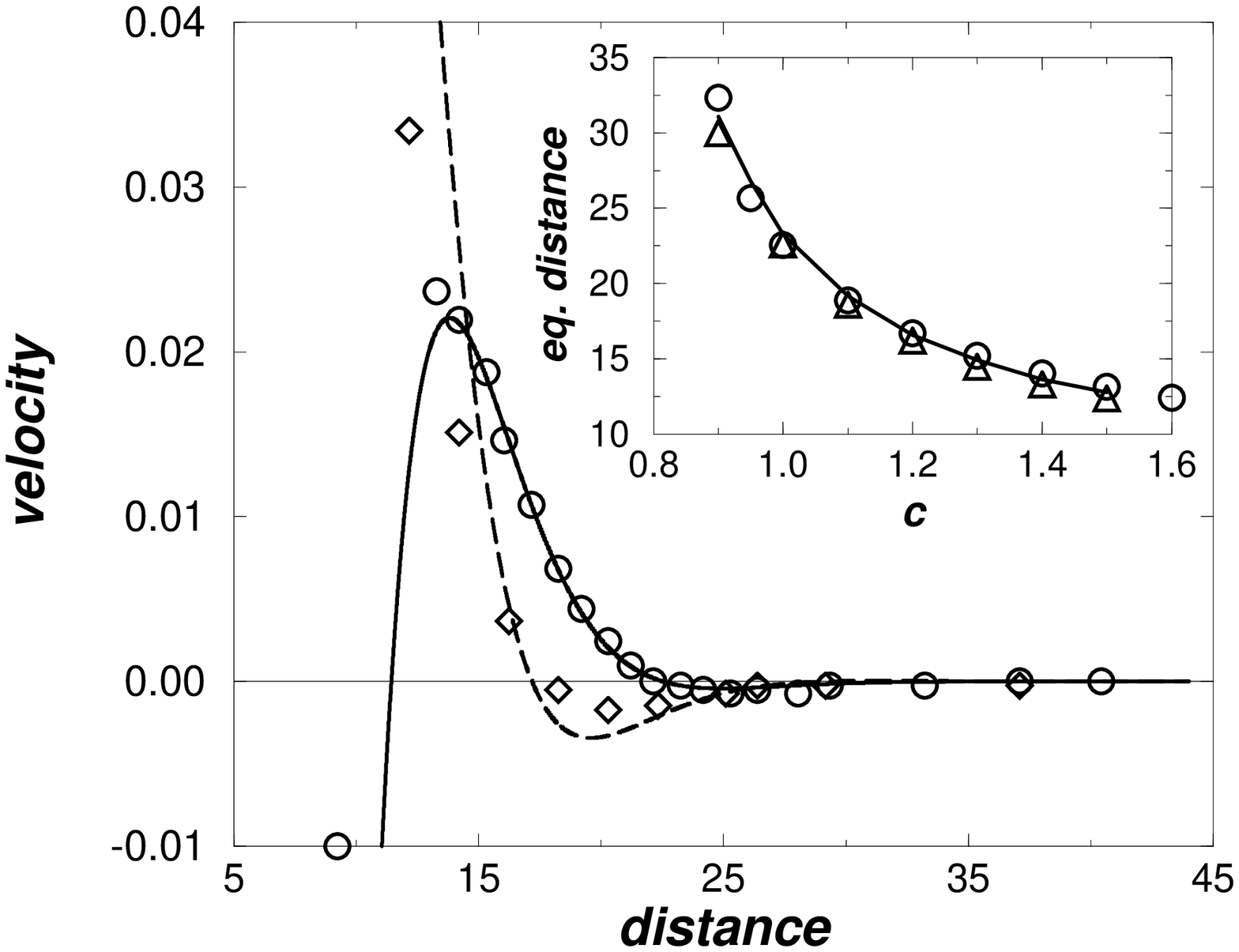,height=2.5in}
b) \psfig{figure=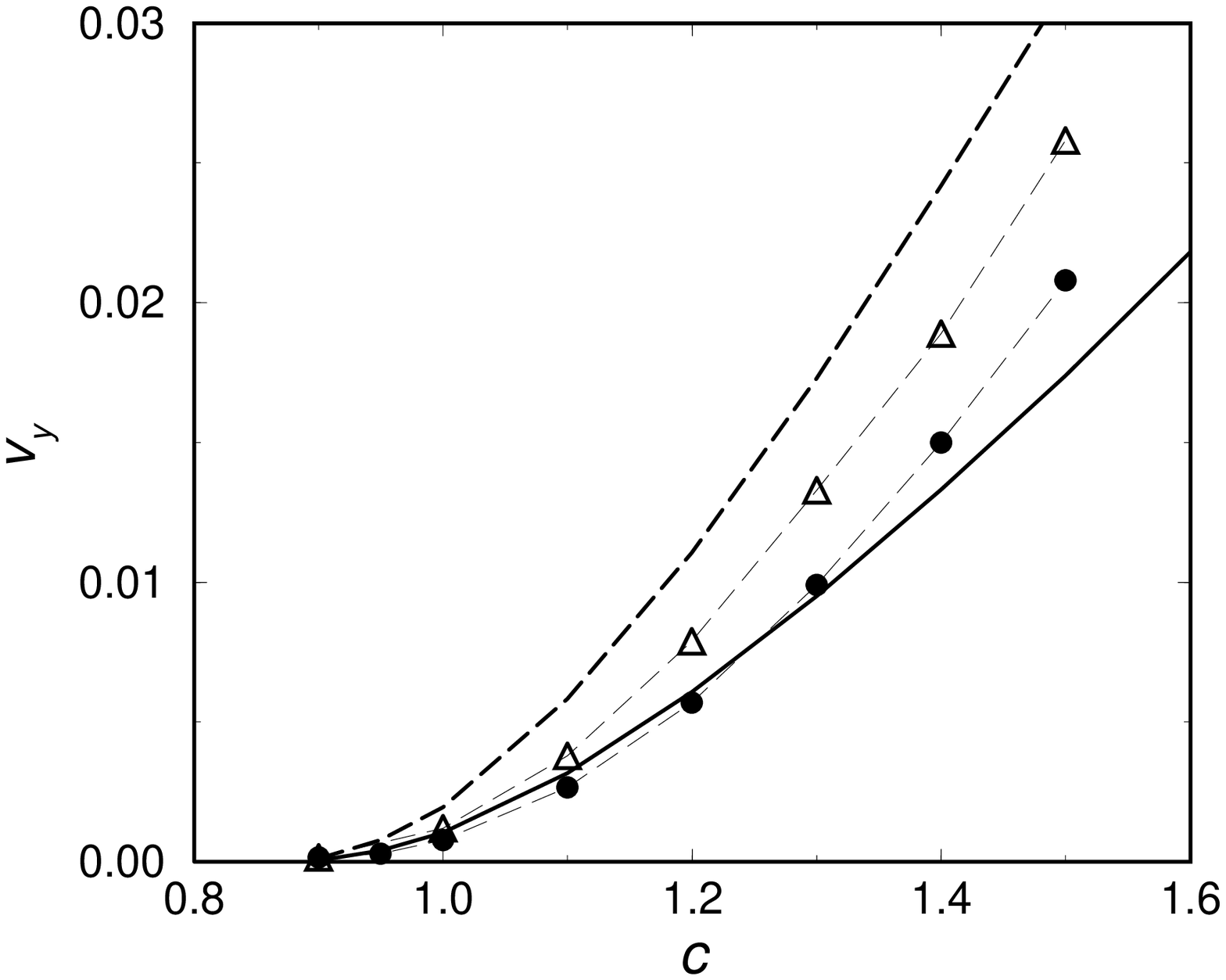,height=2.5in} }
\caption{
(a) Radial velocity ($v_x$) and tangential velocity ($v_y$)
 vs spiral separation $2 X$ for $b=0, c=1$ for oppositely charged pair.
Solid / dashed lines show radial and tangential velocities  obtained
from solution of Eq. (\protect \ref{velxy}) correspondingly.
Symbols ($\circ$ is radial  and  $\diamond$ is tangential velocities)
are obtained from  the simulations.  Inset:
Equilibrium distance $2X_e$ (solid line) given by
Eq. (\protect\ref{eq.dist1}) as function of $c$ for $b=0$,
$\circ$ and $\bigtriangleup$ correspond to numerical results for 
oppositely and like-charged pairs
respectively.
b) Tangential velocity $v_y$ at the equilibrium distance for
$b=0$, $\bullet$ and $\bigtriangleup$ correspond to 
to numerical results for 
oppositely and like-charged pairs
respectively. 
Solid and dashed lines show theoretical results obtained 
using nonlinear
and linear treatments of  the shock correspondingly.
}
\label{fig4}
\end{figure}

\begin{figure}
\leftline{ (a) \psfig{figure=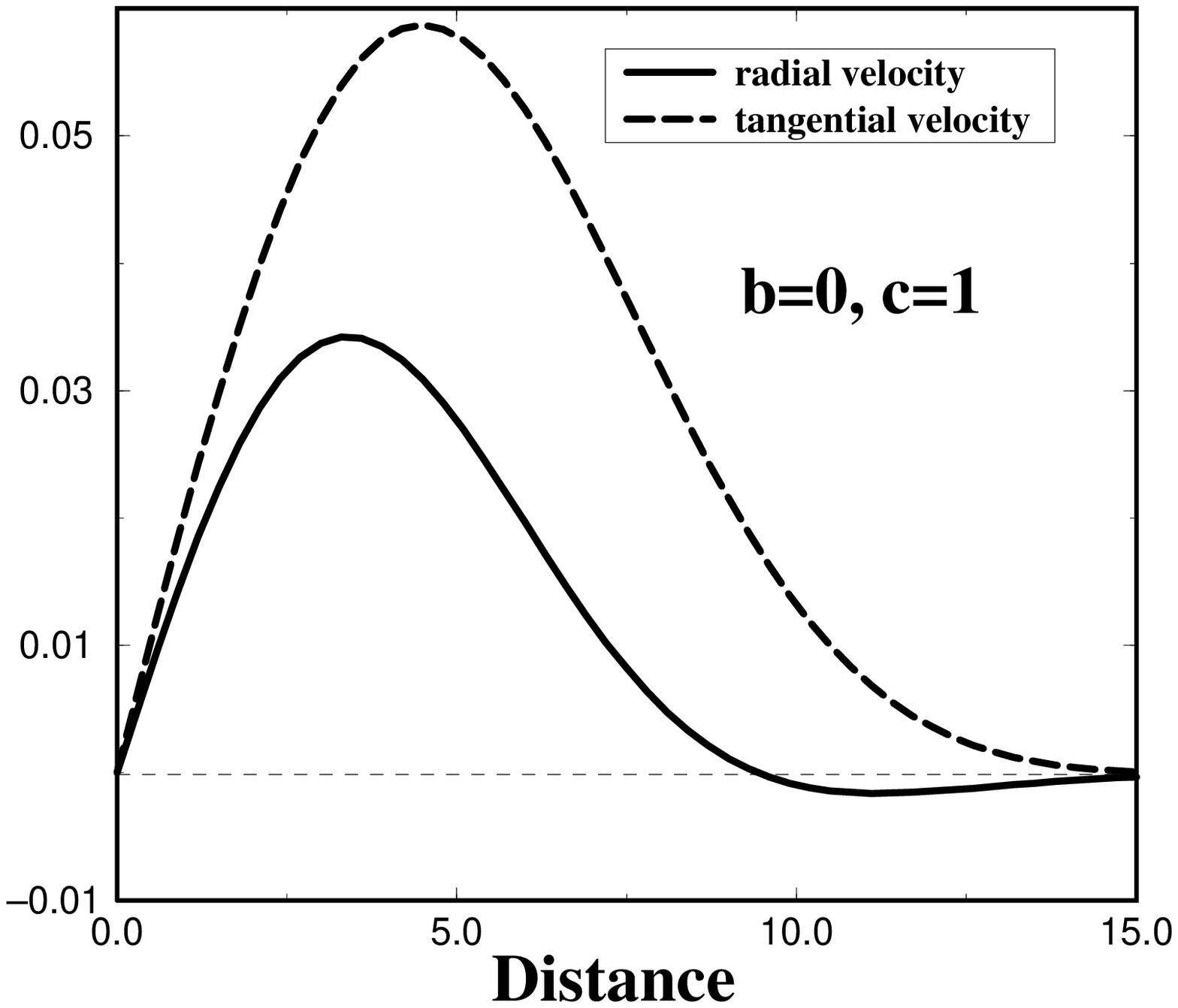,height=3in}
b) \psfig{figure=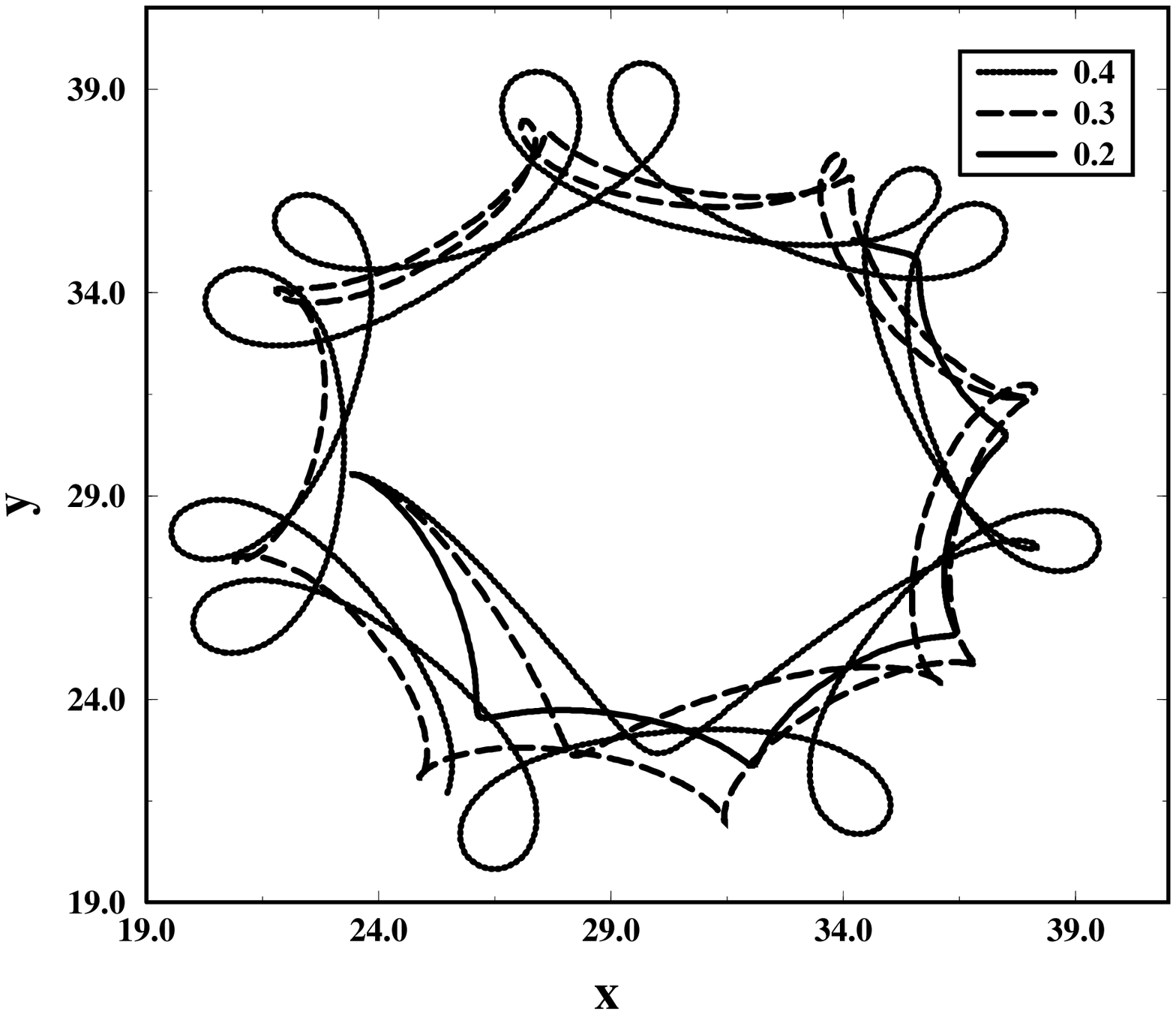,height=2.3in} }
\caption{(a) The tangential and radial velocities of the spiral
versus the distance $ X $
from the inhomogeneity
$\nu(r) = b_0 \exp (-\frac{X^2}{\sigma})$
for $b=0, c=1$ and $b_0=0.3, \sigma=20$.
From the numerical
simulations one finds the radius of
the first stationary orbit $r_0 \approx 9.8$;
(b) The limiting orbits of the spiral core for different values of
inhomogeneity "strength" $b_0= 0.1 , 0.2 , 0.3$ in the ``large $b$ limit''
for $\varepsilon=1/b=0.2, c=1, \sigma=20$
}
\label{fig3_14}
\end{figure}

\begin{figure}
\centerline{  \psfig{figure=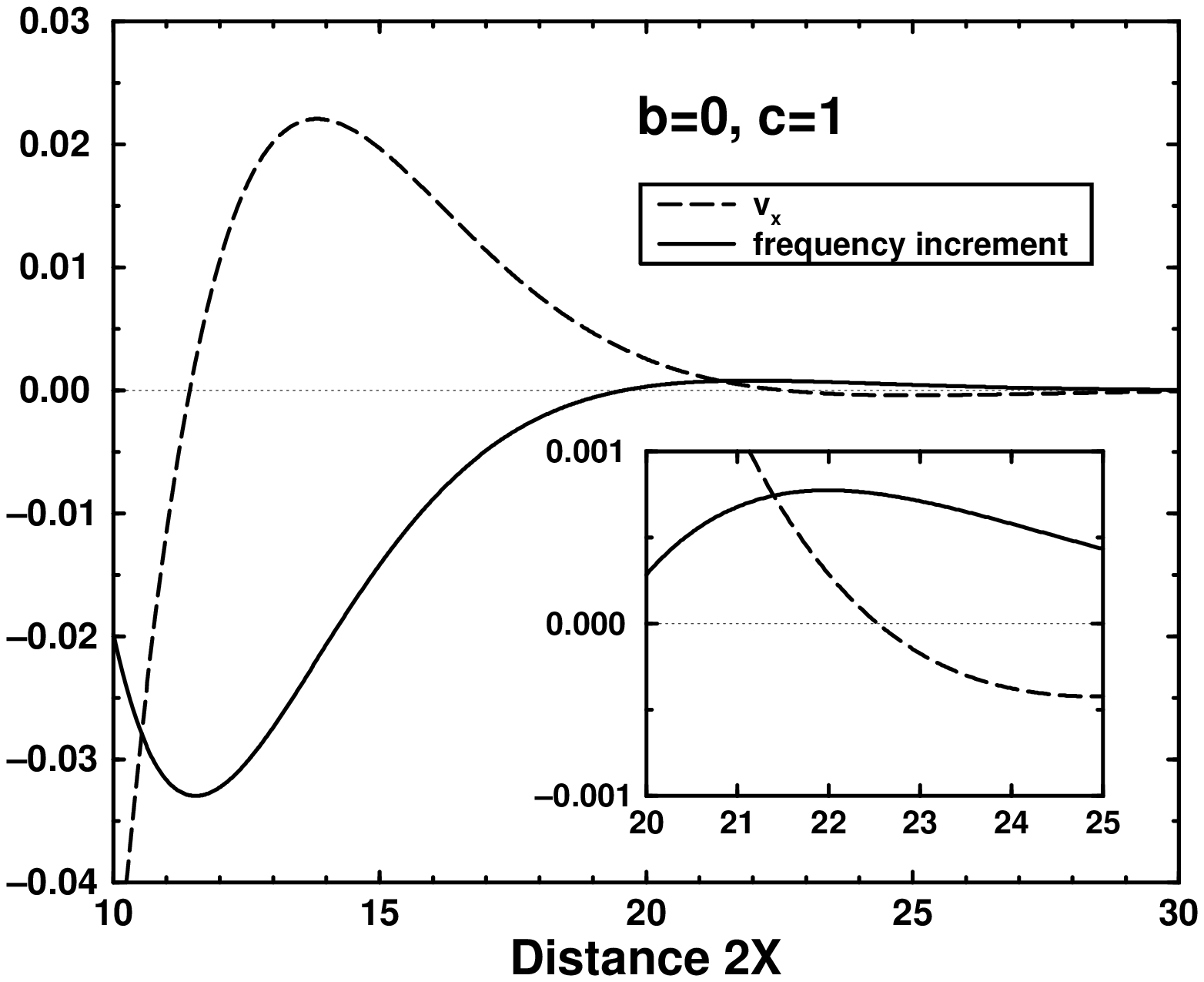,height=3in}}
\caption{
The dependence of radial velocity ($v_x$) and the frequency
increment $\zeta=d (\partial_t \varphi) / d \varphi $ at $\varphi=0$ for
$b=0, c=1$.
}
\label{fig7_b}
\end{figure}

\begin{figure}
\centerline{  \psfig{figure=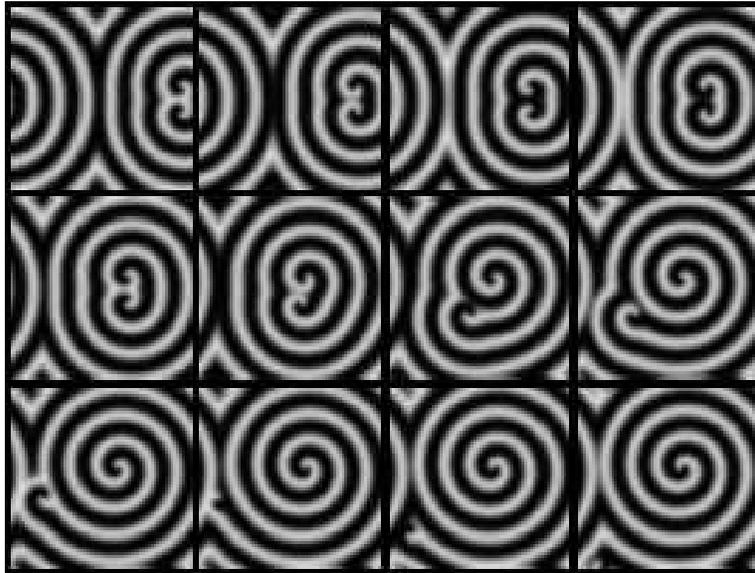,height=3in}}
\caption{
Evolution of an unlike-charged spiral pair
into a stable antisymmetric state, $b=0$, $c=1$,
$L_x=L_y=100$, time lapse between snapshots $=50$
}
\label{fig8}
\end{figure}

\begin{figure}
\centerline{   \psfig{figure=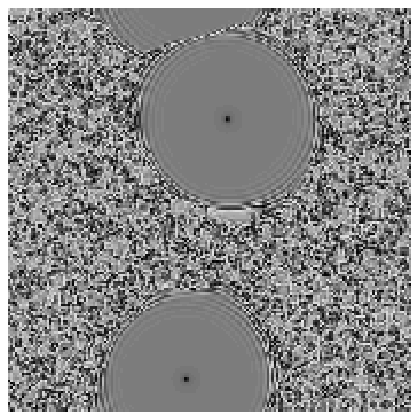,height=3in}
\psfig{figure=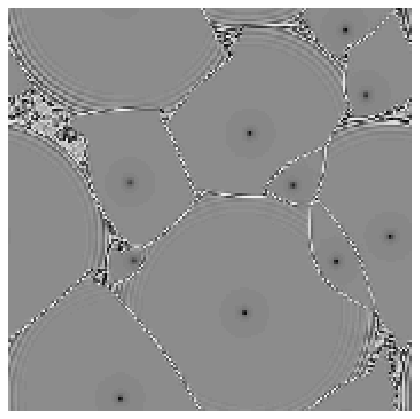,height=3in}}
\caption{
Vortex glass in the convectively unstable range,
$1024 \times 1024 $ lattice,  periodic boundary
conditions (Chat\'e and Manneville, 1996).
Large spirals nucleating in turbulent sea, $c=0.75, b=-2$ (left)
and developed vortex glass for $c=0.7, b=-2$  (right).
}
\label{fig_gr}
\end{figure}

\begin{figure}
\centerline{  \psfig{figure=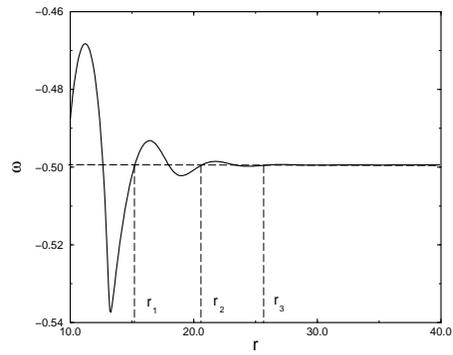,height=3in}}
\caption{
The spiral's frequency $\omega$ as a function of the
domain radius for $c=0.8, b=-1$.
}
\label{fig_om}
\end{figure}

\begin{figure}
\centerline{  \psfig{figure=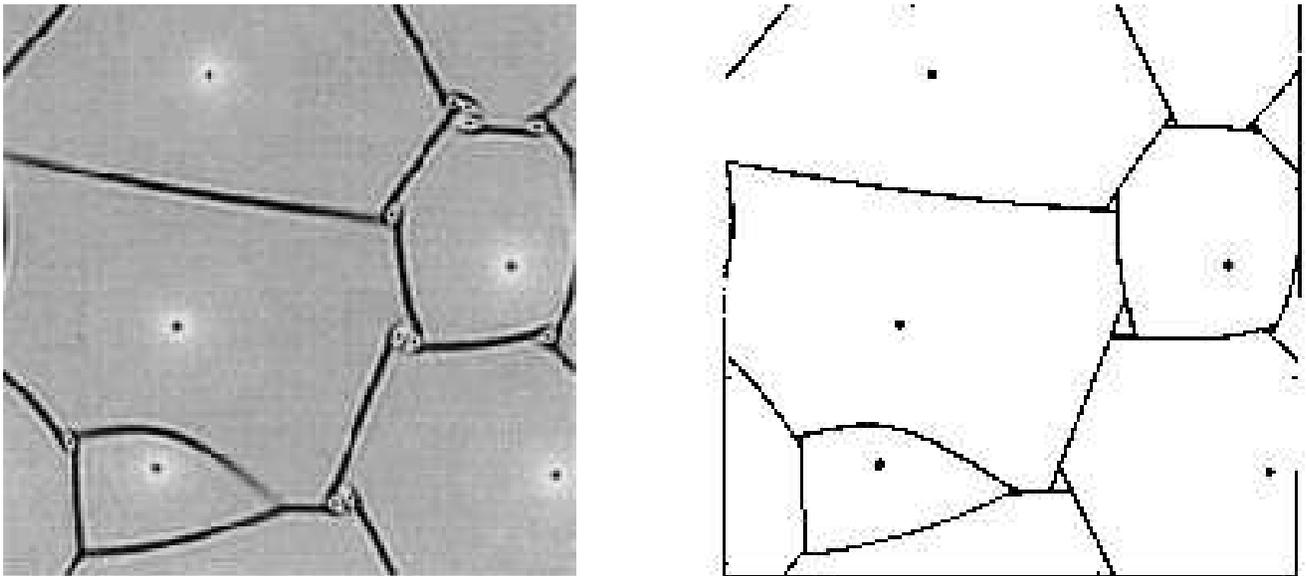,height=3in}}
\caption{
Close-up of the shock  structure (left). Reconstruction of the shock
structure using the hyperbolic approximation (right) (Bohr {\it et al}, 1995).
}
\label{fig3_bohr}
\end{figure}

\begin{figure}
\centerline{  \psfig{figure=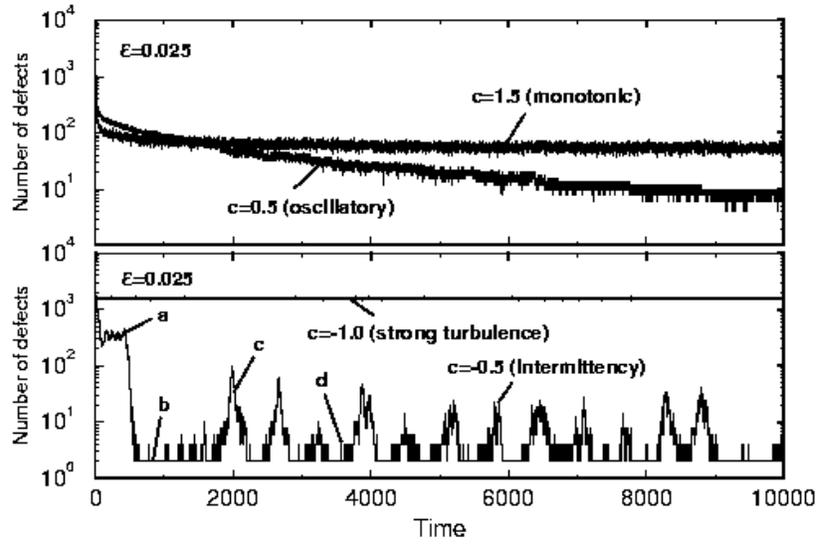,height=3in}}
\caption{The number of defects versus time for four different values
of $c$ and $\varepsilon=0.025$.
Parameters of the simulations are: the domain of integration
$150 \times 150$, the
number of Fourier harmonics $256 \times 256$, the time step 0.02. }
\label{fig11}
\end{figure}

In the monotonic range (Fig. \ref{fig10}, above curve $OR$), 
this turbulence is characterized by  fast motion of the defects and
collisions, which often do not result in  annihilation, in contrast to the 
usual defect chaos discussed above. 
In Fig. \ref{fig11}, the number $n$ of defects 
as a function of time is 
shown for a fairly large system ($150 \times 150$). 
Apart from the rapid fluctuations 
due to creation and annihilation there is an 
extremely slow decrease of $n$.
Crossing $\varepsilon_{c}$ in the monotonic
range the disordered state appears to persist (or is at least very long lived). 
This indicates a hysteretic behavior, which is to be expected from the 
subcritical character of the core instability.

In the oscillatory Eckhaus stable range (Fig. \ref{fig10}, below curve $OR$) 
the behavior is drastically different.  
Starting from random initial conditions one first has the evolution towards
a vortex glass as in the range $\varepsilon>\varepsilon_c$, see Fig. \ref{fig11}
and Sec. \ref{vgs} (Aranson {\it et al} 1993b). 
However the spirals
are unstable with respect to acceleration resulting in
continued dynamics of the spiral and shocks. 
Additional spirals are created very rarely.
One might call this a ``hot vortex glass''.

\begin{figure}
\centerline{  \psfig{figure=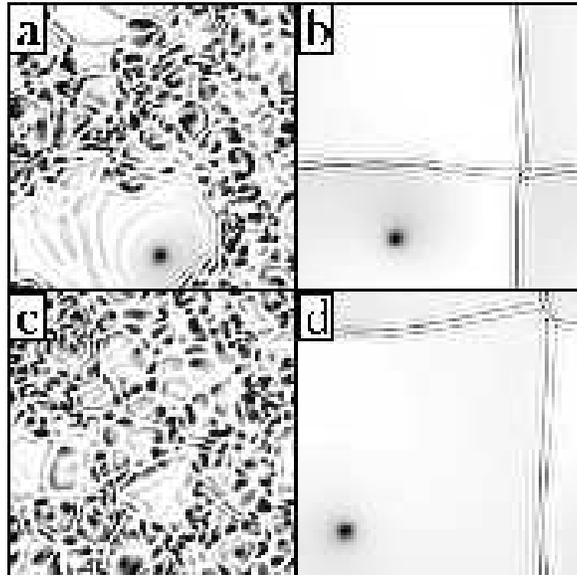,height=3in}}
\caption{Snapshots of $|A(x,y)|$ in the intermittency range
for four consecutive times
($c=-0.5, \varepsilon=0.025$; black: $|A(x,y)|=0$, white: $|A(x,y)|=1$). The
times corresponding to a,b,c,d are indicated in previous
Figure \protect{ \ref{fig11}}.
}
\label{fig12}
\end{figure}

When the Eckhaus instability sets in at the curve $EI$ of Fig. \ref{fig10}
the perturbations produced by the accelerating core 
of the dominant spiral
are amplified away from the core due to the convective character of the
instability.
When some critical level is exceeded, the state
looses stability and many new defects are created throughout the cell.
Then the process repeats (Figs. \ref{fig11} and \ref{fig12}). 
Such phenomena are very similar to
spatio-temporal intermittency of holes observed in the 1D 
CGLe (Chat\'e {\it et al}, 1994; Popp {\it et al}, 1993, 
see Subsec. \ref{NB_boundary}) and
might be called {\it defect-mediated intermittency}.
In a large cell one expects to have 
such processes developing independently in different
places of the cell, so one has persistent chaotic bursts (or spots) on the
background of growing spirals. 

Below the curve $ST$ in Fig. \ref{fig10}, which presumably is the continuation
of the T curve in the range $\varepsilon > \varepsilon_{c}$ 
(Chat\'e  and Manneville,  1996),
the ``strong'' chaotic characteristic of 
the intermittent bursts becomes persistent.
The state is similar to the usual defect chaos.
The curve $ST$ lies somewhat
above the limit of absolute instability, curve AI(see Sec.\ref{abs}).
For $\varepsilon \to 0$ the AI curve tends to $c \approx -1.2$.
The curve was determined by simulating 
Eq.\ (\ref{CGLEL}) with the restricted class of functions (\ref{spiral}) and
boundary conditions $A(0)=0, \partial_r A(L)=0, L\gg 1$,
which is effectively a 1D problem. The curve $ST$ determined in
this manner is consistent with 2D simulations (Aranson {\it et al}, 1994).


\section{Dynamics in 3D} 
\label{sec:3D}

\subsection{Introduction} \label{3D_intr} 
The 3D analog of the 2D vortex or spiral wave is 
called vortex filament or scroll wave. 
The point singularity of the  phase of the complex function $A$ 
at the center of the  
spiral becomes a line singularity in 3D.  
The  
filaments can be open (scrolls), closed (vortex loops and rings), 
knotted or even  interlinked, twisted, 
or entangled. Depending on the parameters of the CGLe the scroll wave
can be stable or develop some instabilities. Remarkably, 3D vortices 
can be highly unstable even in the range of parameters where 
their 2D analog is completely stable. 

CGLe has a stationary solution in the form of a straight vortex with twist
\begin{eqnarray}
A(r,\theta,z) = F(r) \exp i [ \omega t \pm \theta + \psi(r)+k_z z ] \;,
\label{spir}
\end{eqnarray}
Here the axial wavenumber $k_z$ characterizes the twist. 
Curved vortex lines are non-stationary. In 
most of the cases vortices untwist, and the solution with 
$k_z=0$ is  the most stable one. 

Scroll waves have been observed experimentally in
 slime mold (Siegert and  Weijer, 1991), heart tissues 
(Gray and Jalife, 1996),  gel-immobilized 
 Belousov-Zhabotinskii (BZ) reaction (Vilson {\it et al}, 1997).
Long-lived entangled vortex patterns in three-dimensional
BZ reactions were observed
by the group 
of Winfree using optical tomography techniques,
(Winfree {\it et al}, 1996).
Complex vortex configurations have also been 
observed in
numerical simulations of
reaction-diffusion
equations 
(Winfree, 1995, Biktashev, 1998, Fenton and Karma, 1998a,b; 
Aranson and Mitkov,  1998; 
Qu, Xie and  Garfinkel, 1999).  

Theoretical investigation of scroll vortices in reaction-diffusion  
systems was
pioneered  by Keener and Tyson, 1990, 1991, who derived the equation of  
motion for
the filament axis. In particular, it was 
found that vortex rings typically shrink with a rate
proportional
to the local curvature of the filament, leading to collapse  
in finite time. The existence of non-vanishing vortex  
configurations and
expansion of vortex loops,
observed also in numerical simulations of  reaction-diffusion
equations, was interpreted as 
``negative line tension'' of the vortex
filament (Biktashev, 1998). 

In order to characterize the motion of vortex lines in three-dimensional 
space let us consider 
a  curve $C$ at any moment of time $t$ in a parametric form ${\bf X}(s,t)$, 
where $s$ is the arclength. At any point of the curve a local 
orthogonal coordinate basis can be defined ({\it Frenet trihedron}), 
yielding the Frenet-Sorret equations
(see e.g. Pismen (1999))
\begin{equation} 
{\bf X}_s={\bf l}, \;  {\bf l}_s=\kappa {\bf n}, \;
 {\bf n}_s=-\kappa {\bf l} + \tau {\bf b}, \;
 {\bf b}_s=- \tau {\bf n}
\label{fse}
\end{equation}
where ${\bf l, n, b} $ are tangent, normal and bi-normal unit vectors, 
and $\kappa$ and $\tau$ are curvature and  torsion of the curve, correspondingly.

\subsection{Vortex line motion in NLSe} \label{3D_nlse} 

As in 2D, the analysis of vortex motion in the 3D NLSe is based 
on the analogy with the Euler equation for ideal fluids,(see Batchelor,
1967; Pismen, 1999). 
Following the analogy with the vortex lines in an ideal fluid, the local 
``superfluid velocity'' ${\bf v} =  \nabla \phi$, $\phi = 
\arg A$ can be found from 
the Biot-Savart integral 
\begin{equation} 
{\bf v(x)} = - \frac{\Gamma}{2} \oint_C \frac{{\bf R }\times d {\bf l}} {R^3}
\label{bsi} 
\end{equation} 
where $\oint \nabla \phi = 2 \pi \Gamma $, $\Gamma =\pm 1$ is the vorticity 
of the line, ${\bf R }= {\bf x-X }$, $R=|{\bf R }|$. 
The Biot-Savart integral describes the velocity far away from the 
vortex line but does not allow to compute the {\it velocity} of the line
since it diverges at the core. 
In order to get the velocity of line motion one need to perform 
matching of the vortex core field with the far field given by the
Biot-Savart integral (Pismen and Rubinstein, 1991, Pismen, 1999). 
As a result of matching one obtains 
\begin{eqnarray} 
{\bf v}_\perp= {\bf v_s} + \Gamma  \kappa {\bf b} 
\log \frac{ \lambda}{a_0 } 
\label{psirub} 
\end{eqnarray} 
where ${\bf v}_\perp$ is the vortex drift velocity in the local normal plane,  
$\lambda$ is a constant 
dependent on the geometry of vortex, and $a_0 \approx 1.856$ 
for a single-charged 
vortex line. For  a vortex ring of the radius $R$ 
one has  $\lambda=8 R$. 
The last  term  expresses the drift along the bi-normal with the 
speed proportional to the curvature, the first term 
${\bf v_s}$ accounts for a contribution from non-local induction of 
the Biot-Savart integral. Neglecting the first term one recovers the
so-called {\it localized induction approximation}, often used in 
hydrodynamics ( for review see Ricca (1996)).  
The localized induction approximation is in fact a rather crude approximation for vortex motion in the NLSe. 
It has been extensively used due to its mathematical elegance.

In the simplest case of the vortex ring of radius $R$ one obtain from 
Eq. (\ref{psirub}) 
that 
the ring drifts as a whole along the bi-normal, 
i.e. along the axis of symmetry,  
with the speed given by (Pismen, 1999) 
\begin{equation} 
v= \frac{1}{ R} \log( R/R_0) 
\label{v_nlse} 
\end{equation} 
where  $R_0 \approx 0.232$ is the constant obtained from matching with 
the vortex core. Clearly, a similar result can be obtained for 
classical vortices in an ideal fluid with the difference that in 
the latter ones 
there is no  a well-defined core.

\subsection{Collapse of vortex rings in the CGLe} \label{collapse}

In GLe one  finds that a vortex ring (radius $R$)
collapses. Indeed, substituting  the ansatz 
\begin{equation} 
A(r,\theta,z,t) = A_0(r-\int_0^t v(t^\prime) dt^\prime ,z)
\label{an1}
\end{equation} 
($A_0$ is the 2D stationary vortex solution; $r,\ \theta,\ z$ are cylindrical
coordinates)
into the 3D GLe one derives 
\begin{equation} 
-v \partial_r A_0 = 
A_0 (1-|A_0|^2) +  
\partial_r^2  A_0+
\partial_z^2  A_0  + \frac{1}{r} \partial_r A_0. 
\end{equation}
Replacing  the explicit $r$ dependence in the last term  by the radius of the 
ring, one obtains  from the consistency condition that 
\begin{equation} 
v=\frac{d R}{d t} = -\frac{1}{R}. 
\label{vr}
\end{equation}
Solving Eq. (\ref{vr}) one derives $R(t) = \sqrt{R_0^2 - 2 t}$, i.e. 
the ring collapses in finite time.   
Surprisingly, in the case of the GLe the analog of the localized induction approximation produces 
the correct answer.

Gabbay {\it et al}  (1997), (1998a) 
generalized this result for the CGLe where the ansatz Eq. (\ref{an1})
has to be generalized to include the curvature-induced
shift of the filament wavenumber. 
They  showed
that the ring collapses in  finite time
according to the evolution law
 \begin{equation}
\frac{dR}{dt}=-\frac{1+b^2}{R}.
\label{ott}
\end{equation}
In addition, there is no (at least, at first order in $1/R$)
overall drift of the vortex ring in the direction perpendicular
to the collapse motion. The collapse rate (often associated with the ``line
tension'') $\nu=1+b^2$ appears to be
in reasonable agreement with simulations for not too large $|b|$.
This corrects a previous
erroneous estimate  $\nu=1+b c$ (Frisch and Rica, 1992). 

For the evolution of the local twist of a straight vortex one obtains 
the Burgers equation (Gabbay, Ott, Guzdar  1997; 1998b, Nam {\it et al}, 1998)
\begin{equation} 
\partial_t \phi = (b-c) (1-k_0^2) (\partial_z \phi_z )^2 
+(1+bc +(b-c)bk_0^2) \partial_{zz} \phi
\label{twist} 
\end{equation} 
where $ \partial_z \phi =k_z$ and $k_0$ is the asymptotic wavenumber of 
the 2D spiral solution.
A more complicated equation was obtained for a twisted and 
curved vortex filament.


\subsection{Vortex nucleation and reconnection} \label{recon} 

Vortex reconnection in  NLSe was studied
in relation to turbulence in superfluid liquid helium. 
Large-scale computations were performed by Schwarz (1988) using the
localized induction approximation. These computations give an impressive pictures of vortex tangles. 
However, 
it remains unclear if the particular features of the tangle are real or
an artifact of the localized induction approximation. 

\begin{figure}
\centerline{(a)  \psfig{figure=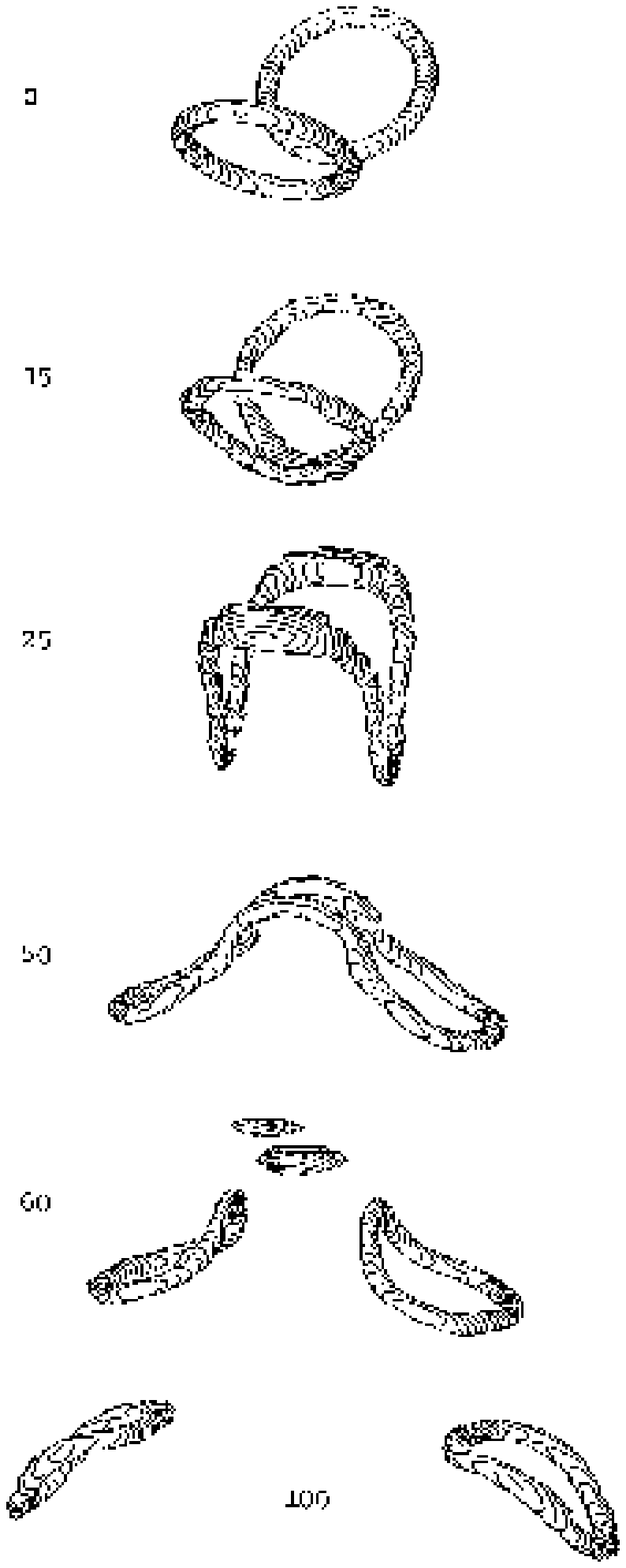,height=2.5in}
\hspace{1.0cm} (b) \psfig{figure=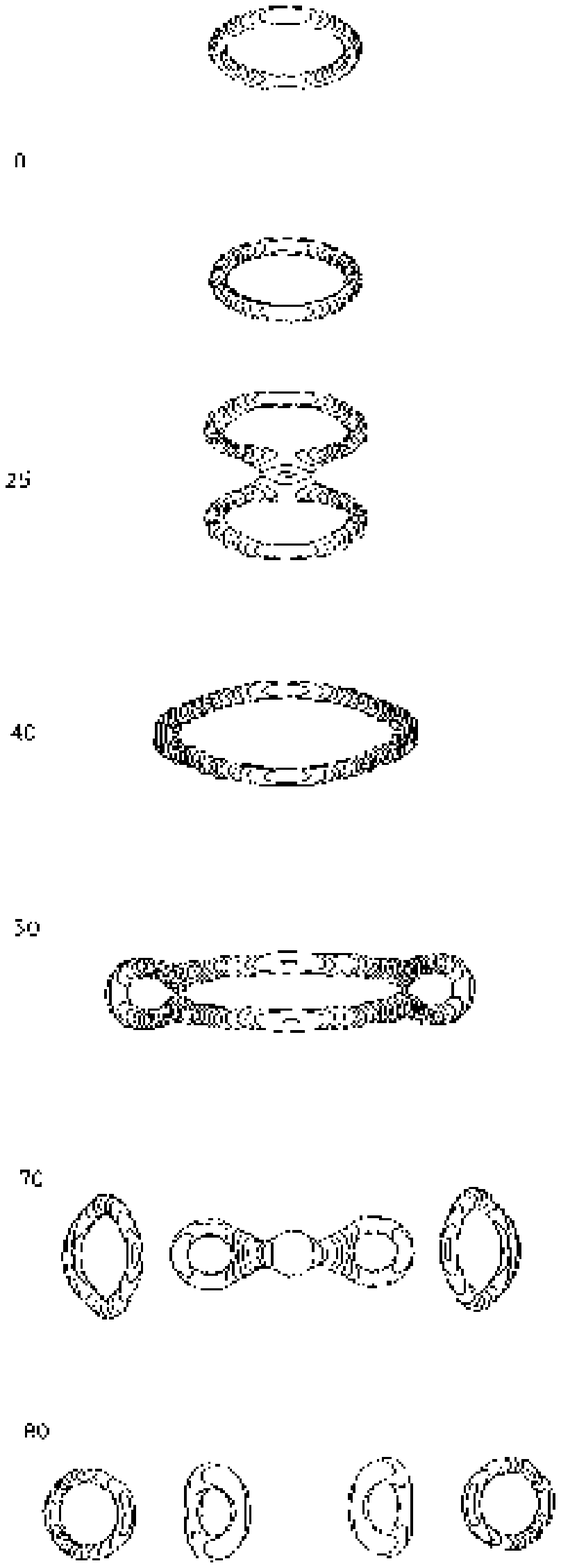,height=2.5in}}
\caption{Dynamic of vortex rings in the NLSe. (a) In-plane collision of two
rings at $90^o$ incidence,
seen form the side; (b)
in-plane collision of two
rings at $120^o$ incidence,
 seen from above,  Koplik and Levine (1996)}
\label{Fig1_3d}
\end{figure}

Obviously, vortex reconnection must be described by the full NLSe. 
Koplik and Levine (1993,1996) find in full numerical simulations of the NLSe 
that vortices reconnect
when they approach within a few core lengths. 
Depending on conditions, the  vortex rings may 
scatter, merge and  break up subsequently, see
Fig. \ref{Fig1_3d} 

Gabbay {\it et al}  (1998b) studied vortex reconnection in the CGLe. 
As a result of the interplay 
between two effects: motion of the filaments towards each-other
due to attraction (two-dimensional effect) and opposite motion 
due to curvature (three-dimensional effect) 
a criterion for vortex reconnection was proposed. 

In the GLe vortex rings ultimately shrink. However, with an additional 
phase gradient $j$ parallel to the ring axis
(e.g. due to rotation of a superfluid or supercurrent in 
superconductors) the vortex rings may expand. The additional force is 
the analog of the Peach-Koehler force on a 2D vortex in a background
wavenumber $Q=j$. In this situation 
one has the following 
equation for the ring radius $R$: 
\begin{equation}
\frac{d R}{d t} = -\frac{1}{R}+ \sigma j.
\label{vr2}
\end{equation}
where $\sigma\approx 2 j /\log(v_0/2j)$ (see Eq. (\ref{mob1})). 
Thus, if $R>1/(\sigma j)$ the vortex ring will expand.  

Expanding vortex rings in the GLe were obtained in simulations 
as a result of nucleation 
after a rapid thermal quench  by Aranson, Kopnin and Vinokur (1999). 
This problem was considered in the context of experiments in superfluid 
liquid $^3$He heated well above the transition temperature 
by absorption of neutrons by  Ruutu {\it et al} (1996, 1998). 
This experiment was designed to verify the 
fluctuation-dominated mechanism for the
formation of topological defects  in the early Universe 
suggested by Kibble (1976)  and Zurek (1985) and elaborated in later work
by Dziarmaga {\it et al} (1999), Antunes {\it et al} (1999) mostly in 
1D and 2D. 

\begin{figure}
\centerline{\hspace{.0cm}  \psfig{figure=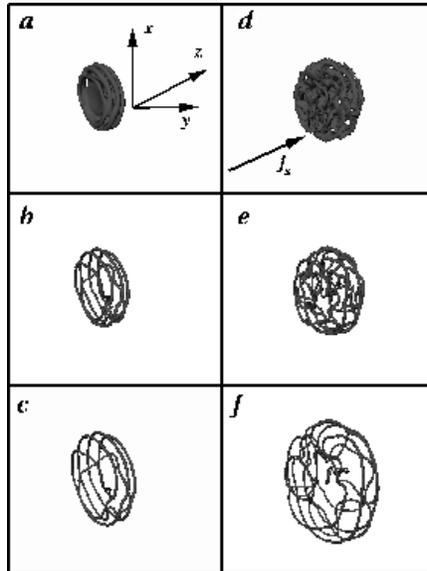,height=3.in}}
\caption{Nucleation and expansion of vortices in the GLe with background 
$k=0.4$. The supercurrent $j_s$ is given by $j_s=k (1-k^2)$.  
Shown are the 
3D isosurfaces of
$|A|=0.4$. 
(a-c) No thermal fluctuations,  images are taken at times $t=36,48,80$.
(d-f), Amplitude of thermal fluctuations $T_f=0.002$, $t=24, 48 , 80$.
(Aranson, Kopnin, Vinokur, 1999)}
\label{Fig2_3d}
\end{figure}

Selected  results are shown in Fig. \ref{Fig2_3d}.
One sees (Fig. \ref{Fig2_3d}a-c) that without fluctuations
the  vortex rings nucleate upon passage of the
thermal front. Not all of the rings survive: the
small ones collapse and only the big ones grow. Although the vortex lines
are centered around the  point of the quench, they exhibit a
certain degree of entanglement. After a long transient period,
most of the vortex rings reconnect and form an
almost axisymmetric configuration.

It turns out that fluctuations have a  strong effect at
early stages:
the vortices nucleate not only at the
normal-superfluid interface, but also in the bulk of the supercooled
region (Fig. \ref{Fig2_3d}d-e).
However,  later on, small vortex rings in the interior collapse and only
larger rings
(primary vortices) survive and expand (Fig. \ref{Fig2_3d}f).

\subsection{Instability of weakly-curved filaments in the large $b$
limit} \label{inst_3D}

Aranson and Bishop (1997), Aranson {\i et al}  (1998) 
have shown that the simple relation for the collapse rate 
Eq. (\ref{ott}) is violated in the large $b$ limit,  
$b=1/\epsilon \gg  1$. 
As a result of an asymptotic expansion for $\epsilon \ll 1$ the equation of 
motion  of the filament takes the form 
\begin{equation}
{\bf \partial_t v} +  \hat K [\epsilon \bf  v - \kappa  \bf n ] =0.
\label{accel2}
\end{equation}
where $  {\bf v}$ is the velocity of the filament,  
 and $\kappa$ the local curvature.
The $2\times 2$ matrix $\hat K$ corresponds to the ``friction''
matrix of the 2D spiral waves, see Sec. \ref{corins}.

Note, that dropping the
acceleration term in Eq.  (\ref{accel2}) one  recovers  the  
result of Eq. (\ref{ott})
for $b \to \infty $, since for the ring $v_N  
=\partial_t R$,
$\kappa = -1/R$. Restoring the original scaling $r \to r/\sqrt b$,  
one obtains
$\partial_t R=-b^2/R$. However,  since in 3D
the local   velocity
in general varies along the vortex line, even small acceleration
may cause severe  instability, 
because the local curvature becomes very large.
Moreover, deviation of the local velocity  from the direction of the
normal will lead to stretching and bending of the vortex line.
Thus the acceleration term, which formally can be considered as
a higher-order correction to the equation of motion, plays a  
pivoting (crucial)
role in the dynamics of a vortex filament.

\subsubsection{Perturbation around a straight vortex}

An almost straight vortex parallel to the $z$ axis can be
parameterized by the position along  the $z$ coordinate: $(X_0(z),Y_0(z))$.
Since in this limit
the arclength $s$ is close to $z$,
the curvature correction to the velocity $\kappa  {\bf  n }$
is simply $\kappa {\bf n  }  =(\partial_z^2 X_{0},\partial_z^2 Y_{0})
=\partial_z^2 {\bf r}$, where ${\bf r}= (X_0, Y_0)$.
Using $\partial_t {\bf r} ={\bf v}$, 
Eq. (\ref{accel2}) reduces to a linear equation
\begin{equation}
\partial_t  {\bf v }+  \hat K [\epsilon   {\bf v} -  \partial_z^2   
{\bf r}] =0, 
\label{accel3}
\end{equation}
The solution can be written in the 
form ${\bf  r}  \sim \exp [ i k z + \lambda(k) t]$, where $\lambda  
$ is the
growth rate. We immediately obtain the following relation for  
$\lambda$:
\begin{equation}
 \lambda^2+\chi (
\epsilon \lambda+ k^2)=0. 
\label{lambda}
\end{equation}
where $\chi=K_{xx} \pm i K_{xy}$ is the ``complex friction'' (
compare 
Sec. \ref{corins}). 
One may consider two cases: $k \ll \epsilon $ and $k \gg  
\epsilon$.
For $k \ll  \epsilon$ from Eq. (\ref{lambda}) one obtains $\lambda =-
\epsilon \chi  + O(k^2)$, i.e. recover the core instability of 
the 2D spiral. 
For $k \gg \epsilon $ one derives
$\lambda \approx
  \pm \sqrt {-(K_{xx}\pm i K_{xy } )} k $.
There always exists a root with a large positive real part:   
$\lambda \sim
k \gg \epsilon$.
Therefore, for finite $k$, the growth rate $\lambda(k)$
may significantly exceed the growth rate  of the acceleration  
instability in
2D (corresponding to $k=0$): $  {\rm Re} \lambda = -\epsilon K_{xx}$.
Hence, the "small-curvature" approximation  
considered above can be valid
only for finite time. 
The  fall off of the growth rate $\lambda$ at large $k$  
is not
captured by the small-curvature approximation used here. 

\subsubsection{Numerical results}

\begin{figure}
\centerline{\hspace{.0cm} \psfig{figure=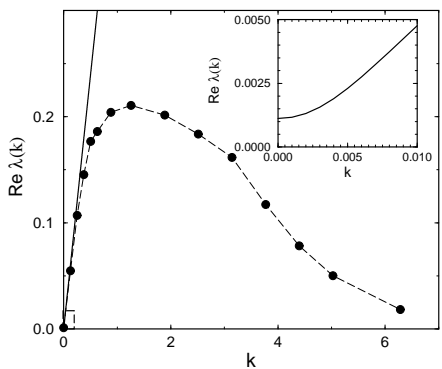,height=3.in}}
\caption{
The  growth rate $Re \lambda(k)$ as function of $k$ for
$\epsilon=0.02, c=.1 $. Solid line is the theoretical result
for $k \ll 1$,
dashed line with symbols the result of numerical solution of 3D
CGLe. Inset: blow-up of small $k$ region.
(Aranson and Bishop, 1997)}
\label{Fig3_3d}
\end{figure}

Comparison of the
theoretical results with numerical simulations 
is shown in Fig. \ref{Fig3_3d}.  
As an initial condition a straight vortex line with  
small periodic modulation
along the $z$-axis was taken.
As one sees from the figure,
the growth rate indeed  increases initially with $k$, and then  
falls off for large $k$. 
The theoretical expression (\ref{lambda}) shows reasonable 
agreement with the simulations for small enough $k$.
The growth rate  at the optimal wavenumber exceeds 
the corresponding growth-rate of the
acceleration instability ($k=0$) by
more then two orders of magnitude.

\begin{figure}
\centerline{\hspace{.0cm} \psfig{figure=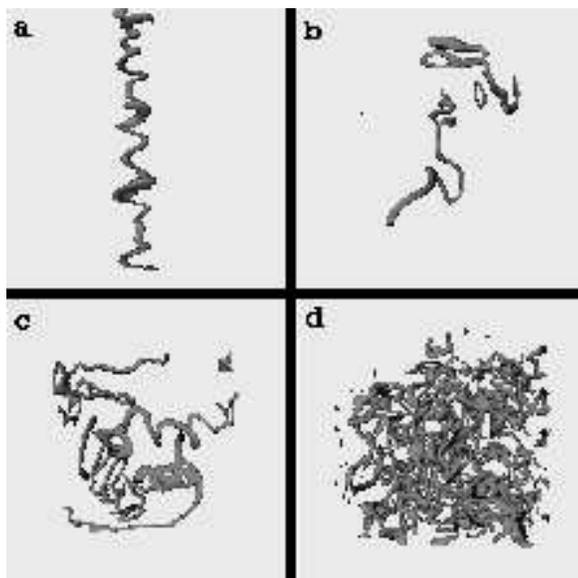,height=3.in}}
\caption{
Instability of a straight vortex filament.
3D isosurfaces of $|A(x,y,z)|=0.1$ for
$\epsilon=0.02$, $c=-0.03 $, shown at  four
times: 50 (a), 150 (b), 250 (c), 500(d).
Similar dynamics is observed also for larger value of $\epsilon$
(Aranson and Bishop, 1997).
}
\label{Fig4_3d}
\end{figure}

\begin{figure}
\centerline{ \psfig{figure=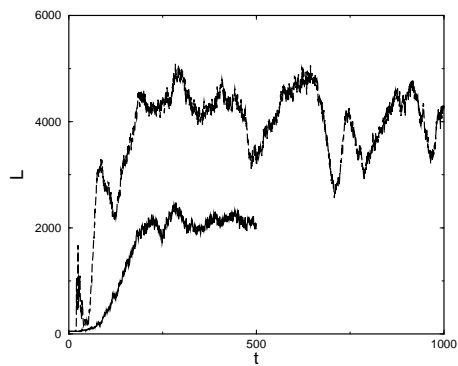,height=3.in}}
\caption{
The dependence of filament length $L$ on time.
Solid line corresponds to
$\epsilon=0.02, c=-0.03 $; dashed line corresponds to
$\epsilon=0.02, c=-0.5$.
}
\label{Fig5_3d}
\end{figure}

The long-time evolution of a perturbed 
straight vortex is shown in Fig. \ref{Fig4_3d}.
As  one sees from the figure,
the length of the vortex line grows.
The dynamics seems to
be very rapidly varying in time, and the line intersects itself  
many times
forming numerous vortex loops. 
The long-time dynamics shows,
however, a saturation when a highly-entangled vortex
state is developed  and the total  length of the line cannot grow  
further due to
a repulsive interaction between closely packed line segments.
The dependence of the line length on time is
shown in Fig. \ref{Fig5_3d}.
One can identify two distinct stages of the dynamics:
first, fast growth of the length;
second, oscillations of the line's length around some
mean value.

\begin{figure}
\centerline{\hspace{.0cm}(a)  \psfig{figure=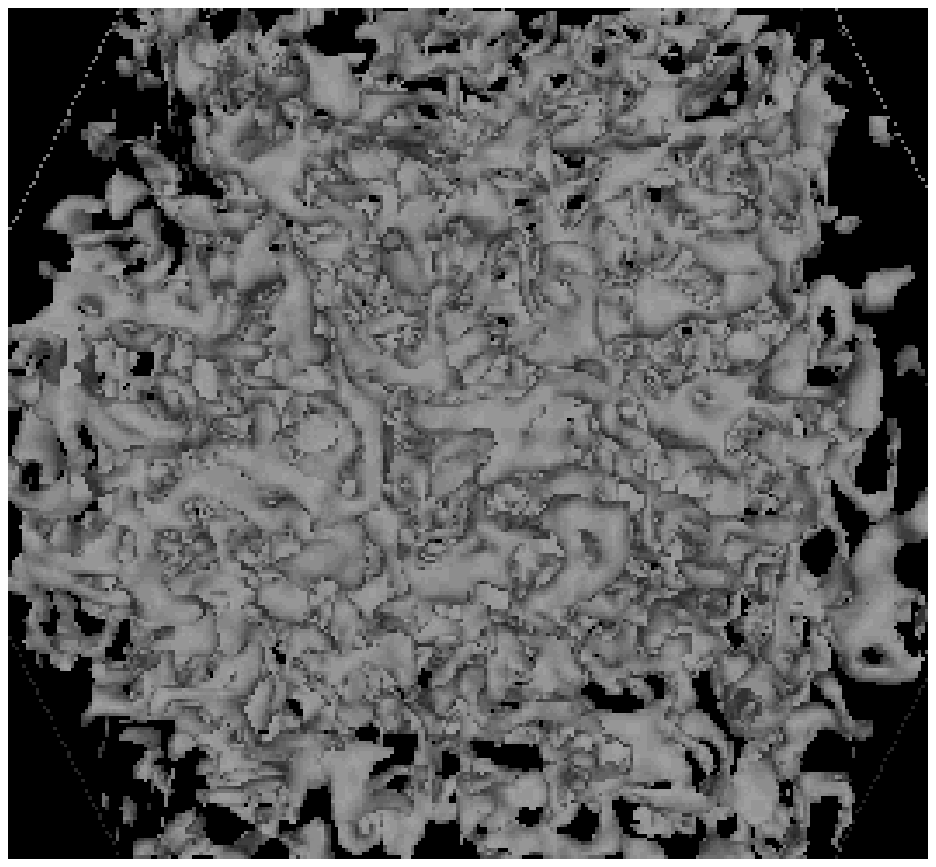,height=3in}
(b) \psfig{figure=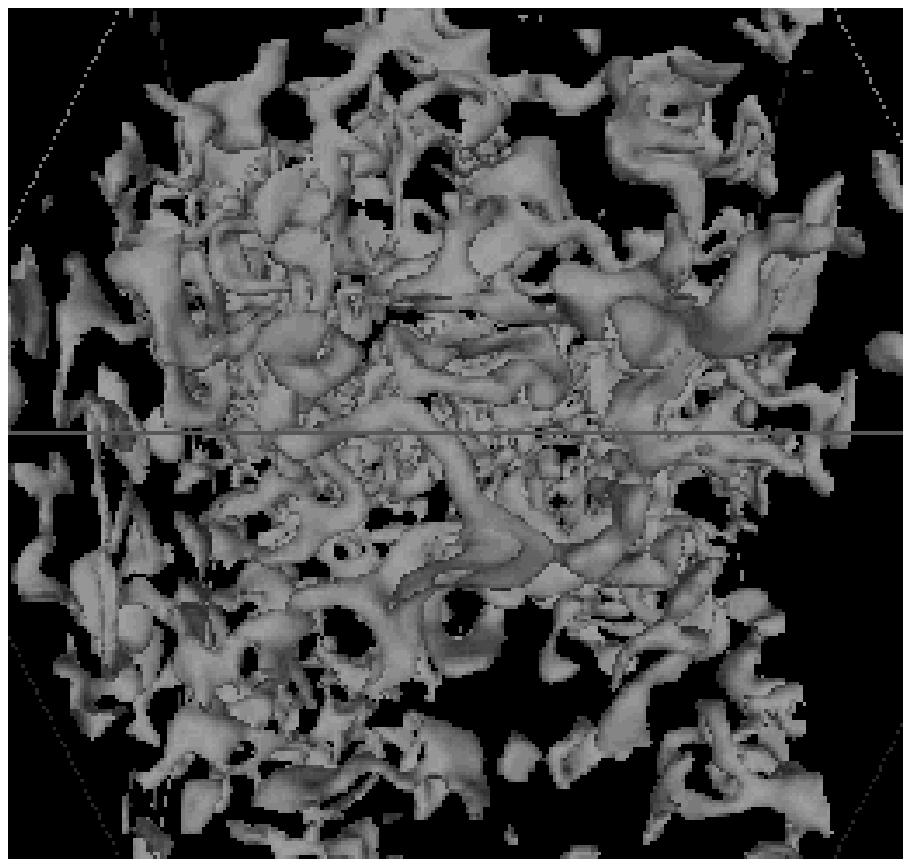,height=3in}}
\caption{
Two snapshots of 3D isosurfaces of $|A|$ taken in the regime of
spatio-temporal intermittency,  $\epsilon=0.02, c=-0.5$.
(a) Left image corresponds to  $t\approx 620$ for Fig. \protect
\ref{Fig5_3d}; (b) right image
corresponds to $t\approx 740$.
(Aranson, Bishop, and Kramer, 1998).
}
\label{Fig6_3d}
\end{figure}

For small enough $\epsilon$,   two  
distinct behaviors
of the total vortex length
depending on the value of $c$ are observed. Above a critical value $c_c $
corresponding
approximately to the convective instability range of the  
2D  spiral 
($c_c \to 0$ for $\epsilon \to 0$),
the total length approaches some equilibrium value and does not exhibit
significant fluctuations. On the contrary, for $c< c_c$, the total  
length exhibits large
non-decaying intermittent fluctuations around the mean value.
Figure \ref{Fig6_3d} presents the 
snapshots illustrating the  
structure  of
the vortex field corresponding to the  
moments of maximum 
and minimum of the length. One sees, that
in this situation some segments of vortex lines start to expand  
spontaneously,
pushing away other vortex filaments and in such a way making substantial
vortex-free holes around them. Then the instability takes over and  
destroys
these almost-straight segments of filament, bringing the system back to a
highly-chaotic state. This dynamics can be considered as a  
3D
spatio-temporal vortex intermittency, which is an extension of
spiral intermittency discussed in Sec. \ref{core-inter}. 
For even smaller values, $c< -1 $, one has the
transition to a highly chaotic state, which is an analog of ``defect
turbulence'' in the 2D  CGLe. In this regime small vortex loops
nucleate and annihilate spontaneously.

\begin{figure}
\centerline{\hspace{.0cm} \psfig{figure=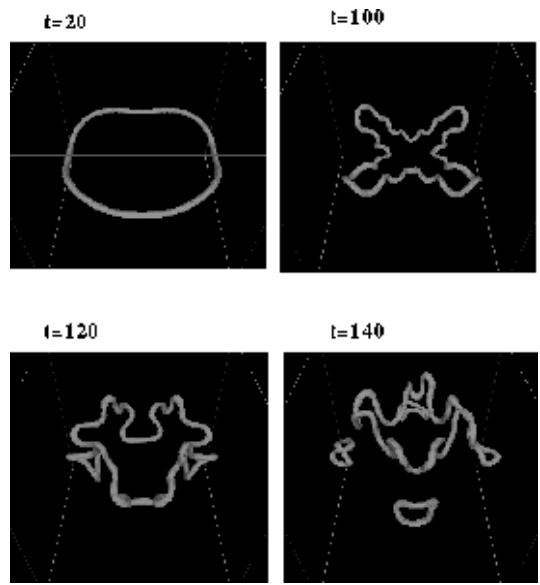,height=3in}}
\caption{
Sequence of snapshots  demonstrating the evolution of a vortex
ring for $\epsilon=0.2$ and $c=0.2$.}
\label{Fig7_3d}
\end{figure}

The evolution of a closed vortex loop is shown in Fig. \ref{Fig7_3d}. 
The  simulations show that the 3D instability may  
prevent the ring from
collapse, causing the stretching of the loop in the direction  
transversal to
the  collapse  motion. However, 
small rings
typically collapse, since then the instability described above does  
not have
time to develop substantial distortions of the ring. Even in this
situation the ring exhibits a few oscillations of the radius.

\subsubsection{Limits of three-dimensional instability}

\begin{figure}
\centerline{\hspace{.0cm} \psfig{figure=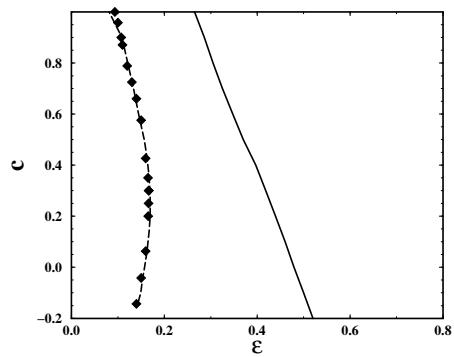,height=3in}}
\caption{
Stability limits in three-dimensions (solid line) and two dimensions
(dashed line), obtained from linear stability analysis. Symbols
represents
the  limit of the two-dimensional instability, obtained by direct
numerical simulation of the CGLe (Aranson, Kramer, Weber, 1994).
Vortex lines and two-dimensional spirals are stable to the right  of
the respective lines.
}
\label{Fig8_3d}
\end{figure}

The previous analysis indicates instability of vortex lines in the  
limit
$\epsilon \to 0$ for all $c$. However, it cannot
describe  the boundary  of the instability for increasing $\epsilon$. 
In order to obtain the stability limit one needs to perform a full
linear stability analysis of a straight vortex solution,  
not limited
to small $k$ and $\epsilon$
(Aranson, Kramer, Bishop, 1998). 
The linear stability analysis shows that   
the 3D instability persists substantially beyond the  
2D core instability. 
The results are shown in  
Fig. \ref{Fig8_3d}.
Moreover,
the typical growth-rate in 3D is much higher than in 2D.

\subsection{Helices, twisted Vortices and supercoiling instability}
Close to the stability boundary of the  3D
instability, the evolution of a 
straight vortex does not necessarily result in spatio-temporal chaos. 
In contrast, the simulations show that the 
instability saturates, resulting in 
a traveling helix solution (Aranson {\it et al}  1998) 
or superposition of two helices with opposite chirality 
(Rousseau {\it et al}  1998). Indeed, since the left and 
right rotating unstable modes of a straight vortex have the 
same growth rate, the resulting configuration is determined by 
the cross coupling coefficient between these modes, 
which is a function of the parameters $b,c$. 

The symmetry between left and right rotating helices can be broken 
by applying an additional twist to the straight filament. 
Studies of twisted filaments were performed by 
Rousseau {\it et al}  (1998) and by 
Nam {\it et al} (1998). The simulations revealed stable helices
and a secondary super-coiling instability. 

Nam {\it et al} (1998)  performed linear stability analysis of 
a straight filament with twist. It was shown that the  twist 
reduces the domain of stability for the straight filament. 

The question arising in this context is how to prepare a vortex 
with twist. One way to proceed is to create 
an inhomogeneity close to the axis of the vortex filament. 
The inhomogeneity
will locally change the frequency of the vortex and will result  in 
persistent twist. This situation was realized experimentally in a
reaction-diffusion system by Mironov {\it et al} (1996).

\section{Generalizations of the CGLe}
\label{sec:general}

The CGLe is a minimal, universal model that cannot be 
further simplified. However, there are many ways to 
generalize it in order to include qualitatively new
features (we are not concerned with additional terms that merely give
quantitative corrections). 
The main trends of generalization can be associated with the different terms 
the  CGLe: 
\begin{itemize}
\item
generalization of the nonlinearity
\item
generalization of the differential operator
\item
generalization of the symmetry group 
\end{itemize}
 
\subsection{Subcritical CGLe} 
\subsubsection{Small Amplitude Solutions in the weakly nonlinear case}
\label{smampsol}
In this subsection we consider the CGLe with a destabilizing nonlinearity
\begin{equation}
\partial_t A = A + (1+i b) \Delta A - (-1+ic) |A|^2 A 
\label{scgle}
\end{equation}
The CGLe was first derived in this form for a physical system by 
Stewartson and Stuart (1971) in the context of 
plane Poiseuille flow where one has a (strongly) subcritical bifurcation. 

Spatially-homogeneous solutions of Eq. (\ref{scgle}) with $A(t=0)=A_0$ diverge
at finite time according the following expression:
\begin{equation}
|A| = \frac{|A_0| e^t} { [1 + |A_0|^2 (1-e^{2 t}) ] ^{1/2} }
\end{equation} 
Also, for $c \ne 0$ the rate of phase winding diverges. 
When $|A|$ is a function of $x$, however, its behavior is more subtle.

At first sight one expects blowup of the solutions which can be avoided 
by adding a fifth-order stabilizing term. 
It was suggested by Hocking and Stewartson (1972) and 
Hocking {\it et al} (1972), that for generic
initial conditions the blowup does not occur in a considerable 
region of the parameter space $(b,c)$. 
By considering the evolution of pulse-like solutions,
the region in ($b,c$) plane was found where  the solution remains bounded. 
Since $|c|$, which is the ratio of nonlinear dispersion over nonlinear growth,
has to be sufficiently large, this may be called the ``weakly subcritical case''.

Then these results were apparently forgotten, and addressed again 
(independently)
by Bretherton and Spiegel (1983) (in the limit $c \to \infty$), and 
Sch\"opf and Kramer (1991). They reproduced much of the results of 
Hocking and Stewartson (1972)  and  found stable periodic solutions 
of Eq. (\ref{scgle}). The analytic work was supported by detailed simulations.
The work was continued by Powell and Jakobsen (1993), Kaplan {\it et al} 
(1994a,b),
Kramer {\it et al} (1995), and Popp {\it et al} (1998). 
Weakly subcritical Hopf bifurcations are found in convection in binary fluids 
(see Moses {\it et al} (1987), Heinrichs {\it et al} (1987), 
Kolodner {\it et al}, 
(1988, 1995, 1999)), 
and presumably in nonlinear optics (Powell and Jakobsen, 
1993, Kramer {\it  et al}, 1995).

Hocking and Stewartson (1972), Bretherton and Spiegel (1983), and 
Sch\"opf and Kramer (1991) have found
bursts of two types depending on the relative signs of $b,c$ 
(by analogy with the NLSe the case of  $bc \ge 0$ will be called the 
focusing case and otherwise defocusing).

Kaplan {\it et al} (1994a,b), Kramer {\it et al} (1995) 
proposed a simple physical mechanism, called the  {\it phase gradient mechanism},
providing arrest of the collapse in Eq. (\ref{scgle}) if $|c|$ is sufficiently
large and $|b|$ not too large. 
The phase gradient effect manifests itself as a fast differential phase rotation 
resulting from the explosive burst amplitude increase. To understand this effect 
it is convenient to rewrite Eq. (\ref{scgle}) in
the variables $A= \sqrt R \exp [ i \theta ]$ (see Eqs. (\ref{modpha})).
For  the sake of simplicity one considers the limit $b \ll 1$.
Introducing the local wavenumber $k = \partial_x \theta$
Eqs. (\ref{modpha})  reduce to 
\begin{eqnarray}
\partial_t R & = & (1 +R^2) R + \partial^2_x R -k^2 R \nonumber\\
\partial_t k &=& c \partial_x R^2 
+\partial_x\left(\frac{\partial_x (R^2 k)}{R^2}\right)
\label{phgrad}
\end{eqnarray} 
From the second equation one sees that a gradient in $R$ drives growth of 
$|k|$, which in turn saturates R via the last term in the first equation. 
If $|c|$ is sufficiently large, this effect overcomes the explosive growth 
manifest in the first term on the right-hand side of the first equation.
Specifically, consider an initial pulse having a small, broad, constant plateau
and decaying away at the edges
Let $k=0$ initially. At the linear stage of the instability
and then in the forthcoming 
blow-up regime the amplitude will remain approximately constant inside the 
plateau region. At the boundaries sharp gradients of $R$ will 
be formed. These gradients will act as sources for the generation of the 
phase gradient, i.e. $k$, in narrow regions. 
There the large value of $k$ will saturate the blow-up.
The net result will be two counter-propagating fronts representing the 
moving plateau boundaries.
The front propagation speed will not be constant but it will grow 
at the blow-up stage. Thus, if $|c|$ is sufficiently large, the pulse will be 
``eaten'' by these fronts moving from the edges to the center.

\subsubsection{Strongly subcritical case} 
The solutions considered above bifurcate from the trivial state supercritically
(in the range of $b,c$ where they remain bounded), in spite of the fact that
the sign of the real part of the nonlinear term signalizes a subcritical 
bifurcation. Thus in that parameter range, but below threshold, the trivial 
state is the global attractor. The scenario is not changed 
qualitatively by the addition of stabilizing higher-order terms. Outside 
this parameter range one needs (at least) quintic terms to saturate the 
explosive instability provided by the cubic term. 
The equation can be written in the form
\begin{equation}
\partial_t A = \epsilon A + (1+ib) \Delta A - (-1+ic) |A|^2 A -(1 + i d) |A|^4 A
\label{scgle1}
\end{equation}
The finite-amplitude
solutions persist stably with respect to amplitude fluctuations below threshold
$\epsilon<0$ in a certain parameter range, where they coexist with the linearly 
stable trivial solution.
There exist moving fronts and - surprisingly - stable localized pulses over
a finite interval of parameters (Thual and Fauve, 1988). This clearly is
a result of the nonvariational nature of Eq. (\ref{scgle1}).
A study of existence, stability and selection of various solutions
is given by van Saarloos and Hohenberg (1992). 
Localized perturbations around the trivial state can either decay
(small $\epsilon$), or evolve into pulses (intermediate $\epsilon<0$),
or develop into fronts that invade a plane-wave state. In certain cases
their velocity is
selected by a ``nonlinear'' marginal stability criterion below some positive
value of $\epsilon$ and by linear marginal stability, as in the supercritical 
cubic CGLe, for larger $\epsilon$. 
In a parameter range with sufficiently small nonlinear dispersion there exists a 
class of fronts that can be expressed in the form (polynomial fronts) 
 \begin{equation}  \label{polyfront}  
  k=k_N+e_0 (R^2- R_N^2),\quad R'=e_1 R(R^2-R_N^2)
 \end{equation}
Very recently steady fronts that cannot be expressed in this form were found 
(Coullet and Kramer, 2001). They exist in particular in a parameter range 
where there are
no polynomial fronts. Whereas the polynomial fronts are sources in their rest
frame the new fronts are sinks. They move in a direction to generate the 
trivial state even for positive values of $\epsilon$, which can be understood 
from the phase gradient mechanism discussed above. They play a significant role
in the dynamics of spatio-temporal chaotic states and for the formation of 
localized structures.

Deissler and Brand (1994, 1995) and 
Akhmediev {\it at al} (2001) have studied   
periodic, quasiperiodic and chaotic pulses of Eq. (\ref{scgle1}) and 
its generalization. 
The interaction of these solutions in the framework of two coupled equations 
Eq. (\ref{scgle1}) shows that the result depends sensitively 
on the initial conditions. 
Convective and absolute instabilities in the subcritical 
CGLe are investigated by Colet {\it al} (1999).

Deissler and Brand (1991) found 
in 2D localized particle-like solutions of Eq. (\ref{scgle1}). 
Moreover, one expects localized solutions possessing a topological charge
coexisting  with extended (conventional) spirals known for CGLe  
(Malomed and Rudenko, 1988). 
Recently, the properties of spiral waves and other localized solutions 
in the cubic-quintic 
CGLe were studied by Crasovan {\it et al}, 2001.

\subsection{Complex Swift-Hohenberg equation}
The complex Swift-Hohenberg equation  in the form 
\begin{equation} 
\partial_t A = r A - (1+ic ) |A|^2 A + i a \Delta A -(\Omega + \Delta)^2 A 
\label{cshe}
\end{equation}
where $r \ll 1$ is the control parameter, $a$ characterizes the diffraction properties 
of the active media, 
was derived asymptotically in the context of large aperture lasers (class 
A and B) with small detuning 
$\Omega$ between the atomic and cavity frequencies (see Staliunas, 1993; Lega,
Moloney and Newell, 1994, 1995). 
This equation is believed to be relevant also for oscillatory convection
in binary fluids, however, it can not be derived asymptotically from appropriate
Navier-Stokes equations. Eq. (\ref{cshe}) is a generic equation in the vicinity of 
a  
codimension-2 bifurcation  
where the coefficient 
in front of the diffusive term is allowed to change sign 
(see, e.g. 
Coullet and Repaux, 1987).  
 
Clearly one should distinguish between the real Swift-Hohenberg equation
(see, e.g. Cross and Hohenberg, 1993) 
and  Eq. (\ref{cshe}).  The 
Swift-Hohenberg equation is a phenomenological model and can not be derived from the original 
equations \footnote {The Swift-Hohenberg equation is obtained by  keeping only one nonlinear term.
However, nonlinear terms involving also derivatives have formally the 
same order, and, strictly speaking, cannot be neglected}. 
In contrast, the complex Swift-Hohenberg equation 
 in the form (\ref{cshe}) is derived asymptotically rigorous
in the limit of $\Omega \to 0$, i.e. in the so-called ``long-wavelength'' limit.
In this limit the differential nonlinearities have formally higher order 
and, therefore, can be dropped. In this case the ``wavenumber selecting term''
$(\Omega + \Delta)^2 A$ is just a small correction to the ``diffraction term''
$i a \Delta A $. Thus, the complex Swift-Hohenberg equation can be treated as 
a perturbed CGLe. 
In contrast to the real Swift-Hohenberg equation,  Eq. (\ref{cshe}) has two 
independent wavenumber selection mechanisms: the first is related to the 
$(\Omega + \Delta)^2 $ term, 
providing maximal amplification for the plane wave 
$\exp[ i (\omega t + k x)]$ with the optimal wavenumber $k = \sqrt \Omega$;
the second is related to the selection of wavenumber by topological defects 
(e.g. spirals, holes) in the CGLe and relies on the diffraction $i a \Delta A$. 
For this case the last term in Eq. (\ref{cshe}) changes slightly the wavenumber selected 
by the defects  (Aranson,  Hochheiser and  Moloney, 1997).

The stability of plane waves in Eq. (\ref{cshe}) was studied in details by
Lega {\it et al} (1994). A new feature  is the zig-zag 
(transversal) instability of plane waves for wavenumbers away from the
band center.
 
In 2D Eq. (\ref{cshe}) possesses, as the  CGLe, topological defects in the form of 
spiral waves. Aranson,  Hochheiser and  Moloney, 1997 show that in 
the limit of small $r$
these spiral waves undergo a core instability, leading to stable meandering. 
Another feature of Eq. (\ref{cshe}) in 2D is the existence of domain boundaries 
between traveling waves with different orientation, usually called
zipper states. 
Furthermore, in 2D, the question of wavenumber selection
can be transformed into one of {\em wavevector} selection,
since the domain wall can adjust the direction of ingoing or
outgoing waves. The domain wall itself may no longer be stationary,
but may move in a certain direction, if
there is no reflection symmetry of the wave pattern with respect to
the domain wall axis. The second spatial dimension
(along the domain wall) opens the possibility for additional
instabilities of the wall, as it was observed experimentally in convection
in binary mixture by
Moses {\it et al}  (1987), La Porta and Surko (1997). 
Aranson and Tsimring (1995) have shown that near threshold the 
active (emitting waves) zipper states are alway unstable with respect 
to transversal undulation. The nonlinear stage of this instability leads 
to creation of a chain of 
topological point defects (spirals) which themselves are unstable. 
The latter appears to be analogous to
the famous Kelvin-Helmholtz instability  of a tangential discontinuity
of shear flows. Passive (absorbing) zipper states turn out to be stable.

\subsection{CGLe with broken gauge invariance}

It is interesting to break the global gauge invariance of the 
CGLe. This corresponds in particular to a situation where a system undergoing 
a Hopf bifurcation with the frequency $\omega_c$ is parametrically forced
(modulated) at a frequency near $\omega_c/2$. This leads to 
\begin{equation} 
\partial_t A = A(\epsilon+i \omega) + (1+i b )  \Delta A-(1+i c) |A|^2 A + \gamma A^*
\label{cglec}
\end{equation}
Obviously, instead of global gauge  
invariance   $ A \to A e ^{i \Phi}$  one is left with the discrete symmetry 
$A \to -A$. 
Here $\omega$  is the 
detuning,  $\gamma>0$ is the amplitude of forcing and 
$\epsilon $ describes the distance from the threshold of instability.

\subsubsection{From oscillations to bistability ($\epsilon>0$)}

For $\epsilon >0$,
\footnote{Note that for $\epsilon >0$, by rescaling $t, {\bf r}, \omega, 
\gamma, A$ in Eq. (\ref{cglec}), $\epsilon$ can be replaced by $1$.} 
depending on the values of the other parameters,  Eq. (\ref{cglec})
describes an oscillatory or bistable situation. In the latter case the system is in 
general  (i.e. for $b,c,\omega \ne 0$) of excitable nature. 
In the  context of ferromagnets in a static magnetic field it describes 
domain walls separating two stable 
states  (domains with opposite spins). The
domain walls exhibit a transition involving a spontaneous 
breaking of chirality (in ferromagnets Ising walls become Bloch walls as the 
strength of crystal anisotropy is reduced, Lajzerowicz and Niez, 1978, 1979). 
In such an equilibrium situation the imaginary coefficients in (\ref{cglec}) 
vanish and Eq. (\ref{cglec}) can be cast into variational form:

\begin{equation}
\partial_t A = - \frac{\delta F}{\delta A^*}
\label{varf}
\end{equation} 
where the ``free energy'' functional is of the form
\begin{equation}
F = \int \left[ - \epsilon |A|^2 +\frac{|A|^4}{2} +|\nabla A|^2 -
\frac{\gamma}{2}  ((A^*)^2 +A^2) \right] dx dy 
\label{lfunc}
\end{equation}
Equation (\ref{varf}) possesses stationary kink-like solutions 
connecting stable homogeneous equilibria $A = \pm \sqrt{1+\gamma}$ for $\epsilon=1$: 
\begin{eqnarray}
A_I &=& \pm \sqrt{1+\gamma} \tanh(\sqrt{1+\gamma} x) \label{ising} \\ 
A_B &=& \pm \sqrt{1+\gamma} \tanh(\sqrt{2 \gamma} x)
\pm i \frac{\sqrt{1-3 \gamma}} {\cosh(\sqrt{2 \gamma} x)} \label{bloch} 
\end{eqnarray}
The first solution, called by Coullet {\it et al} (1990) Ising wall, is stable 
when $\gamma > \gamma_c = 1/3$ and the second solution, which is stable
for $\gamma<\gamma_c$ and breaks chirality, is called Bloch wall. 
The order parameter $A$ vanishes at the core of the Ising wall but not at 
the core of the Bloch wall.
For $\gamma =1/3$ an exchange of stability 
(pitchfork bifurcation) occurs between  these two
solutions.

Coullet {\it et al} (1990) investigated the behavior of Ising and Bloch walls 
in nonequilibrium conditions, when at least one of the coefficients 
$b,c$ or $\omega$ is nonzero. They found that nonpotential effects in general 
lead to motion of the Bloch walls, and not of the Ising walls (
see also Sakaguchi, 1992; Coullet and Emilson , 1992; 
Mizuguchi and  Sasa, 1993; Chat\'e {\it et al}, 1999).
Coullet  {\it et al} (1991) applied  this concept to the description of the 
Ising-Bloch transition in ferromagnets in a rotating magnetic field.
 
Frisch {\it et al} (1994) have studied  Eq. (\ref{cglec}) in 2D in the context 
of a homeotropically aligned nematic liquid crystal in a rotating magnetic field
in the vicinity of the electric Fredericks transition. 
They have found that Bloch walls  containing a defect separating the two variants 
assume the form of rotating
spiral waves.
Those spiral waves had been studied experimentally  also  by Migler and 
Meyer (1994). 
Frisch {\it et al} (1994) have shown that 
spiral waves in Eq. (\ref{cglec}) 
combines properties of spirals in oscillatory (as CGLe) and excitable 
spirals (as in reaction-diffusion systems, Tyson and Keener, 1988). 
The problem of the spiral's frequency selection
was solved by Aranson (1995) for small $\omega$ and $b=c=0$ . 
In particular, it was shown that the frequency is selected by the local curvature
of the moving Bloch wall.
For $\gamma, \omega \ll 1$ the spiral wave solution can be described
in the phase approximation by the overdamped sine-Gordon equation
($\phi=\arg A$)
 
\begin{equation}
\partial_t \phi = \omega -\gamma \sin (2 \phi) + \Delta \phi
\end{equation}

Korzinov {\it et al} (1992) have studied Eq. (\ref{cglec}) in the context of 
periodically forced convection.  
Hanusse and Gomez-Gesteira (1994) considered it in the context of 
chemical systems.

Frisch and Gilli (1995) as well as Coullet and Plaza (1994) considered 
the more general equation:
\begin{equation}
\partial_t A = A(1+i \omega) + (1+i b )  \Delta A-(1+i c) |A|^2 A + \gamma A^* + \gamma_0
\label{cglec1}
\end{equation}
The term $\gamma_0$ is responsible for the effect of a tilted magnetic field. 
In contrast to Eq. (\ref{cglec}) the spiral waves in the framework of 
Eq. (\ref{cglec1}) exhibit a diverse variety of behaviors from rigid rotation 
to meandering and even hypermeandering. 
The Ginzburg-Landau equation with 
more complicated forcing terms was studied by Gilli and Gil (1994).

\subsubsection{Parametric excitation of waves in the GLe} 
Eq. (\ref{cglec}) for $\epsilon<0$ is a phenomenological model of 
parametric excitation of surface waves in fluids (Gollub and Langer, 1999). 
In the absence of parametric driving 
$\gamma A^*$ the system always relaxes towards the trivial state
$A=0$. However, if the parametric driving exceeds the critical value
$\gamma_c=(\omega-\epsilon b)/\sqrt{1+b^2}$, the  trivial state 
$A=0$ 
becomes unstable with respect to standing waves  with wavevector 
$k_c^2=(b \omega + \epsilon)/(1+b^2) $ of arbitrary orientation. 
In this sense Eq. (\ref{cglec}) is reminiscent of the Swift-Hohenberg 
equation (Cross and Hohenberg, 1993). 

In 2D 
a variety of nontrivial static and dynamic states are possible. 
Theoretical studies of spatio-temporal 
chaos  in Eq. (\ref{cglec})
in the context of surface  waves in fluids   were conducted by 
Zhang and Vi\~nals (1995).
Spiral waves were studied experimentally and theoretically by
Kiyashko {\it et al}  (1996).

Tsimring and Aranson (1997),  Aranson {\it et al.}  (1999)
used Eq. (\ref{cglec}) coupled to an additional field  to describe pattern formation 
in a thin layer of vibrated granular materials 
in connection with experimental studies 
(Melo {\it et al} 1994, 1995, Umbanhowar {\it et al} 1996).
In this case the parameters 
$\gamma$ and $\omega$ can be associated with the amplitude and the frequency 
of external periodic driving. Depending on the values of $\gamma$ and $\omega$, 
a variety of stable solutions ranging from localized {\it oscillons} and 
{\it interfaces} separating domains of opposite polarity to periodic stripes, 
squares and hexagons were found. The topology of the transition lines between
different types of the solutions turns out to be in agreement with the 
experiments.

\subsection{Anisotropic CGLe in 2D} \label{subsec:aniso}
In many physically relevant situations the 2D CGLe is essentially anisotropic
(see Sec. \ref{sec:intro}, class iii), i.e. it
is of the form 

\begin{equation}
\partial_t A = A + (1+i b_1 )  \partial_x^2 A + (1+ib_2) 
\partial_y^2 A -(1+i c) |A|^2 A 
\label{cglea}
\end{equation}
with $b_1 \ne b_2$. The equation in this from was 
studied by Weber {\it et al} (1991),
Brown {\it et al} (1993), Roberts {\it et al} (1996).  
The stability of
plane waves depends on the orientation of
the wavevector. In particular, one has the situation 
when the wave is stable only in one direction. 
Under such condition
Weber {\it et al.} (1991)
found stable lattices of defects.

New features of 
phase and defect chaos in the 2D  anisotropic CGLe 
were found  by Faller and Kramer (1998) and (1999).
The phase-chaotic states exist in a broader parameter
range than in the isotropic case, often even broader than in one dimension. 
They  may represent the global
attractor of the system. There exist two variants of
 phase chaos: a quasi-one dimensional and a two-dimensional
solution. The transition to defect chaos is of intermittent type.

\subsection{Coupled Ginzburg-Landau Equations} 
When both, the critical wavenumber $q_c$, and the 
critical frequency $\omega_c$ are non-zero at the bifurcation
(class iii) with reflection symmetry, Sec. \ref{sec:intro}), 
the primary modes are traveling waves 
which in 1D or in the presence of anisotropy are described by 
two coupled complex Ginzburg-Landau equations. 
The physical fields in the  weakly nonlinear regime are of the form 
$\sim A_R \exp [-i (\omega_c t -q_c x)] +
 A_L \exp [-i (\omega_c t +q_c x)] + c.c$, where $A_R$ and $A_L$ are the 
complex amplitude of right/left traveling waves. In 1D the 
coupled CGLes are given by
\begin{eqnarray} 
\partial_t A_R + s 
\partial_x A_R & =&  A_R + (1+i b) \partial_x^2 A_R - 
\left((1+i c) |A_R|^2 + (1+i d ) g | A_L|^2 \right) A_R \nonumber \\
\partial_t A_L - s 
\partial_x A_L & =&  A_L + (1+i b) \partial_x^2 A_L - 
\left((1+i c) |A_L|^2 + (1+i d ) g | A_R|^2 \right) A_L
\label{coupled} 
\end{eqnarray} 
where $s$ is the linear group velocity, $(1+id) g$ is the complex coupling 
coupling coefficient between 
the two modes (Cross and Hohenberg, 1993). 

In addition to the usual GLe parameters $b$ and $c$ one  here  has 
three relevant  parameters $s$, $d$ and $g$. Elaborate surveys of the 
various regimes occurring in Eqs. (\ref{coupled}) are given 
by van Hecke {\it et al} (1999),  Riecke and Kramer (2000), 
see also, 
Amengual {\it et al} (1996,1997), Neufeld {\it et al} (1996).

The case of strong suppression corresponds to $g>1$. In this situation dual-wave 
solutions with $A_R=A_L\ne 0$ are unstable. In contrast, single-wave solutions 
$A_R\ne0$, $A_L=0$ or vise versa can be stable. 
For $g>1$ a variety of source and sink solutions in counter-propagating waves 
were analyzed by Malomed (1994),
Alvarez {\it et al},   (1997),  van Hecke {\it et al}  (1999).

\subsection{Complex defects in vector Ginzburg-Landau equation} 
The vector CGLe (VCGLe) can be viewed as a particular case 
of Eq. (\ref{coupled}) for 
$s=0$: 
\begin{eqnarray}
\partial_t A_\pm 
 =  A_\pm + (1+i b) \nabla^2 A_\pm -
(1+i c) \left( |A_\pm|^2 + (1+i d ) g | A_\mp|^2 \right) A_\pm 
\label{vgle}
\end{eqnarray}
The problem of nonlinear dynamics of a complex vector field 
arises most naturally in the context of nonlinear optics, 
where the order parameter is the electric field envelope  
in the plane normal to the direction of propagation; the 
fields $A_\pm$ can be identified with the two circularly 
polarized waves of opposite sense (Gil, 1993; Pismen, 1994a,b,
Haelterman and Sheppard, 1994, San Miguel, 1995,
Hernandez-Garcia {\it et al}, 1999, Hoyuelos {\it et al}, 1999, 
Byryak {\it et al},  1999).

A distinguished feature of the
VGLE is the  possibility of a transition between two
``phases'', which can be characterized by either ``mixing'' or
``separation'' of two ``superfluids''. Defects (vortices) can exist in both
``superfluids'', and transitions between alternative core structures are
possible (Pismen, 1994, 1999); 
in this sense the real case (VGLe) could be viewed as a toy model
for the 9-component description of superfluid $^3$He, dressed down to 
two components.
Pismen (1994)a,b introduced the notation of ``vector'' and ``scalar''
defects in the VGLe. 
Vector defects have topological charge (and, therefore,  zeros of $A_\pm$ ) in 
both fields, whereas scalar defects have non-zero charge only in one of the 
fields $A_\pm$. 

Simulations of the 
VCGLe  by Hern\'andez-Garc\'ia {\it et al} (1999,2000) 
have shown spiral wave patterns with an exceptionally
rich structure where both separated (but closely packed) ``scalar'' defects 
in the two fields and ``vector'' defects with a common core
could be seen. 

A particularly intriguing possibility, suggested  by Pismen (1994)a, is
the formation of a
bound pair of defects in the two fields, i.e. a vortex ``molecule'' with
dipole structure. 
Aranson and Pismen (2000), have 
shown that such a ``molecule'' requires complex coefficients. 
Analytical calculations were  conducted in the limit of small coupling $g$ 
between
two complex fields. As was shown, the  interaction
between a well-separated pair of defects in two different fields is
always long-range (power-like), in contrast to the  interaction
between defects in the same field which falls off exponentially as in a
single CGLE (Aranson, Kramer, Weber, 1993a). In a certain region of
parameters of the VCGLe stable rotating bound states of
two defects -- a ``vortex molecule'' - are found.

\subsection{Complex oscillatory media}
Two-dimensional oscillatory media exhibit a wide variety of wave phenomena, 
including spiral waves, phase and defect turbulence etc. 
The CGLe describes the dynamics of oscillatory media in the vicinity of 
a primary Hopf 
bifurcation. 
Brunnet {\it et al} (1994), Goryachev and Kapral (1996) 
have shown that spiral waves may also exist in  systems with more 
complex local dynamics,  e.g.  period-doubling bifurcations 
or  chaos. 
In this situation the rotational symmetry of
 spiral waves may be broken by  line synchronization defects. 

Goryachev {\it et al} (1999) studied transitions to line defect turbulence 
in complex oscillatory media supporting spiral waves. 
Several types of line defect turbulence 
were found in a system where the local dynamics is described by a
chaotic R\"ossler oscillator. Such complex periodic spirals and line-defect 
turbulence were observed experimentally in chemical systems 
by Park and Lee (1999).


\section{Concluding Remarks}

In this work  we attempted to overview  a wide variety of dynamic
phenomena described by the CGLe in one, two and three dimensions.  
The CGLe exhibits  in many respects similar behavior in all dimensions, 
e.g. active and passive 
defects, distinct chaotic states, convective and 
absolute instabilities etc. Surprisingly, quantitative aspects of 
instabilities and transitions are  different. 
In particular, the core (acceleration) 
instability of 1D Nozaki-Bekki holes, 2D spirals 
and 3D vortex filaments 
has different manifestations: Nozaki-Bekki holes undergo a stationary 
instability, spirals exhibit unsaturated Hopf bifurcations 
and 3D vortex filaments show a supercritical Hopf bifurcation. 
One observes a general trend in the region of occurrence: 1D defects 
have the smallest stability domain, 2D spirals are stable in a much wider 
range of parameters and the stability domain again decreases for 3D vortex 
filaments.

The unique combination of all these features in one equation 
stimulates continuous interest in this topic in a broad scientific
community. The insights obtained from the CGLe over the last decades 
had an enormous impact on the physics of non-equilibrium systems, 
pattern formation, biophysics etc. and will be  usefull 
for further progress in the physics of complex systems.

Let us discuss briefly some open problems in the CGLe world.

\begin{itemize}
\item
Description of turbulent states  in all dimensions.
Although considerable work was done to identify the stability
limits and transition lines between various turbulent states in
the CGLe, surprisingly little is known on the statistical properties
of these states.
The main obstacle is the lack of appropriate analytical tools, since the
traditional methods of statistical physics are not suitable for the
description of spatio-temporal chaos.
\item
The structure and statistical properties of the
vortex glass.  The ``glassy'' properties of this state
(such as power-like decay of correlations, hierarchy of
the relaxation times etc.) are not exposed yet.
\item
The revolutionary development in computers will make possible
detailed
investigation of the dynamics in  3D CGLe.
\end{itemize}

Hopefully, the last two question will be elaborated on during the
next decade. However, the problem of spatio-temporal chaos and turbulence
will possibly require considerable  time and effort.

\section*{ Acknowledgements}

We would like to thanks all our colleagues who
have assisted us in preparing this review over the
number of years: Len Pismen, Hugues Cha\'te, Pierre Coullet,  Alan Newell,
Hermann Riecke, Yves Pomeau, Herbert Levine, David Kessler,
Lev Tsimring, Pierre Hohenberg, Helmut Brand, Harry Swinney,
Jerry Gollub, Valerii Vinokur, Alan Bishop, Victor
Steinberg,  Wim van Saarloos, Martin van Hecke, Marcus B\"ar, Igor Mitkov,
Maxi Sam Miguel, Eberhard Bodenschatz, Andreas Weber, Mikhail
Rabinovich, Michael Cross, Alex Abrikosov, Nikolay Kopnin,
Werner Pesch, Lutz Brusch, Alessandro Torcini, Antonio Politi,
Walter Zimmermann and many others.

This work was supported by the United States of America  Department
of Energy at Argonne National Laboratory under contracts W-31-109-ENG-38.

\newpage

\end{document}